\pdfoutput=1
\documentclass[letterpaper,twocolumn,10pt]{article}
\usepackage[table]{xcolor}
\usepackage{usenix-2020-09}

\usepackage{makecell}
\usepackage{hyperref}
\usepackage{cleveref}
\usepackage{multirow}
\usepackage{graphicx}
\usepackage{subcaption}
\usepackage{caption}
\usepackage{booktabs}
\usepackage{array}
\usepackage{amssymb}
\usepackage{tabularx}

\usepackage{enumitem}

\begin{document}

%
\title{The (Un)suitability of Passwords and Password Managers in Virtual Reality}

\author{
{\rm Emiram Kablo}\\
\href{mailto:ekablo@mail.upb.de}{ekablo@mail.upb.de}\\
Paderborn University
\and
{\rm Yorick Last}\\
\href{mailto:ylast@mail.upb.de}{ylast@mail.upb.de}\\
Paderborn University
\and
{\rm Patricia Arias Cabarcos}\\
\href{mailto:pac@mail.upb.de}{pac@mail.upb.de}\\
Paderborn University\\KASTEL Research Labs
\and
{\rm Melanie Volkamer}\\
\href{mailto:melanie.volkamer@kit.edu}{melanie.volkamer@kit.edu}\\
Karlsruhe Institute of Technology
}

\maketitle

\begin{abstract}
As Virtual Reality (VR) expands into fields like healthcare and education, ensuring secure and user-friendly authentication becomes essential. Traditional password entry methods in VR are cumbersome and insecure, making password managers (PMs) a potential solution. To explore this field, we conducted a user study (n=126 VR users) where participants expressed a strong preference for simpler passwords and showed interest in biometric authentication and password managers. On these grounds, we provide the first in-depth evaluation of PMs in VR. We report findings from 91 cognitive walkthroughs, revealing that while PMs improve usability, they are not yet ready for prime time. Key features like cross-app autofill are missing, and user experiences highlight the need for better solutions. Based on consolidated user views and expert analysis, we make recommendations on how to move forward in improving VR authentication systems, ultimately creating more practical solutions for this growing field.
\end{abstract}

\section{Introduction}
The emerging field of Virtual Reality (VR) is expected to have an increasing impact on our lives. Apart from being used for entertainment, use of virtual reality technology is gaining traction in domains such as military applications \cite{pallavicini2016virtual}, healthcare \cite{moline1997virtual}, industry \cite{ma2011virtual, naranjo2020scoping}, education \cite{Virtualtrip}, and culture \cite{MuseumVR}. Therefore, it is crucial to implement user authentication mechanisms that make sure only authorized individuals have access to their VR services and related data. But the most prevalent authentication method in VR, passwords, has reached a breaking point due to their lack of usability \cite{stephenson2022sok}.
Entering secrets using VR controllers or free-hand interactions is cumbersome, slow, and error-prone, particularly when dealing with complex and lengthy character combinations. From a security perspective, these gestures lack shoulder-surfing resistance, leaving them vulnerable to external observation attacks \cite{lange2024vision, george2017seamless, yu2016exploration}.
 
To approach the VR authentication problem, current research is mostly focused on proposing solutions that leverage VR capabilities to support immersive and user-friendly mechanisms \cite{noah2023proposal, mathis2020knowledge, yu2016exploration, mathis2021fast, riyadh2024usable, lange2024vision}. But the process from research to real-world deployment will take time. As it happened in other scenarios like authentication for web services or smartphones, moving away from passwords completely becomes challenging due to inertia and their easiness to deploy as a quick-and-dirty solution \cite{bonneau2012quest}.

To facilitate this transitory period dealing with passwords in VR, we research: 1) the experience and desires of users for alternative authentication mechanisms; and 2), the suitability of password managers (PMs) to support VR users. We investigate PMs since they are a recognized approach to enhance both security and usability in dealing with passwords \cite{arias2016comparing, pearman2019people, lyastani2018better}. Our survey reveals a growing interest in integrating PMs into VR, a trend increasingly reflected in public discussions on online forums \cite{bitwarden2024, redditrequest}. This demand is expected to grow given that web browsing is a non-negligible use-case in VR, and that passwords remain the most frequent authentication method encountered by VR users, as shown in our survey and previous studies \cite{stephenson2022sok}. Furthermore, with the increasing VR adoption trends 
\cite{statista_vr_stats}, the ability to synchronize passwords across users' personal devices becomes essential. Indeed, the recent releases of VR-specific PMs \cite{lastpass2024} highlight the increasing relevance of PMs in the VR ecosystem. However, no prior studies explore their applicability to VR. 
Our research questions and contributions are summarized below.

\begin{itemize}[leftmargin=0pt,labelsep=0pt]
    \item[] 
   
\textbf{RQ1 What are user experiences with authentication in VR? What are users' password manager behaviors and suggested improvements regarding authentication in VR?}

We conducted a survey-based user study (n=126) with VR users to explore their interactions with authentication in VR, covering used mechanisms and their usability and security perceptions, partially building on Stephenson et al.'s user study \cite{stephenson2022sok} on authentication experiences. We extended the scope to cover user desires for authentication in VR and considering a more diverse sample (less biased towards males and using VR for purposes other than gaming). \\

Our findings confirm previous research \cite{stephenson2022sok} indicating that users  choose simpler passwords for convenience due to the cumbersome input methods in VR. New insights revealed challenges in account creation, with limited authentication options and no automated password generation. Automated password creation and input, could help streamline this process.
Of 126 participants, 34 used PMs for VR accounts, while 67\% of non-users considered adopting a PM on their headset to simplify logins.

    \item[] 
    \textbf{RQ2 How suitable are current password manager solutions for VR as evaluated by experts?}
    
    In response to our survey results highlighting demand, we conduct the first comprehensive review of password managers' suitability and features in VR.
    Specifically, we conduct expert reviews of seven popular PMs on two VR headsets, including the few managers available that are specifically designed for VR, and covering standalone, browser-based, extension-based, and web-based versions. We analyzed the usability through 91 cognitive walkthroughs (7 PMs across 13 tasks) \cite{lazar2017research}, heuristic evaluations using Nielsen's rules \cite{nielsenheuristics, nielsen1990heuristic}, and VR-specific usability principles \cite{murtza2017heuristic}. We further expanded the evaluation with the VR-UDAS (Usability, Deployability, Accessibility, and Security) framework, incorporating 20 criteria from \cite{stephenson2022sok} and 9 additional security criteria from \cite{bonneau2012quest}.
    
    Password managers in VR reduce the need to switch between real and virtual worlds by eliminating the hassle of recalling passwords. Compared to traditional VR password input \cite{stephenson2022sok}, they better meet key usability criteria but face significant limitations. The only standalone VR PM lacks features like autofill, password editing, and secure configuration. No PM could autofill across apps, highlighting the need for better integration. A VR-specific browser-based PM had vulnerabilities exposing usernames and stored site data despite password protection. Common design flaws included nested menus, icon overuse, and poor password strength indicators. 
\end{itemize}

Our findings on users' authentication needs in VR align with the research roadmap proposed in \cite{stephenson2022sok}, confirming that user expectations match the direction of ongoing research. Most participants rely on passwords, often exhibiting risk-prone behaviors due to the VR context. Users recognize the potential of password managers in VR and desire broader availability, especially on devices. We recommend improvements to authentication support, including PMs, to address gaps in the VR authentication research agenda.
 
\section{Related Work}
Virtual reality immerses users in a computer-generated environment via a headset, allowing interaction with virtual objects. Initially focused on entertainment and gaming, VR now extends to culture \cite{MuseumVR}, industrial training \cite{naranjo2020scoping}, education \cite{Virtualtrip}, therapy \cite{bordeleau2022use, difede2022enhancing}, and military applications \cite{pallavicini2016virtual}. 

\textbf{Authentication methods in VR.}
Previous studies proposed modified or novel authentication schemes for VR, developed in the lab and evaluated for performance and usability. However, many of these prototypes were not based on prior user studies and did not address actual user needs. We complement this line of work by surveying users and confirming that ongoing research indeed aligns to user needs. 

In 2017, George et al. \cite{george2017seamless} conducted a lab study with 25 participants to evaluate the usability and security of PINs and unlock patterns as VR authentication methods. They found these traditional mechanisms to be usable and secure when adapted to VR. However, 18\% of 400 entered secrets were guessed correctly within three attempts. Despite the VR display being hidden from observers, vulnerabilities such as visible controller inputs and users' unawareness of physical observation due to immersion were highlighted.

Länge et al. \cite{lange2024vision} analyzed the shoulder-surfing resistance of knowledge-based authentication methods, including their own schemes designed to resist observation attacks. While the classic 4-digit PIN excelled in usability and authentication performance, it proved highly vulnerable to shoulder surfing. Other schemes showed slightly lower performance but greater resilience against observational attacks. 

Sadik and Ruoti \cite{sadik2024large} surveyed 999 participants to examine devices used for password entry and associated challenges. They found that without password managers, users often weaken passwords for ease of entry. Although not focused on VR, a third of participants criticized virtual keyboards. The authors recommend device-aware password generation algorithms to address these challenges.

Our study confirms that traditional authentication methods like PINs and patterns achieve high usability, whereas passwords present challenges for users when typing in virtual environments.

Several studies have proposed novel authentication schemes, predominantly biometric-based \cite{mathis2020knowledge, mathis2020rubikauth, suzuki2023pinchkey, pfeuffer2019behavioural}. Our research confirms that this trend aligns with user needs, as we included a more diverse sample compared to previous studies \cite{jones2021literature}.

In 2021, Jones et al. \cite{jones2021literature} reviewed VR authentication literature, emphasizing studies with user participation. They found a lack of diversity, with more male participants than female, often without gender reporting, and a significant absence of elderly individuals.

Stephenson et al. \cite{stephenson2022sok} evaluated authentication mechanisms in AR and VR, analyzing novel research and existing methods based on threat models, deployability, usability, accessibility, and security. They also surveyed 139 AR/VR users and developers, including 37 VR developers and 132 VR users, to assess current methods. Their findings highlight that traditional passwords are impractical in VR, particularly for users with disabilities, leading to workarounds. This underscores the need for alternative methods, such as sensor-based intuitive authentication or federated logins. Operating systems often lack support for non-password authentication methods, raising concerns about data leakage from sensor access. To ensure secure VR authentication, robust arbitration and protection mechanisms are essential.

Bonneau et al. \cite{bonneau2012quest} developed a framework to evaluate authentication schemes based on Usability, Deployment, and Security criteria. Their analysis of 33 schemes and two desktop password managers highlighted the challenges of replacing passwords, as no scheme fully meets all criteria. Password managers, however, showed significant advantages, aiding users in managing passwords effectively. While the framework has been adapted for mobile authentication \cite{kunda2021survey}, Stephenson et al. \cite{stephenson2022sok} modified it for AR and VR, incorporating accessibility criteria to address the unique interaction methods of these environments. They used it to systematize authentication schemes for AR and VR. We have incorporated the framework for our expert evaluation of password managers in VR.

\textbf{Password Managers.}
While no prior research has specifically addressed the usability and adoption of password managers in virtual reality—and they therefore remain underexplored— many researchers have studied password managers in other contexts such as desktop PCs \cite{arias2016comparing, simmons2021systematization}, mobile phones \cite{seiler2019don}, or educational institutions \cite{mayer2022users}, and among different user groups \cite{pearman2019people, ray2021older}. Pearman et al. \cite{pearman2019people} found built-in manager users prioritize convenience, while separate manager users value security. Mayer et al. \cite{mayer2022users} surveyed 277 US university participants, highlighting usability and convenience as key to password manager adoption.  Ray et al. \cite{ray2021older} noted older adults mistrust cloud storage but adopt password managers due to recommendations. Other studies explored credential audits \cite{kablo2024m, checkups} and setup behaviors \cite{amft2023would}. Munyendo et al. \cite{munyendo2023} examined user behavior when switching password managers. Key drivers included usability issues, costs, and distrust from security breaches, with nearly 20\% citing breaches as a primary reason. Such breaches erode trust in both the affected and other password managers.

Simmons et al. \cite{simmons2021systematization} categorized password manager use cases and design paradigms, using cognitive walkthroughs to evaluate desktop PMs. They found extension-based PMs challenging to set up, browser-based ones lacking functionality and a locking feature, and interface issues like nested menus and excessive icon use. We build on their approach to review seven password managers in VR, comparing findings to gain insights into their usability in this context.

A significant amount of research exists focusing on the security of password managers \cite{fahl2013hey, oesch2020, oesch2021, carr2020revisiting, silver2014password, stock2014protecting}. Fahl et al. \cite{fahl2013hey} analyzed the security of 22 mobile password managers, including LastPass and 1Password, revealing that Android apps could exploit clipboard functionality to access copied credentials without requiring permissions. They found all PMs used Advanced Encryption Standard (AES) for vault protection. Oesch et al. \cite{oesch2020} later showed that generated passwords in browser- and extension-based PMs remain vulnerable to guessing attacks. By 2021, Oesch et al. \cite{oesch2021} reported mobile PMs to be less secure than desktop versions. We evaluated the security of password managers in virtual reality as part of the applied VR-UDAS framework.

\section{User Survey Methodology}\label{sec:method}

\textbf{Study Design.} We have designed a survey to capture VR authentication experiences, and needs, structured in four groups of questions. The full questionnaire is attached in Appendix \ref{app:mainsurvey_questionnaire}, including a section to collect basic demographic information at the end of the survey. The survey covers:

\begin{itemize}
\item \textbf{General experience with authentication in VR.} Participants should state when they had to authenticate, which methods they have used, and how satisfied they were with the process in general (Questions \ref{q1}-\ref{q4}).
\item \textbf{Perceived usability and security.} Participants who used a specific authentication method in VR were asked to rate its ease of use and security on a Likert scale from 1 (very difficult/not secure) to 5 (very easy/very secure). They were also asked to provide reasons for their ratings (Questions \ref{q5}-\ref{q8}).
\item \textbf{Account creation with passwords for VR.} We asked participants about their experience in creating accounts that require passwords on VR devices, focusing on challenges related to password management. Specifically, we aimed to determine if participants chose different secrets for VR accounts (Questions \ref{q9}-\ref{q14}).
\item \textbf{Improvements and password managers.} 
These questions aimed to uncover users' preferences and suggestions for improving VR authentication. We also explored whether VR users would adopt password managers, recognizing their role in securely managing passwords in non-VR scenarios \cite{arias2016comparing, mayer2022users} (Questions \ref{q15}-\ref{q21}).
\end{itemize}

The first two blocks of questions on general authentication experiences, usability, and security are based on Stephenson et al.'s script \cite{stephenson2022sok}. Our extensions provide deeper insights into user-centered authentication improvements and the potential use of password managers for VR users.

\textbf{Recruitment and Pilot.}
Participants were recruited via Prolific \cite{prolific, palan2018prolific} and social media platforms like Reddit, Discord, and Facebook. Eligibility required participants to be over 18, fluent in English, and VR users. A screening questionnaire (\Cref{app:screening_questionnaire}) assessed VR usage and authentication experience, with only eligible participants invited to the main survey. Participants were randomly selected from four VR usage categories (Education, Health and Wellness, Productivity, Entertainment), with 30 per category, ensuring diversity and aligning with qualitative research sample size recommendations \cite{guest2006how}. This approach expanded beyond the typical gaming-focused participant pool. We conducted a pilot test with six participants to refine question formulations and verify completion time estimates. Some participants reported using smartphones with headsets like Google Cardboard \cite{googlecardboard} for basic VR experiences. However, these setups lack the interactivity and capabilities of dedicated VR systems, so we added a question to the screening questionnaire to exclude participants with only smartphone-based VR experience.

\textbf{Survey Data Analysis.} Our analysis was exploratory, focusing on capturing under-explored perspectives in VR authentication, so we limited quantitative data analysis to descriptive statistics without statistical testing. Open-ended questions were analyzed using an iterative, inductive coding approach \cite{inductiveCoding}. One researcher developed a codebook, which was then applied to the dataset by two researchers independently. Inter-coder reliability, assessed using Cohen’s Kappa \cite{cohen1960coefficient}, ranged from 0.71 to 0.93, indicating substantial to almost perfect agreement \cite{mchugh2012interrater}. The final codebook, including Kappa values, is available in \Cref{app:complete_codebook}.

\textbf{Ethics.} The user study was approved by our university's Institutional Review Board (IRB). The survey complied with GDPR regulations \cite{GDPR}, and all data collection was conducted using a LimeSurvey \cite{limesurvey} instance hosted on university servers. Participants were fairly compensated at an hourly rate of 14 EUR, receiving 1 EUR for the 5-minute screening questionnaire and 4 EUR for the main 20-minute survey.
\section{Survey Results}\label{sec:results}

The survey was conducted between June 2023 and April 2024, with 797 participants completing the screening. Of these, 195 (24.5\%) reported never having authenticated. Non-smartphone VR users were asked why they had not authenticated, with common reasons including authentication being unnecessary for the device/app (44\%) or not owning the device (41\%), confirming the hypothesis from Stephenson et al. \cite{stephenson2022sok}. Many devices were owned by friends, family, employers, or businesses like gaming arcades. Additionally, 16\% of users had someone else authenticate for them, likely for convenience or because they didn’t own the account or device. Allowing others to authenticate poses security risks, such as unauthorized access. Interestingly, 4 participants stated that authentication was unnecessary because they owned the device or used it privately. 

After excluding participants who did not use authentication or used smartphones for VR, the sample was reduced to 484. Using the sampling strategy in \Cref{sec:method}, we finalized a group of 126 users, evenly distributed across VR usage categories (see \Cref{tab:metacategories}). Demographics and participant backgrounds are detailed in \Cref{tab:demographic}. Our analysis found no significant differences in users' qualitative feedback across the different categories, as discussed in the following sections.

\subsection{RQ1.1 - VR Authentication Experiences}

\textbf{VR Apps.}\label{sec:authposs}
We asked participants to list their three most frequently used VR apps to understand their authentication exposure. We then tested these apps on VR devices (see \Cref{tab:devices} in the Appendix) to investigate their authentication methods. Most apps require textual passwords, paired accounts, or software-based tokens. For example, YouTube VR on Meta Quest devices requires a 6-symbol code and Google account login on a secondary device, potentially involving password managers or biometrics depending on the device. Authentication in VR browsers varies by website, using methods like passwords, one-time passwords (OTP), single sign-on (SSO), and federated identity systems. Overall, passwords were the most common authentication method, consistent with prior research \cite{stephenson2022sok}.
  
\begin{figure}[ht]
    \centering
     \begin{minipage}[b]{0.45\textwidth}
        \includegraphics[width=\textwidth]
        {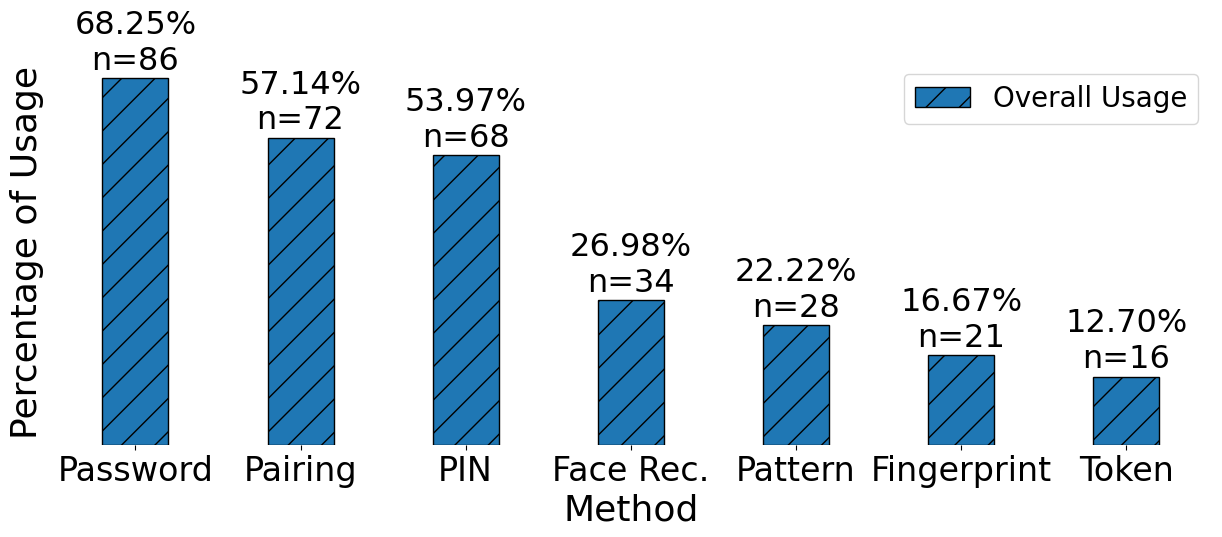}
        \caption{Authentication methods used by participants (in percentages and absolute numbers), sorted from highest to lowest.}
        \label{fig:methods}        
    \end{minipage}
    \hspace{4pt}
    \begin{minipage}[b]{0.45\textwidth}
        \includegraphics[width=\textwidth]{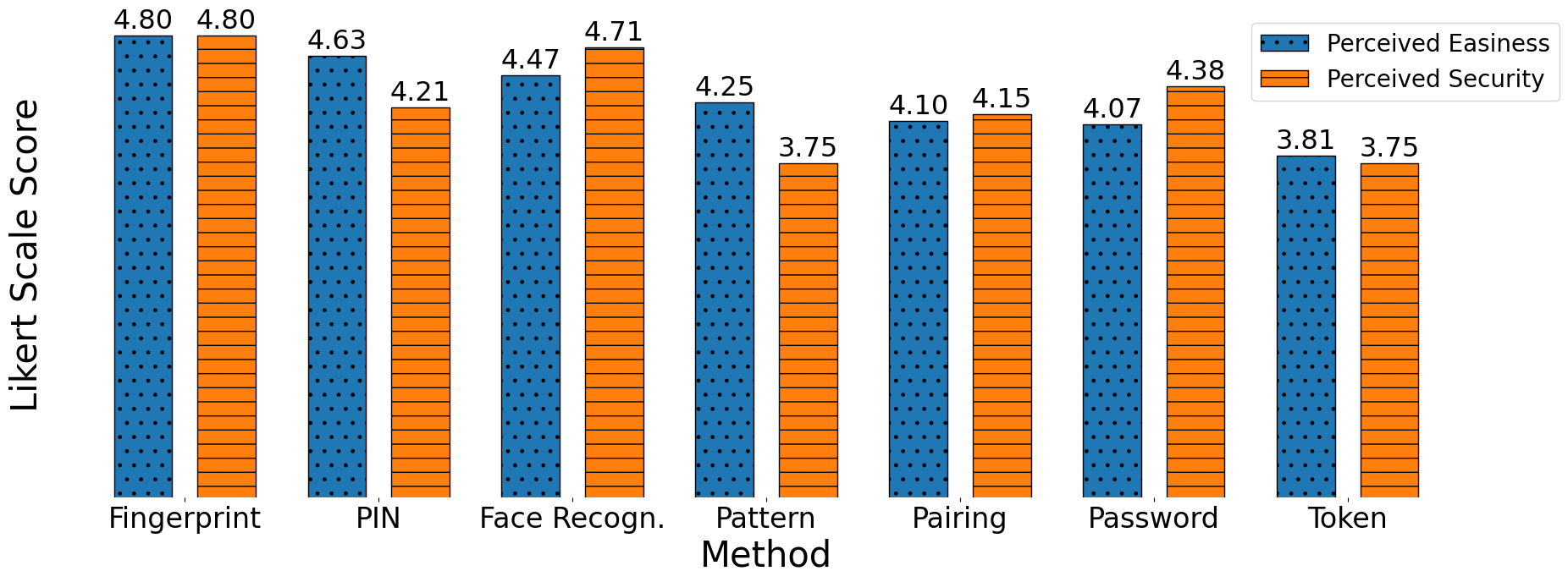}
        \caption{Perceived ease and security of authentication methods rated by participants on a Likert scale from 1 (very hard to use/not secure) to 5 (very easy to use/very secure), sorted by ease scores (highest to lowest).}
        \label{fig:overall}
    \end{minipage}   
   
\end{figure}

\textit{Authentication Exposure and Methods Used.} From the 126 participants, 80\% reported that authentication was required during the initial device setup. Additionally, 55\% mentioned having to authenticate either while setting up or after opening an app for the first time. 29\% percent needed to authenticate before making a purchase, 17\% had to authenticate each time they opened an app, and 13\% required authentication before every device usage. 
The most used authentication method  in VR  are passwords, which were used by 68\% of {the participants}. The second most used method is pairing with another device, followed by entering a PIN.
The frequency of used methods is shown in {\Cref{fig:methods}}. {Some} respondents reported using the mechanisms face and fingerprint recognition, which are not directly available in VR headsets but may be part of the authentication flow when device pairing or token-based authentication involves a secondary device, such as a smartphone. {We have decided to include these results in our study as they offer valuable insights into perceptions of authentication (see \Cref{sec:limitations} for more details).}\\

\textbf{Perceived Usability and Security.}
Next to the frequency of use, \Cref{fig:overall} illustrates the mean values of the perceived easiness and perceived security for each authentication method our participants used in VR. 
 All methods, except for token-based authentication, are rated as somewhat easy to very easy. The  highest scores are {given to} Fingerprint (m=4.8, SD=0.51), PIN (m=4.63, SD=0.67), and Face Recognition (m=4.47, SD=0.9). The open-text responses revealed that fingerprint authentication was indeed considered easy to execute (n=14) and fast (n=5). PINs were considered easy to input in VR (n=24), easy to remember (n=16), and fast (n=14), according to the open questions. Face recognition was deemed easy due to its automatic tracking (n=11), though some participants reported issues with the enrollment process and lighting conditions. Regarding passwords, 63\% of the participants who used them found the mechanism easy in VR, citing reasons such as familiarity (n=9) and easiness of input (n=12). In contrast, 34\% of participants who used passwords in VR reported difficulties, such as challenges with input (n=19) and remembering complex passwords (n=11). Despite 29 participants being negative, the overall easiness score for passwords was relatively high (m=4.07, SD=1.14).
 Token-based authentication had the lowest perceived easiness score (m=3.81, SD=1.17). While 56\% cited ease of use, the stronger negative scores come from users that encountered  challenges like code expiration and the potential loss of the token. 
 The highest ranked methods regarding security, are fingerprint (m=4.8, SD=0.51), face recognition (m=4.71, SD=0.46) and password (m=4.35, SD=0.92). The high perceived security of fingerprint authentication is attributed to: uniqueness (n=6), the perception that nobody can get access in general (n=5), and participants' belief that the fingerprint cannot get stolen, copied or hacked (n=3). Participants expressed similar concerns about face recognition, with two noting the risk of attacks using photos or AI to clone faces. Regarding passwords, 21 participants felt strong passwords provided very high security, and 10 believed no one could access their accounts. Some felt secure due to storing passwords safely or using unique ones for each account (2 each).\\

\textbf{Account Creation with Passwords.}
Roughly 40\% of the participants, 51 out of 126, had  created an account for or on VR. These accounts were created for various purposes such as gaming (33\%), setting up the device (22\%), or gaining access to specific applications (14\%). Regarding password selection, the average satisfaction with account creation on/for VR is slightly positive (m=3.76, SD=1.01), mostly due to familiarity with the process on  other devices. However, the answers surface important reasons for dissatisfaction, including controller annoyance (n=8), time-consuming process (n=7), difficult process on VR with too many or too complex steps (n=3) and no other authentication options available. In fact, 5 participants reported opting for simpler passwords for convenience, to ease input in VR environments (see \Cref{subsec:easierpw}), and 18 participants indicated that they use the same or similar passwords across all accounts for ease of memorization.

\subsection{RQ1.2 - Password Management and Improvements}
This section analyzes users' password management practices and suggestions for improvement.\\

\textbf{Password Managers.} Out of the 126 respondents, 81\% indicated that they use a password manager in general. The primary drivers behind adoption were convenience, cited by 88.2\%, followed by security at 27.4\%. To be specific, 44.1\% of password manager users found it convenient to eliminate the need to remember passwords, while around 13.7\% respectively mentioned that using a password manager made the login process easier and saved time (11.7\%). In terms of security benefits, the most mentioned aspect for users is the secure storage of passwords, mentioned by 10.8\% of respondents. Other reasons for using a password manager include it being enabled by default or mandated by the individual's company.

The main reasons for not using password managers were trust issues (62.5\%), including concerns about security breaches, fear of losing access, and reluctance to use them for critical accounts. Three participants specifically avoided using PMs for sensitive credentials like banking. Additionally, 29.2\% preferred their own methods, such as memorizing passwords or storing them elsewhere. Other reasons included lack of awareness, cost, and time constraints for setup.

Interestingly, most non-PM users are aware of password managers but choose not to use them due to distrust, often stemming from recent security breaches. Our findings align with prior research on PM adoption and non-adoption \cite{pearman2017let, pearman2019people, mayer2022users, ray2021older, kablo2024m}. On the other hand, one participant expressed concerns about data leaks, which has prompted them to consider adopting password managers.

Regarding the devices used with password managers, 34 individuals reported using them for or on VR systems.  We asked the remaining 92 participants, including password manager non-adopters, if they would consider using a PM for their credentials used in VR systems. In this regard, 67.4\% of participants currently not using a PM for VR are considering adopting one (n=62). Among this group, 66.1\% (41 out of 62) express a preference for using a password manager directly on the VR device. Reasons for this perspective are {easiness (n=12)}, the convenience of automated login processes (n=8), {to save time (n=7)}, for synchronization (n=6), no need to remember (n=5), for convenience (n=4), and as a support for the frustrating password typing processes (n=3). Others (n=3) stated a positive experience from previous PM usage, or like the feature of storing passwords (n=2). Two participants mentioned that a reason was to avoid the need to take the headset off and go back to the main device. Another person claimed interest in using a password manager if it was available on their device. 
21 participants preferred using a password manager for VR authentication on a device outside the VR environment. Reasons included easier use (n=5), greater attention to security on non-VR devices (n=4), better input methods on other devices (n=1), and a perceived increase in security. Two participants expressed trust issues with using a PM directly on the VR headset, while one preferred smartphone authentication, and another felt the VR device didn’t have enough passwords to warrant a PM.

Despite the benefits of password managers, some users remain hesitant due to concerns about recent data breaches (n=10), a preference for personal security practices like memorizing passwords (n=4), or a lack of perceived need for a PM (n=11). One participant mentioned difficulty typing a primary password in VR and expressed a preference for entering individual passwords instead, possibly because they use fewer passwords in VR.\\

\textbf{Suggested Improvements.}
When asked if VR authentication could be improved, 61\% (n=77) of participants responded affirmatively. Of these, 30\% (n=23) expressed interest in biometric solutions, such as iris scanning, fingerprint, or facial recognition, alongside improvements to existing methods like body movement-based authentication. Seven participants specifically mentioned the need for VR-friendly solutions:\\

\small{\textit{``Current methods often rely on traditional inputs like passwords or gestures, which may not fully leverage VR's capabilities. Introducing more immersive and biometric authentication methods tailored to VR could enhance security and user experience.''} - P768}\\

\normalsize
Indeed, 13 participants 
  recognized VR as an innovative technology that is still evolving and thus requires improvement, including the development of novel solutions:\\

\small
    \textit{``Right now, most VR platforms rely on usernames and passwords to verify users' identities, you'd think we could come up with something more innovative.''} - P209\\

\normalsize
Regarding alternatives, enhanced security was important for 15 
participants, who suggested using more secure mechanisms, implementing multifactor authentication, or generally enhancing security measures. 9\% of the participants (n=7) expressed a preference for other alternative authentication methods such as voice-based authentication, token-based authentication, or simply more options beyond passwords. Two participants stated to prefer PINs and patterns because those are easier to input. Additionally, 28 participants highlighted the importance of improving usability. 
 Two participants mentioned frustration, while six emphasized the need for faster authentication. Others reported wishing for increased accuracy for specific methods, better accessibility, and improved input methods or a larger virtual keyboard. Three participants specifically mentioned the importance of integrating password managers into the VR environment. Another participant emphasized the need for greater transparency in the user authentication process in VR.

In summary, participants highlighted the need for improved authentication in VR, emphasizing usability. They are open to novel methods, particularly biometric authentication, as long as they are secure and user-friendly. They recognize VR's unique requirements and support the use of tailored solutions, including password managers. While biometrics are well-researched, the potential of password managers in VR remains underexplored, motivating our expert analysis in Section \ref{sec:reviewmethod}.

\section{Expert Review Methodology}\label{sec:reviewmethod}
To address the survey findings on password manager interest and improving authentication flows, we examined the availability and features of PMs for VR. Our research includes what we believe to be the first usability testing of password managers in a VR environment. We conducted expert reviews to evaluate how well current solutions help users manage passwords in VR. We considered four types of password managers:\\

\noindent\textbf{VR-standalone PM}. A VR-native app, i.e., an application specifically designed for virtual reality environments, which can be downloaded from the official app store and installed for use with the headset.\\
\textbf{Browser-based PM.} Password managers that are directly integrated into VR web browsers (built in).\\
\textbf{Extension-based PM.} Browser extensions that can be downloaded and installed in a VR browser to enable password management functionality.\\
\textbf{Web-based PM.} Password managers accessible through a web interface on any device with a browser. \\


We selected a range of password managers for review, in total seven, including the only VR-standalone manager (LastPass for Meta Quest \cite{lastpassvr}), a browser-based manager (Meta Quest browser password manager), the sole extension-based manager for Meta Quest (LastPass Extension), and four web-based managers (LastPass \cite{lastpasswebtested}, 1Password \cite{1passwordtested}, Dashlane \cite{dashlanetested}, Bitwarden \cite{bitwardentested}), all of which are popular according to rankings \cite{cnet, wiredranking}. We opted for the premium versions of LastPass, 1Password, and Dashlane, as their free versions do not support cross-device use, which is important for most users. We focused on officially available PMs, excluding those requiring sideloading or workarounds. Screenshots of the selected PMs can be found in \Cref{fig:pms_grid} in the Appendix.

\textit{Cognitive Walkthrough tasks.} 
We evaluate the selected managers on a common set of tasks that are relevant when dealing with passwords. We based on the tasks from the analysis of desktop password managers conducted by Simmons et al. in \cite{simmons2021systematization}, tailoring them to our VR scenarios. Specifically, we cover 13 tasks: setup, registering a password, login with stored and non stored credentials, updating a password and login with it, removing a password from the manager, login with the PM installed in a secondary device, creating accounts, (un-)locking the PM, ensuring safe settings, and storing data other than passwords in the PM.  While the reference study evaluated password managers across 17 different tasks, we have removed four tasks which do not fit into our context, namely: login from multiple subdomains, set up and sync on a secondary device, complete a credential audit, and recover access to the manager. We expect these tasks to be more complex or infrequent, and therefore likely to be carried out on the user's primary device(s), rather than in VR. A fully detailed description of the 13 tasks we considered in our analyses can be found in \Cref{app:cw_tasks}.

\textit{Heuristics.} During the cognitive walkthroughs, we used usability heuristics to evaluate the password managers. We applied Nielsen's 10 usability heuristics, which are also relevant for VR \cite{nielsenheuristicsvr}, and selected three additional heuristics from Murtza et al. \cite{murtza2017heuristic} that apply to 2D apps like password managers. The full list of heuristics used in our evaluation is provided in \Cref{app:heuristics}.

\textit{VR-UDAS Framework.} To extend the evaluation of password managers beyond interface usability, we have employed the evaluation framework to assess and compare  authentication mechanisms initially developed in \cite{bonneau2012quest} and subsequently adapted for AR/VR in \cite{stephenson2022sok}. The VR refinement includes 20 Deployability, Usability, Accessibility, and Security criteria focusing on benefits relevant for users as collected in a user survey. 
We added 9 security criteria from the original framework in order to provide a more solid analysis of PMs' security. We call this framework VR-UDAS and apply it to compare how well PMs perform in VR with respect to relevant authentication methods available in VR, passwords and iris scanning, as evaluated in \cite{stephenson2022sok}. The complete description of the framework criteria is provided in \Cref{app:criteria}.

\textit{Review Execution.} 
The expert review regarding PM interfaces was conducted independently by three researchers, following the cognitive walkthrough approach \cite{nielsencw} and Nielsen's recommendations 
\cite{nielsen1990heuristic} for usability heuristic assessment. This approach balances effectiveness and efficiency, as three reviewers are sufficient to identify most usability issues. The evaluators brought diverse expertise with backgrounds on usable security, and VR security and privacy. 
 The team documented heuristic violations and compliance issues during task completion, using the think-aloud method to record verbal comments and VR device screens for post-task analysis.

The second part of the evaluation, followed the VR-UDAS framework criteria for a semi-structured assessment of password managers, based on guidelines from \cite{stephenson2022sok, bonneau2012quest}. The analysis combined insights from the cognitive walkthrough, heuristic evaluation, and hands-on testing with the device. The first two authors independently conducted the review, scoring each scheme separately. They then discussed their findings to address any disagreements. When necessary, a third researcher facilitated the discussion to help reach a consensus.

We conducted the password manager reviews on Meta Quest devices, the most common among our survey participants. Testing was done using the Meta Quest Pro (Meta Horizon OS version 66.0) and Meta Quest 2 (version 68.0) ({see Appendix \ref{app:pmversions} for PM version details}).

\section{Expert Review Results}\label{sec:reviewresults}
In this section, we present the results of the expert reviews  to address our second research question on the suitability of PMs for VR.
\begin{table*}[ht]
    \centering
    \footnotesize 
    \caption{Password manager feature comparison based on cognitive walkthrough tasks.}
    \label{tab:pms}
    \rowcolors{2}{gray!25}{white}  
    \newcolumntype{C}{>{\raggedright\let\newline\\\arraybackslash\hspace{0pt}}m{22mm}}
    
    \begin{tabular}{p{3.2cm}|p{2.7cm}|c|c|c|c|c|c|c}
        \textbf{Task} & \textbf{Feature} & \textbf{\makecell{LastPass\\ VR app¹}} & \textbf{\makecell{Meta\\browser¹}} & \textbf{\makecell{LastPass\\extension²}} & \textbf{\makecell{LastPass\\web³}} & \textbf{\makecell{Bitwarden\\web³}} & \textbf{\makecell{Dashlane \\web³}} & \textbf{\makecell{1Password\\web³}} \\ 
        \hline
        \textbf{1}: First time setup & \makecell[l]{\textcolor{black}{Streamlined Setup}} &  & $\checkmark$ &  & o & o & o & o \\ 
        \textbf{2}: Registering a credential & Adding credentials &  & o & $\checkmark$ & $\checkmark$ & $\checkmark$ & $\checkmark$ & $\checkmark$ \\ 
        \textbf{3}: Login with unstored credential & \makecell[l]{Auto-saving credentials} &  & $\checkmark$ & $\checkmark$ &  &  &  &  \\ 
        \textbf{4}: Updating a credential & Editing credentials & & & $\checkmark$ & $\checkmark$ & $\checkmark$ & $\checkmark$ & $\checkmark$ \\ 
        \textbf{5}: Login with updated credential & \makecell[l]{Auto-updating \\credentials} &  & $\checkmark$ & $\checkmark$ & &  &  & \\ 
        \textbf{6}: Removing a stored credential & Removing credentials &  & $\checkmark$ & $\checkmark$ & $\checkmark$ & $\checkmark$ & $\checkmark$ &$\checkmark$  \\ 
        \textbf{7}: Login with stored credential & Autofill credentials & o & $\checkmark$ & $\checkmark$ & o & o & o & o \\ 
        \textbf{8}: Login without PM & \makecell[l]{Using PM on another \\device} & $\checkmark$ & & $\checkmark$ & $\checkmark$ & $\checkmark$ & $\checkmark$ & $\checkmark$ \\ 
        \textbf{9}: Creating an account & Password generation on account creation & & & $\checkmark$ & o & o & o & o \\
        ~& {\makecell[l]{Strength feedback \\of generated pw }} & & & \textcolor{black}{$\checkmark$} & \textcolor{black}{$\checkmark$} &  & \textcolor{black}{$\checkmark$} & \textcolor{black}{$\checkmark$} \\
        \textbf{10}: Locking PM & Locking &  &  & o & $\checkmark$ & $\checkmark$ &  & $\checkmark$ \\ 
        \textbf{11}: Unlocking PM & Unlocking &  &  & $\checkmark$ & $\checkmark$ & $\checkmark$ &  & $\checkmark$ \\ 
        \textbf{12}: Ensuring safe settings & \makecell[l]{Item access protection} &  & $\checkmark$ & $\checkmark$ & $\checkmark$ & $\checkmark$ & {$\checkmark$} &  \\ 
        \textbf{13}: Storing {\& autofilling} non-credential data & \makecell[l]{Storing {\& autofilling} \\other data} &  & & $\checkmark$ & o & o &o & o \\ 
    \end{tabular}

    \vspace{4mm}
    $\checkmark$ : feature available | o : feature partially available | \textit{empty} : feature not available.    ¹Native application | ²Browser extension | ³Web application
\end{table*}

\subsection{RQ2 - Password Manager Usability in VR}\label{sec:reviewresults_if}
We begin with a comparative overview of supported features in \Cref{tab:pms}, then address critical usability issues that could impact security or efficiency, followed by less significant concerns.\\

\textbf{Supported Features - The Good and the Bad.}
At first glance, we can observe that the only VR-native application, LastPass VR app, offers the fewest features. The app provides a basic view of the vault, allowing users to launch websites and copy credentials. It does not support autofill, editing or adding items, or ensuring secure settings.
The main feature it offers is to manually copy the credentials and paste them into a website or an application for authentication. These credentials should be previously added in another device or through the web interface so they are synchronized and appear in the VR LastPass App. As illustrated in \Cref{fig:flow_lastpass} in the Appendix, this involves considerably more steps compared to the autofill features as seen in \Cref{fig:flow_browser} for the Meta Quest browser PM and in \Cref{fig:flow_extension} for the LastPass browser extension (both in the Appendix).
It is also possible to lock the password manager by using OS-based functionality to lock the app itself, helping mitigate vulnerabilities (see \Cref{subsec:lock}). The app suffered from technical issues, leading to credentials not being shown or updated in the vault, and requiring us to switch to another account for the review. It should also be noted that, as of writing, the app has not been updated since October 2022 (according to data listed in the Meta Quest store).

While all the web-based password managers we reviewed offer similar features up to task 9, Bitwarden is the only one that does not provide feedback on credential strength, Dashlane does not fully feature (un)-locking capabilities and 1Password does not support individual  secure settings for critical credentials. None of the managers provide autofill functionality for other websites, which is unsurprising, as websites typically cannot access content across different browser tabs or domains.

In conclusion, the LastPass browser extension provided the best user experience in VR. All tasks were completed with minimal effort, and the extension included all necessary features. The locking mechanism, which requires activation in settings, was the only partially fulfilled feature. One option (auto logout on browser close) did not work in the VR browser, but the time limit option functioned correctly. The UI was intuitive and easy to navigate. A major limitation of all reviewed managers is the lack of autofill functionality for apps outside the browser, requiring users to manually copy and paste credentials. The Meta Quest browser password manager is an exception, supporting autofill for progressive web apps (PWAs) installed via the Meta Quest store \cite{pwa}. Extending this functionality to non-PWAs would require access to application rendering pipelines, necessitating significant architectural changes by OS developers.\\

\textbf{Usability Issues During Setup.}
The setup process for all password managers, except for the Meta Quest browser manager, was cumbersome. Web-based managers require entering the primary password and confirming authorization on the VR device or primary device, creating additional steps. 1Password further complicates the process by not allowing the primary password field to be unmasked and requiring a 40-character secret key. The flow for installing extensions is misleading: 1Password incorrectly identifies Meta Quest as a Chrome browser, and while LastPass provides a download link, the extension must be found through a non-standard process in settings. Once located, installation is straightforward. LastPass's VR app is easy to find and install but requires typing the primary password. The Meta Quest browser manager is the simplest, requiring no installation or primary password, but is the least secure. It also interacts with other managers, offering to store their primary passwords. Overall, typing complex passwords in VR could lead users to choose weaker secrets or rely on less secure built-in managers, which negatively impacts security \cite{hu2024unmasking}. The interference of multiple password managers and their security implications, seen in non-VR studies \cite{oesch2020, simmons2021systematization}, warrant further research, especially regarding how using less secure managers may undermine the benefits of adopting an external one.\\

\textbf{Usability Issues Related to PM Locking.}
\label{subsec:lock} 
Locking functionality is essential for VR devices, especially when shared, to prevent unauthorized access to credentials. While all managers can benefit from the Meta device's OS-level access controls (app and headset PIN-based protection), this requires extra configuration and does not count as built-in support. Four out of seven managers offer vault locking options within the PM, including time settings for inactivity, browser closure, or a manual lock button. Bitwarden and 1Password have auto-locking enabled by default.

Protecting specific credentials with a primary password in the vault or when autofilling those, adds an extra security layer, even when the main password manager is unlocked. Five out of seven managers provide this feature, with options to protect all credentials or individual ones. LastPass extension, LastPass web, Bitwarden, and Dashlane offer these options. The Meta Quest browser also includes a security feature that locks saved passwords and prompts for a passcode before autofilling credentials.   

The absence of options to secure individual credentials (LastPass App, Bitwarden) reduces flexibility and efficiency, leaving users with higher security needs unsupported. Additionally, password managers with locking features often provide unclear or difficult-to-find documentation on vault protection, leading to suboptimal configurations. For example, Dashlane does not clarify that its lock feature activates only after 5 minutes, which can create mistrust or perceptions of unreliability. Secure defaults and in-context setup documentation are critical to prevent such errors, especially for managers relying on system-wide access controls \cite{seiler2019don}. For instance, while the Meta Quest browser manager allows configuring protection for individual items with a passcode, usernames, email addresses, and website details remain visible in browser settings. Without further safeguards, anyone with browser access can view these details or disable protection. Simmons et al. \cite{simmons2021systematization} noted similar issues in browser-based password managers, where credentials remained accessible even after logging out.\\

\textbf{Password Generation and Feedback Usability Issues.}
Automated password generation is a feature to guarantee the security benefits of having a password manager \cite{lyastani2018better} and one of the most valued functionalities by users \cite{seiler2019don}. Neither the Meta Quest browser manager nor LastPass app support automated password generation, significantly limiting the usability of choosing secure passwords. Users would have to choose passwords themselves (potentially leading to less secure secrets \cite{sadik2024large}) or take additional steps to use external generators (high usability friction). 

All web-based PMs and the LastPass extension include password generation features, though their functionality presents differences in integration and parametrization options. Bitwarden, Dashlane and 1Password offer a generator directly integrated in the credential creation pane. While Bitwarden lacks any customization option in the integrated generator, a more complete tool can be found in advanced settings. Dashlane and 1Password both feature fully integrated and highly customizable generators. Among the possibilities for customization, the user can create ``memorable'', ``easy-to-use'', ``easy-to-say'' passwords, tune the length, and/or exclude special or ambiguous characters. LastPass web and extension offer mixed experiences: the web interface hides the generator in nested settings, while the browser extension conveniently displays a customizable generator directly on the registration form. Therefore, LastPass extension fulfills complete support for password generation. In terms of usable security, for managers that cannot interact with websites, integrated parametrizable passwords are the best option (only covered in 2 PMs). Separate tools can be difficult to locate, require additional steps (find, generate, copy, switch to creation pane, paste) and prevent users from creating strong passwords easily. Similarly, the lack of customization might be negative when it is necessary to comply with specific password policy requirements, leading users to manually edit them in a less secure, more predictable fashion \cite{ur2015added}. 

Password strength meters, often shown as a colored bar with labels, are one method to raise user awareness of weak passwords. Four out of the seven evaluated managers—1Password, LastPass extension, LastPass web, and Dashlane—feature these meters during password generation (see \Cref{tab:pms}). All use the \textit{zxcvbn} library \cite{zxcvbn,lastpasswhite,1password_strength,dashlanewhite} to assess password strength, although the user-facing feedback lacks detailed explanations on what constitutes a strong password.\\

\textbf{General Interface Usability Issues.}
Our examination of heuristic violations highlighted both strengths and weaknesses in password managers. Consistent user feedback, essential for system status visibility, was well-implemented in most tools. Notifications informed users when items were added or modified in the vault, though this feature was absent in 1Password and the Meta Quest browser manager. Helpful features like LastPass’s download progress bar and password strength checkers supported this heuristic by keeping users informed and aiding in secure password selection, as also noted by Simmons et al. \cite{simmons2021systematization}. However, adding explanatory labels to strength checkers could further enhance their usability. On the downside, some managers failed to provide warnings or confirmation prompts before credential deletion, increasing the risk of accidental password removal.

Relying solely on small, unlabeled icons can confuse users, as noted by Simmons et al. \cite{simmons2021systematization}. For example, Bitwarden uses three unlabeled icons next to the password field in the add/edit item window. Their functions (checking password breaches, generating a new password, or counting symbols) become clear only after clicking, which is cumbersome in VR. Hidden features further complicate usability, such as LastPass’s password generator buried under multiple menu layers or Meta Quest browser PM’s credential delete option hidden in a three-dots menu. Deeply nested settings, also criticized by Simmons et al., increase search time. Input assistance is particularly valuable in VR, yet the Meta Quest browser manager and LastPass VR app only store usernames and passwords, unlike web-based PMs, which support additional fields like names and addresses. Only LastPass extension can both store and autofill these fields.

Many managers offer help buttons next to icons that link to documentation pages. While useful, this is less critical in VR, as users likely access the manager on their primary devices and are already familiar with its features.

\begin{table*}[ht]
\captionsetup{justification=centerlast}
\footnotesize
\caption{Evaluation of the password managers based on the framework by Bonneau et al. \cite{bonneau2012quest} and Stephenson et al. \cite{stephenson2022sok}, including {their} evaluations of passwords and iris scanning as incumbent methods for comparison. }
\raggedright
\begin{tabular}{>{\arraybackslash}m{2.7cm}>{\centering\arraybackslash}b{0cm}>{\centering\arraybackslash}b{0cm}>{\centering\arraybackslash}b{0cm} >{\centering\arraybackslash}b{0cm}|>{\centering\arraybackslash}b{0cm}>{\centering\arraybackslash}b{0cm}>{\centering\arraybackslash}b{0cm}>{\centering\arraybackslash}b{0cm}>{\centering\arraybackslash}b{0cm}
>{\centering\arraybackslash}b{0cm}>{\centering\arraybackslash}b{0.2cm}|>{\centering\arraybackslash}b{0cm}>{\centering\arraybackslash}b{0cm}>{\centering\arraybackslash}b{0cm}
>{\centering\arraybackslash}b{0cm}>{\centering\arraybackslash}b{0.2cm}|>{\centering\arraybackslash}b{0cm}>{\centering\arraybackslash}b{0cm}>{\centering\arraybackslash}b{0cm}
>{\centering\arraybackslash}b{0cm}>{\centering\arraybackslash}b{0cm}>{\centering\arraybackslash}b{0cm}>{\centering\arraybackslash}b{0cm}>{\centering\arraybackslash}b{0cm}>{\centering\arraybackslash}b{0cm}>{\centering\arraybackslash}b{0cm}>{\centering\arraybackslash}b{0cm}>{\centering\arraybackslash}b{0cm}>{\centering\arraybackslash}b{0cm}>{\centering\arraybackslash}b{0cm}}
 & \multicolumn{4}{p{1.5cm}}{\textbf{Deployability}} & \multicolumn{7}{c}{\textbf{Usability}} & \multicolumn{5}{c}{\textbf{Accessibility}} & \multicolumn{13}{c}{\textbf{Security}} \\ 
\cline{2-30}

\multicolumn{1}{l}{\textbf{Password Manager}}& \multicolumn{1}{p{0cm}}{\scriptsize\rotatebox{65}{OS-Supported}} & \multicolumn{1}{p{0cm}}{\scriptsize\rotatebox{65}{Platform-Agnostic}} & \multicolumn{1}{p{0cm}}{\scriptsize\rotatebox{65}{Mature-for-VR}} & \multicolumn{1}{p{0cm}}{\scriptsize\rotatebox{65}{Low-Power-Consumption} } & \multicolumn{1}{p{0cm}}{\scriptsize\rotatebox{65}{Efficient-to-Use}} & \multicolumn{1}{p{0cm}}{\scriptsize\rotatebox{65}{Physically-Effortless}} & \multicolumn{1}{p{0cm}}{\scriptsize\rotatebox{65}{Memorywise-Effortless}} & \multicolumn{1}{p{0cm}}{\scriptsize\rotatebox{65}{Easy-to-Learn}} & \multicolumn{1}{p{0cm}}{\scriptsize\rotatebox{65}{Nothing-to-Carry}} & \multicolumn{1}{p{0cm}}{\scriptsize\rotatebox{65}{Infrequent-Errors}} & \multicolumn{1}{p{0cm}}{\scriptsize\rotatebox{65}{Acceptable-in-Public}} & \multicolumn{1}{p{0cm}}{\scriptsize\rotatebox{65}{Accessible-Visual}} & \multicolumn{1}{p{0cm}}{\scriptsize\rotatebox{65}{Accessible-Hearing}} & \multicolumn{1}{p{0cm}}{\scriptsize\rotatebox{65}{Accessible-Speech}} & \multicolumn{1}{p{0cm}}{\scriptsize\rotatebox{65}{Accessible-Mobility}} & \multicolumn{1}{p{0cm}}{\scriptsize\rotatebox{65}{Accessible-Cognitive}} & \multicolumn{1}{p{0cm}}{\scriptsize\rotatebox{65}{Resilient-to-Physical-Observation}} & \multicolumn{1}{p{0cm}}{\scriptsize\rotatebox{65}{Protects-User-Privacy}} & \multicolumn{1}{p{0cm}}{\scriptsize\rotatebox{65}{Multi-Factor}} & \multicolumn{1}{p{0cm}}{\scriptsize\rotatebox{65}{ {Resilient-to-Targeted-Impersonation}}}& \multicolumn{1}{p{0cm}}{\scriptsize\rotatebox{65}{ {Resilient-to-Internal-Observation}}}& \multicolumn{1}{p{0cm}}{\scriptsize\rotatebox{65}{ {Resilient-to-Leaks-from-Other-Verifiers }}}&\multicolumn{1}{p{0cm}}{\scriptsize\rotatebox{65}{ {Resilient-to-Phishing}}}&\multicolumn{1}{p{0cm}}{\rotatebox{65}{ {Resilient-to-Theft}}}&\multicolumn{1}{p{0cm}}{\scriptsize\rotatebox{65}{ {No-Trusted-Third-Party}}}&\multicolumn{1}{p{0cm}}{\scriptsize\rotatebox{65}{ {Requiring-Explicit-Consent}}}&\multicolumn{1}{p{0cm}}{\scriptsize\rotatebox{65}{ {Unlinkable}}}&\multicolumn{1}{p{0cm}}{\scriptsize\rotatebox{65}{ {Resilient-to-Throttled-Guessing}}}&\multicolumn{1}{p{0cm}}{\scriptsize\rotatebox{65}{ {Resilient-to-Unthrottled-Guessing}}}\\
\hline
LastPass VR app & $-$ &  {$\circ$} &  {$\circ$} & $\bullet$ & $\circ$ & $-$ & $\circ$ & $\bullet$ & $\bullet$ & $\bullet$ & $\bullet$ & $-$ & $\bullet$ & $\bullet$&  {$\circ$} &  {$\circ$} & $-$&$\bullet$ &$-$&$-$ &$-$&$\circ$& $-$ & $-$ &$-$& $\bullet$& $\bullet$ & $\bullet$ & $\circ$ \\
Meta Quest browser PM & $\bullet$ & $-$ & $\bullet$ & $\bullet$ & $\bullet$ & $-$ & $\bullet$ & $\bullet$ & $\bullet$ & $\bullet$ & $\bullet$ & $-$ & $\bullet$ & $\bullet$&  {$\circ$} &$\circ$ &$-$ & $\bullet$ &$-$ & $-$ & $-$&$-$& $\bullet$& $-$&$\bullet$& $\bullet$& $\bullet$ & $-$ & $-$ \\
LastPass extension & $-$ &  {$\bullet$} &  {$\bullet$} & $\bullet$ & $\bullet$ & $-$ & $\circ$ & $\bullet$ & $\bullet$ & $\bullet$ & $\bullet$ & $-$ & $\bullet$ &$\bullet$ &  {$\circ$}&  {$\circ$} & $-$ & $\bullet$ & $-$& $-$&$-$ & $\circ$& $\bullet$ & $\circ$&$-$& $\bullet$& $\bullet$ & $\bullet$ & $\circ$ \\
LastPass web & $-$ & $\bullet$ & $-$ & $\bullet$ & $\circ$ & $-$ & $\circ$ & $\bullet$ & $\bullet$ & $\bullet$ & $\bullet$ & $-$ & $\bullet$ & $\bullet$ &  {$\circ$}&$-$ & $-$& $\bullet$&$-$& $-$&$-$&$\circ$& $-$ & $\circ$&$-$& $\bullet$& $\bullet$ & $\bullet$ & $\circ$ \\
Bitwarden web& $-$ & $\bullet$ & $-$ & $\bullet$ & $\circ$ & $-$ & $\circ$ & $\bullet$ & $\bullet$ & $\bullet$ & $\bullet$ & $-$ & $\bullet$ & $\bullet$ &  {$\circ$}&$-$ & $-$& $\bullet$&$-$& $-$&$-$&$\circ$& $-$ & $\bullet$&$\bullet$& $\bullet$& $\bullet$ & $\bullet$ & $\circ$ \\
Dashlane web & $-$ & $\bullet$ & $-$ & $\bullet$ & $\circ$ & $-$ & $\circ$ & $\bullet$ & $\bullet$ & $\bullet$ & $\bullet$ & $-$ & $\bullet$ & $\bullet$ &  {$\circ$}&$-$& $-$& $\bullet$&$-$& $-$&$-$&$\circ$& $-$ & $\circ$&$\circ$& $\bullet$& $\bullet$ & $-$ & $\circ$ \\
1Password web & $-$ & $\bullet$ & $-$ & $\bullet$ & $\circ$ & $-$ & $\circ$ & $\bullet$ & $\bullet$ & $\bullet$ & $\bullet$ & $-$ & $\bullet$ & $\bullet$ &  {$\circ$}&$-$ & $-$& $\bullet$&$-$& $-$&$-$&$\circ$& $-$ & $\bullet$ & $\circ$& $\bullet$& $\bullet$ & $\bullet$ & $\circ$ \\
\midrule
\midrule
Password  {\cite{stephenson2022sok, bonneau2012quest}} & $\bullet$ & $\bullet$ & $\bullet$ & $\bullet$ & $-$ & $-$ & $-$ & $\bullet$ & $\bullet$ & $-$ & $-$ & $-$ & $\bullet$ & $\bullet$ & $-$&$-$ &$-$ & $\bullet$&$-$& $\circ$&$-$&$-$ &$-$& $\bullet$& $\bullet$& $\bullet$& $\bullet$ & $-$ & $-$ \\
Iris Scan  {\cite{stephenson2022sok, bonneau2012quest}} & $\bullet$ & $-$ & $\circ$ & $\bullet$ & $\bullet$ & $\bullet$ & $\bullet$ & $\bullet$ & $\bullet$ & $\bullet$ & $\bullet$ & $\circ$ & $\bullet$ & $\bullet$ & $\bullet$&$\bullet$ &$\bullet$ & $\circ$&$-$& $-$&$-$ &$-$ &$-$& $-$& $\bullet$& $\circ$& $-$ & $\bullet$ & $\bullet$ \\ \\
\end{tabular}

\makebox[\textwidth][c]{
$\bullet$ = PM fulfills criterion; $\circ$ = PM quasi-fulfills criterion; $-$ = PM does not fulfill criterion.}
\label{table:framework}
\end{table*}

\subsection{RQ2 - Extended Analysis: UDAS in VR}\label{sub:framework}
In the following sections, we present key findings from the VR-UDAS framework evaluation, organized by \textbf{usability, deployability, accessibility,} and \textbf{security}, with a summary of criteria fulfillment in \Cref{table:framework}.\\

\textbf{Usability.}
Autofill is the most crucial usability feature for VR password manager users, so we assess \textit{Efficient-to-Use} based on its availability. As shown in \Cref{tab:pms}, only the Meta Quest browser PM and LastPass extension support autofill, enabling logins in 1–3 seconds, depending on whether an additional click is required. Other PMs offer copy-paste functionality, simplifying the process but remaining only \textit{quasi-Efficient-to-Use}, as noted in our cognitive walkthrough. This method takes about 30 seconds, involving copying the username and password separately between applications (see \Cref{fig:flow_lastpass} for steps). Manual input, depending on complexity and memory, takes 40–60 seconds. These estimates, based on our walkthroughs, may vary with system familiarity.

None of the password managers were \textit{Physically-Effortless} due to input methods relying on hand or head movements. All PMs require users to remember a primary password, making them \textit{quasi-Memorywise-Effortless}, except for the Meta Quest browser PM. This manager does not require a primary password unless a passcode is set, as noted in our cognitive walkthrough, making it \textit{Memorywise-Effortless}, at the cost of less protection. All password managers are \textit{Easy-to-Learn}, as determined through cognitive walkthroughs and heuristic evaluations, despite the absence of user studies. They are also convenient, with \textit{Nothing-to-Carry}, \textit{Infrequent-Errors} during authentication, and being \textit{Acceptable-in-Public}, as they avoid unusual movements—apart from entering the primary password, which we assume has already been input.\\

\textbf{Deployability.}
The Meta Quest browser password manager meets the \textit{OS-Supported} criterion as it is pre-installed on the device and requires no additional installation. The LastPass VR app, LastPass extension, and Meta Quest browser password manager are specifically designed for VR. Due to earlier technical issues (\Cref{sec:reviewresults_if}), LastPass VR is classified as \textit{quasi-Mature-for-VR}, whereas the Meta Quest browser password manager {and the LastPass extension} are fully \textit{Mature-for-VR}. The Meta Quest browser PM is not \textit{Platform-Agnostic}, as it is unavailable on other platforms, while the rest can be used on other devices and browsers. The LastPass VR app is \textit{quasi-Platform-Agnostic} since it supports other platforms but not this specific version. Iris scanning is designated as \textit{quasi-Mature-for-VR} because it is currently available for mixed and augmented reality \cite{appleID, stephenson2022sok} but not yet for VR devices. All options meet the \textit{Low-Power-Consumption} requirement, as running an app within the VR environment requires minimal power.\\

\textbf{Accessibility.}
All password managers perform the same with regard to accessibility criteria in the framework. None are \textit{Accessible-Visual}, as users need to see the interface. {All PMs} quasi-fulfill \textit{Accessible-Mobility}, as interacting with them requires physical movement of controllers, head, or hands {but no more than general interaction with the device would}. This criterion is similar to the criterion \textit{Physical-Effortless} but allows for speech-interaction. Additionally, {most} password managers are {not} \textit{Accessible-Cognitive}, as users must remember a password. The Meta Quest browser password manager partially fulfills this criterion with an optional passcode. Similarly, the LastPass VR app and extension require the primary password only once, keeping users logged in thereafter.

In response to the improvement question on our survey, two participants requested more accessible options but did not provide specific details.\\

\textbf{Security.}
The primary password is evaluated as not \textit{Resilient-to-Guessing}, in line with findings by Bonneau et al. \cite{bonneau2012quest} and Stephenson et al. \cite{stephenson2022sok}, which show that users often struggle to choose secure passwords. In the Meta Quest browser manager, using a secret (4-16 digit passcode) is optional, and even when enabled, it is not \textit{Resilient-to-Guessing}. This vulnerability also applies to \textit{Resilient-to-Physical-Observation}, as passwords or passcodes entered on a virtual keyboard are vulnerable to shoulder surfing. However, since passwords are not considered sensitive user data, all password managers meet the \textit{Protects-User-Privacy} criterion. While most managers can store non-credential information (task 13 of the cognitive walkthrough), this is not necessary for the PM to function.

We interpret the \textit{Multi-Factor} criterion as applying an additional security layer to the initial login into an existing PM account. It is fulfilled if enforced by the PM. While web password managers offer biometric or token-based authentication as additional factors, these options are unavailable in VR, reducing security when transitioning to this platform.

Following the argument in \cite{bonneau2012quest}, all password managers are classified as not \textit{Resilient-to-Targeted-Impersonation}. Attackers could target the primary password, and no multifactor option is enforced by default. The Meta Quest browser manager has no primary password requirement, so an attacker with physical access could obtain credentials if no additional access controls are set. None of the PMs are \textit{Resilient-to-Internal-Observation}, as attackers can intercept user input within the VR device or eavesdrop on cleartext communication between prover and verifier. 

Among the examined PMs, the Meta Quest Browser PM and LastPass extension store the correct URL with credentials, ensuring autofill only occurs on legitimate websites. Assuming sites follow best practices and implement TLS \cite{bonneau2012quest}, we consider these PMs \textit{Resilient-to-Phishing}. In contrast, PMs without autofill rely on users to manually store URLs and initiate the login process. While this reduces phishing risk if followed correctly, it places responsibility on users to act consistently. Therefore, we classify these PMs as not \textit{Resilient-to-Phishing}, as their effectiveness depends on user behavior.
We classify the Meta Quest browser PM and LastPass app, which offer optional system-wide locking, as not \textit{Resilient-to-Theft}. In contrast, Bitwarden and 1Password auto-lock vaults by default after inactivity, making them \textit{Resilient-to-Theft} in case the VR device is stolen. Other PMs do not auto-lock by default but allow configuration or session closure, classifying them as \textit{quasi-Resilient-to-Theft}. 

All PMs with cloud-stored vaults use client-side, "zero-knowledge" encryption, eliminating reliance on third parties. However, verifying this is difficult for closed-source clients like LastPass and 1Password. Given LastPass's history of security incidents \cite{lastpass2015, lastpass2021, lastpass2022}, we align with Bonneau et al. \cite{bonneau2012quest} in exercising caution with its client.
All PMs meet \textit{Requiring-Explicit-Consent}, as initiating authentication requires user input of a secret. The Meta Quest browser PM also fulfills this criterion, requiring user action (e.g., pressing submit) to complete autofill. Following Bonneau et al. \cite{bonneau2012quest}, we assume sites salt passwords, ensuring they remain \textit{unlinkable}. Consequently, password managers meet this criterion, providing security comparable to passwords.

We evaluated the criteria \textit{Resilient-to-Throttled-Guessing} and \textit{Resilient-to-Unthrottled-Guessing} considering the primary passwords. Three PMs—LastPass, Bitwarden, and 1Password—throttle guessing by limiting password entries after repeated failures. LastPass locks the manager for 5 minutes after 8 failed attempts, while Bitwarden and 1Password slow submission processes without strict limits. Dashlane allows unlimited attempts. Resilience to unthrottled guessing depends on the strength of primary passwords. Except for the Meta Browser PM (no primary password) and 1Password (only a length requirement), all tested PMs enforce password rules and detect weak passwords. However, only randomly generated passwords ensure resilience. Among the tested PMs, only LastPass suggests using a randomly generated primary password, though it is not mandatory.

\section{Discussion}
In this section, we discuss our findings, comparing them to previous works and pointing at future works and limitations.

\subsection{VR Users Prefer Simpler Passwords}\label{subsec:easierpw}
Our study found that a significant number of individuals tend to choose simpler passwords for convenience, particularly due to input limitations when using passwords in VR systems. Across all survey questions, 14 participants reported in the open questions that they chose weak secrets. Five participants opted for a simple PIN for its ease and speed of input, as well as easy memorization. Similarly, four participants chose passwords for these reasons and for easier error correction, while two participants mentioned using unlock patterns for the same purposes. Five participants reported opting for an easier password when creating an account, used for VR purposes, so it is easier to enter.\\

\small    \textit{``I tend to use much simpler passwords with no symbols and zero capital letters and numbers, because its a slog typing in a long automatically generated password.''} - P581\\

\normalsize
Particularly, users are encountering challenges in inputting information with controllers, citing difficulties like extended input times and complexity of pointing at individual keys on a virtual keyboard, leading to imprecise interactions. Choosing weak secrets for the sake of convenience poses significant security risks, highlighting the urgent necessity for more user-friendly authentication solutions in VR environments. These findings align with Stephenson et al.'s \cite{stephenson2022sok} survey results, and we confirm them in a more diverse population, as their sample was predominantly male (92\%). Interestingly, we observed that our participants found passwords to be more usable (m=4.07) than those in their study on the same scale (m=2.9) \cite{stephenson2022sok}. Some positive responses mentioned using autofill features, either through a PM in VR or on a smartphone. This suggests participants assessed the ease of inputting passwords based on the entire VR authentication process, including other devices and software. 
This signals that PMs (with autofill) already do a good job in usability, and it is especially noticeable that more than 80\% of the
users in our sample are PM adopters; and that it is difficult to understand perceived usability when the authentication
steps and configurations might vary (see Section \ref{sec:limitations}). Familiarity with these schemes may also contribute to high usability ratings, as noted in prior studies \cite{zimmermann2020password}.

\subsection{Promising but Not Yet Ready} 
Our analysis shows that VR-specific password managers, like the Meta Quest browser PM and LastPass VR app, lack key features such as password generation, credential editing, and non-credential data storage. Overall, extension-based and web-based PMs demonstrated better usability. 
Comparing our evaluation of PMs with the assessment of manually entered passwords in VR using the UDAS framework by Stephenson et al. \cite{stephenson2022sok}, we find that password managers in VR, despite current limitations, still fulfill important usability criteria, whereas traditional passwords fall short in these areas (see \Cref{table:framework}). {The only usability aspects that passwords fulfill are \textit{Easy-to-Learn}, due to their familiarity and non-complexity (in the sense that people are ubiquitously exposed to passwords and know how they work), and \textit{Nothing-to-Carry}.}

PMs offer a viable interim or complementary solution for managing passwords in VR authentication flows. The immersive nature of VR and reliance on spatial interactions demand minimizing users’ physical effort. PMs with autofill capabilities reduce effort by eliminating the need to switch between real and virtual worlds, avoiding secondary devices for password entry—an inconvenient and time-consuming process. This also enhances security by enabling strong, auto-generated passwords without manual entry, mitigating risks like shoulder-surfing or keystroke inference from VR hand motions \cite{yang2024can}. Security requirements for PMs in VR largely mirror those for 2D interfaces \cite{oesch2020}. Frequent primary password entry could be replaced by passwordless authentication methods. In the following, we discuss this and other actionable recommendations to improve PM integration in VR towards achieving a seamless and secure authentication experience, based on our findings.

\subsection{Improving VR Authentication}\label{subsec:opp}
Our participants, as in previous research \cite{stephenson2022sok}, show a preference for the introduction of biometrics in VR. They also acknowledge the (currently unexploited) potential of VR to provide tailored solutions to prove their identity. In this regard, we believe future work could focus on synergistic approaches, for example leveraging VR-specific biometrics to complement password managers. This could allow for a seamless experience where the user wears their device, gets in by biometric proof (e.g., through iris scan {\cite{appleID, john2019eyeveil, microsoft_iris}}, eye-tracking \cite{luo2020oculock}, EEG recognition \cite{arias2023performance, fallahi2023brainnet, rose2023overcoming}) and unlocks access to their PM autofilling any required passwords. This will solve the usability and security limitation of having to type a primary password in VR and current issues with locking/unlocking the PM safely (see Section \ref{sec:reviewresults}).

Seamless functionality requires better integration of password managers across apps and browsers. Currently, iris scanning is the only deployed biometric mechanism in mixed reality. Due to its maturity, it could be an immediate complementary solution for VR password managers.

Moreover, passkeys, cryptographic key pairs based on the FIDO standard \cite{fido2020, fido_web}, are gaining traction as a passwordless authentication method. They are being adopted as a primary password replacement, an alternative to One-Time Passwords for two-factor authentication, and for website logins. Many password managers support passkeys on desktop and mobile devices \cite{lastpass_passkey, dashlane_passkey, 1password_passkey, bitwarden_passkey}, but they are not yet available for VR—a feature worth integrating.

Risk-based (also: adaptive) authentication, which adjusts authentication requirements based on the risk of a user's login attempt, could help address frequent credential entry in VR. By evaluating contextual factors, it selects the best authentication method, balancing security and usability \cite{arias2019survey}. Investigating this approach in VR could streamline authentication while maintaining security, reducing the effort required, and aligning with participants' preferences for more authentication options.

\subsection{Recommendations and Future Work}
\textbf{Researchers.} Our survey on VR authentication experiences highlighted the limitations of online surveys in capturing the complexities of user interactions. VR authentication involves diverse devices, capabilities, and account connections, which can confuse participants about the specific flow being referenced. Surveys lack the granularity to distinguish these variations, leading to potential misconceptions. To address this, we recommend supplementing surveys with observational studies to better understand contextual factors and user behaviors, providing deeper insights into VR authentication challenges.

User studies are essential to explore usability challenges of password managers in VR, identify user needs, and test current solutions. System testing based on expert insights may not reflect typical user behavior \cite{nielsen1990heuristic}. However, our study tasks and the usability issues highlighted in our expert review provide a foundation for designing such studies. Metrics like System Usability Scale (SUS) scores \cite{brooke1996sus} and mental load assessments during PM tasks can help benchmark usability. Future research could examine how usability challenges contribute to insecure behaviors, offering actionable recommendations. Key areas for exploration include the impact of using multiple PMs in VR on security and usability, as well as addressing accessibility issues to improve the overall user experience. Continuing research on increasing awareness and adoption of password managers is as well crucial, as positive experiences with PMs can encourage users to try them on other devices.

\textbf{Developers and Designers.}
Users highlighted the need for more secure and user-friendly authentication solutions, particularly biometrics, which are already common on other devices. Apple’s introduction of iris scanning in the Apple Vision Pro, known as Apple Optic ID \cite{appleID}, could offer significant benefits to other VR devices equipped with similar sensors and cameras.

Our survey found that VR users often allow others to authenticate on shared or non-owned devices, risking exposure of personal information. We recommend secure solutions for shared devices, like easy multiprofile management or a guest mode with quick profile switching. These options offer better security without requiring a full multi-account setup, which may be too effortful for short sessions \cite{multiaccount}.

Our evaluation of password managers in VR revealed missing key features like autofill, password generation, and credential management. These features are crucial to address usability challenges, such as switching between devices and frequent, cumbersome inputs. None of the password managers could autofill across the system, including within applications or the app store. Platform providers should integrate password managers into the OS, enabling both first and third-party managers to autofill password fields across applications, similar to solutions in iOS and Android \cite{apple_password_autofill, android_password_autofill}. Additionally, password managers should recognize when users are in VR and offer tailored configurations to balance usability and security. Currently, VR-specific managers replicate 2D interfaces, missing the unique affordances of VR. 
 We recommend exploring how novel interactions, feedback mechanisms, and visualizations can be designed to better support users in
effectively managing passwords in VR. Credential audits could serve as a starting point, as their implementation and
visualization on other devices are often perceived as overwhelming, leading to user inaction \cite{kablo2024m}.

\subsection{Limitations}\label{sec:limitations}
\textbf{Survey.}  Our study relies on self-reported data from the survey, which is a limitation. To improve data quality, we applied attention checks to filter out invalid responses.

In asking participants about VR authentication mechanisms, we provided a list of both current and research-based options. We noticed some misconceptions, such as participants reporting the use of eye-tracking (n=36) or movement-based authentication (n=36), despite these mechanisms not being supported by their reported headsets. Participants may have confused eye and movement tracking, available for other features, with authentication. We excluded these mechanisms from our usability and security comparison, but exploring how biometric data in VR is perceived as identifying could be an interesting future research topic.

Some participants reported using fingerprint and face recognition for VR, likely referring to their use of secondary devices like smartphones during the authentication flow. We believe participants rated these biometric methods based on their experiences with non-VR devices, offering insights into whether such features are desired for VR. 

We purposely sampled participants to explore diverse perspectives, capping the number per category. This may have underrepresented dominant use cases, such as Entertainment, and overrepresented smaller categories like Health and Wellness. The overlap in usage—83\% of participants reported using VR for gaming—may have contributed to the lack of qualitative differences observed across domains. This overlap highlights the challenge of disentangling user experiences in multifaceted contexts and may limit the ability to draw domain-specific insights. Future studies could refine this approach to better capture these nuances.

\textbf{On-device testing.} Our app analysis was conducted in July 2024. Since apps and app stores are regularly updated, authentication methods and interfaces may change over time.
We tested password managers on Meta devices, the most popular among our survey participants \cite{vrpopular}, but did not assess them on other standalone devices like PICO, which could provide additional insights. Moreover, evaluators' personal use of managers may introduce bias. It is also to mention that the extended VR-UDAS framework analysis is based on heuristics and assumptions, and as noted by Bonneau et al. \cite{bonneau2012quest}, weighting criteria by context could offer a more accurate evaluation. Additionally, the ratings do not account for factors like vulnerability to specific attacks.

\section{Conclusion}
The focus of current research lies predominantly on proposing solutions to improve the VR authentication experience, without fully considering real-world user needs. Understanding user experiences, and integrating their suggested improvements, is vital before implementing novel VR-specific authentication methods. To address this, we conducted a user study to gain a comprehensive understanding of users' challenges and needs in VR authentication. Our contribution fills a gap in the literature by explicitly examining user experiences with authentication, account creation, and the adoption of password managers in VR. Following this, we evaluated seven password managers to assess how well they support password management within VR. Insights from our study can guide researchers and practitioners in advancing authentication practices in VR, thereby ultimately enhancing user security and safety in virtual environments.

\bibliographystyle{plain}
\bibliography{references}

\begin{thebibliography}{10}

\bibitem{1password_strength}
1Password.
\newblock 1password password quality algorithm.
\newblock \url{https://1password.community/discussion/130158/1password-password-quality-algorithm}, 2022.
\newblock Visited on 21/11/2024.

\bibitem{1passwordtested}
{1Password}.
\newblock 1password.
\newblock \url{https://my.1password.com/}, 2024.
\newblock Accessed: 2025-03-20.

\bibitem{1password_passkey}
1Password.
\newblock Passkeys in 1password: The future of passwordless authentication.
\newblock \url{https://1password.com/product/passkeys}, 2024.
\newblock Visited on 27/11/2024.

\bibitem{amft2023would}
Sabrina Amft, Sandra H{\"o}ltervennhoff, Nicolas Huaman, Yasemin Acar, and Sascha Fahl.
\newblock " would you give the same priority to the bank and a game? i do $\{$Not!$\}$" exploring credential management strategies and obstacles during password manager setup.
\newblock In {\em Nineteenth Symposium on Usable Privacy and Security (SOUPS 2023)}, pages 171--190, 2023.

\bibitem{arias2023performance}
Patricia Arias-Cabarcos, Matin Fallahi, Thilo Habrich, Karen Schulze, Christian Becker, and Thorsten Strufe.
\newblock Performance and usability evaluation of brainwave authentication techniques with consumer devices.
\newblock {\em ACM Transactions on Privacy and Security}, 26(3):1--36, 2023.

\bibitem{arias2019survey}
Patricia Arias-Cabarcos, Christian Krupitzer, and Christian Becker.
\newblock A survey on adaptive authentication.
\newblock {\em ACM Computing Surveys (CSUR)}, 52(4):1--30, 2019.

\bibitem{arias2016comparing}
Patricia Arias-Cabarcos, Andr{\'e}s Mar{\'\i}n, Diego Palacios, Florina Almen{\'a}rez, and Daniel D{\'\i}az-S{\'a}nchez.
\newblock Comparing password management software: toward usable and secure enterprise authentication.
\newblock {\em IT Professional}, 18(5):34--40, 2016.

\bibitem{bitwarden_passkey}
Bitwarden.
\newblock Passkey management - manage your passkeys securely.
\newblock \url{https://bitwarden.com/passwordless-passkeys/}, 2024.
\newblock Visited on 27/11/2024.

\bibitem{bitwarden2024}
{Bitwarden Community}.
\newblock Meta quest virtual reality support.
\newblock \url{https://community.bitwarden.com/t/meta-quest-virtual-reality-support/45303}, 2024.
\newblock Accessed: 2024-12-09.

\bibitem{bitwardentested}
{Bitwarden Inc.}
\newblock Bitwarden web vault.
\newblock \url{https://vault.bitwarden.eu}, 2024.
\newblock Accessed: 2025-03-20.

\bibitem{bonneau2012quest}
Joseph Bonneau, Cormac Herley, Paul~C Van~Oorschot, and Frank Stajano.
\newblock The quest to replace passwords: A framework for comparative evaluation of web authentication schemes.
\newblock In {\em 2012 IEEE symposium on security and privacy}, pages 553--567. IEEE, 2012.

\bibitem{bordeleau2022use}
Martine Bordeleau, Alexander Stamenkovic, Pier-Alexandre Tardif, and James Thomas.
\newblock The use of virtual reality in back pain rehabilitation: a systematic review and meta-analysis.
\newblock {\em The Journal of Pain}, 23(2):175--195, 2022.

\bibitem{brooke1996sus}
J~Brooke.
\newblock Sus: A quick and dirty usability scale.
\newblock {\em Usability Evaluation in Industry}, 1996.

\bibitem{MuseumVR}
Rebecca Carlsson.
\newblock How virtual reality is bringing historical sites to life.
\newblock \url{https://www.museumnext.com/article/how-virtual-reality-is-bringing-historical \\ -sites-to-life/}, 2023.
\newblock Accessed: 2023-11-17.

\bibitem{carr2020revisiting}
Michael Carr and Siamak~F Shahandashti.
\newblock Revisiting security vulnerabilities in commercial password managers.
\newblock In {\em ICT Systems Security and Privacy Protection: 35th IFIP TC 11 International Conference, SEC 2020, Maribor, Slovenia, September 21--23, 2020, Proceedings 35}, pages 265--279. Springer, 2020.

\bibitem{cohen1960coefficient}
Jacob Cohen.
\newblock A coefficient of agreement for nominal scales.
\newblock {\em Educational and psychological measurement}, 20(1):37--46, 1960.

\bibitem{zxcvbn}
Dropbox~Inc. Daniel Lowe~Wheeler.
\newblock zxcvbn tests.
\newblock \url{https://lowe.github.io/tryzxcvbn/}, 2016.
\newblock Visited on 21/11/2024.

\bibitem{dashlane_passkey}
Dashlane.
\newblock Passwordless login for your dashlane account.
\newblock \url{https://support.dashlane.com/hc/en-us/articles/10975547141266-Passwordless-login-for-your-Dashlane-account}, 2024.
\newblock Visited on 27/11/2024.

\bibitem{dashlanetested}
{Dashlane Inc.}
\newblock Dashlane.
\newblock \url{https://app.dashlane.com/}, 2024.
\newblock Accessed: 2025-03-20.

\bibitem{difede2022enhancing}
JoAnn Difede, Barbara~O Rothbaum, Albert~A Rizzo, Katarzyna Wyka, Lisa Spielman, Christopher Reist, Michael~J Roy, Tanja Jovanovic, Seth~D Norrholm, Judith Cukor, et~al.
\newblock Enhancing exposure therapy for posttraumatic stress disorder (ptsd): a randomized clinical trial of virtual reality and imaginal exposure with a cognitive enhancer.
\newblock {\em Translational Psychiatry}, 12(1):299, 2022.

\bibitem{vrpopular}
Joseph D'Souza.
\newblock Virtual reality headset statistics by per unit sales, geography, shipments and usages.
\newblock \url{https://www.coolest-gadgets.com/vr-headset-statistics}, 2024.
\newblock Visited on 22/08/2024.

\bibitem{GDPR}
EU.
\newblock General data protection regulation, 2016.
\newblock Available at: \url{https://gdpr-info.eu/} (accessed: 15.09.2022).

\bibitem{fahl2013hey}
Sascha Fahl, Marian Harbach, Marten Oltrogge, Thomas Muders, and Matthew Smith.
\newblock Hey, you, get off of my clipboard: On how usability trumps security in android password managers.
\newblock In {\em Financial Cryptography and Data Security: 17th International Conference, FC 2013, Okinawa, Japan, April 1-5, 2013, Revised Selected Papers 17}, pages 144--161. Springer, 2013.

\bibitem{fallahi2023brainnet}
Matin Fallahi, Thorsten Strufe, and Patricia Arias-Cabarcos.
\newblock Brainnet: Improving brainwave-based biometric recognition with siamese networks.
\newblock In {\em 2023 IEEE International Conference on Pervasive Computing and Communications (PerCom)}, pages 53--60. IEEE, 2023.

\bibitem{fido_web}
fido Alliance.
\newblock Passkeys.
\newblock \url{https://fidoalliance.org/passkeys/}, 2024.
\newblock Visited on 27/11/2024.

\bibitem{lastpass2021}
Sergiu Gatlan.
\newblock Lastpass users warned their master passwords are compromised.
\newblock \url{https://www.bleepingcomputer.com/news/security/lastpass-users-warned-their-master-passwords-are-compromised/}, 2021.
\newblock Visited on 14/11/2024.

\bibitem{george2017seamless}
Ceenu George, Mohamed Khamis, Emanuel von Zezschwitz, Marinus Burger, Henri Schmidt, Florian Alt, and Heinrich Hussmann.
\newblock Seamless and secure vr: Adapting and evaluating established authentication systems for virtual reality.
\newblock NDSS, 2017.

\bibitem{fido2020}
Sanam Ghorbani~Lyastani, Michael Schilling, Michaela Neumayr, Michael Backes, and Sven Bugiel.
\newblock Is fido2 the kingslayer of user authentication? a comparative usability study of fido2 passwordless authentication.
\newblock In {\em 2020 IEEE Symposium on Security and Privacy (SP)}, pages 268--285, 2020.

\bibitem{wiredranking}
Scott Gilbertson.
\newblock The best password managers to secure your digital life.
\newblock \url{https://www.wired.com/story/best-password-managers/}, 2024.
\newblock Visited on 16/08/2024.

\bibitem{limesurvey}
LimeSurvey GmbH.
\newblock Limesurvey.
\newblock \url{https://www.limesurvey.org/}, 2024.
\newblock Visited on 25/01/2024.

\bibitem{lastpass2015}
Dan Goodin.
\newblock Hack of cloud-based lastpass exposes hashed master passwords.
\newblock \url{https://arstechnica.com/information-technology/2015/06/hack-of-cloud-based-lastpass-exposes-encrypted-master-passwords/}, 2015.
\newblock Visited on 14/11/2024.

\bibitem{android_password_autofill}
Google.
\newblock Autofill framework.
\newblock \url{https://developer.android.com/identity/autofill}, 2024.

\bibitem{googlecardboard}
Google.
\newblock Google cardboard.
\newblock \url{https://arvr.google.com/cardboard/}, 2024.
\newblock Visited on 30/08/2024.

\bibitem{guest2006how}
Greg Guest, Arwen Bunce, and Laura Johnson.
\newblock How many interviews are enough? an experiment with data saturation and variability.
\newblock {\em Field Methods}, 18(1):59--82, 2006.

\bibitem{hu2024unmasking}
Yuqi Hu, Suood Alroomi, Sena Sahin, and Frank Li.
\newblock Unmasking the security and usability of password masking.
\newblock In {\em Proceedings of the 2024 ACM SIGSAC Conference on Computer and Communications Security (CCS '24)}, page to appear, Salt Lake City, UT, USA, 2024. ACM.
\newblock Distinguished Paper Award.

\bibitem{checkups}
Adryana Hutchinson, Collins~W. Munyendo, Adam~J Aviv, and Peter Mayer.
\newblock An analysis of password managers’ password checkup tools.
\newblock In {\em Extended Abstracts of the 2024 CHI Conference on Human Factors in Computing Systems}, CHI EA '24, New York, NY, USA, 2024. Association for Computing Machinery.

\bibitem{Virtualtrip}
ImmersionVR.
\newblock Vr for education - what is virtual reality learning?
\newblock \url{https://immersionvr.co.uk/about-360vr/vr-for-education/}, 2023.
\newblock Accessed: 2023-11-17.

\bibitem{appleID}
Apple Inc.
\newblock Optic id for apple vision pro.
\newblock \url{https://support.apple.com/en-us/118483}, 2024.
\newblock Visited on 11/11/2024.

\bibitem{apple_password_autofill}
Apple Inc.
\newblock Password {AutoFill}.
\newblock \url{https://developer.apple.com/documentation/security/password-autofill}, 2024.

\bibitem{dashlanewhite}
Dashlane Inc.
\newblock Dashlane’s security principles \& architecture.
\newblock \url{https://www.dashlane.com/resources/dashlane-zero-knowledge-security}, 2024.

\bibitem{john2019eyeveil}
Brendan John, Sanjeev Koppal, and Eakta Jain.
\newblock Eyeveil: degrading iris authentication in eye tracking headsets.
\newblock In {\em Proceedings of the 11th ACM Symposium on Eye Tracking Research \& Applications}, pages 1--5, 2019.

\bibitem{jones2021literature}
John~M Jones, Reyhan Duezguen, Peter Mayer, Melanie Volkamer, and Sanchari Das.
\newblock A literature review on virtual reality authentication.
\newblock In {\em Human Aspects of Information Security and Assurance: 15th IFIP WG 11.12 International Symposium, HAISA 2021, Virtual Event, July 7--9, 2021, Proceedings 15}, pages 189--198. Springer International Publishing, 2021.

\bibitem{nielsenheuristicsvr}
Alita Joyce.
\newblock 10 usability heuristics applied to virtual reality.
\newblock \url{https://www.nngroup.com/articles/usability-heuristics-virtual-reality/}, 2021.
\newblock Visited on 23/08/2024.

\bibitem{kablo2024m}
Emiram Kablo, Katharina Kader, and Patricia Arias-Cabarcos.
\newblock " i'm actually going to go and change these passwords": Analyzing the usability of credential audit interfaces in password managers.
\newblock In {\em Extended Abstracts of the CHI Conference on Human Factors in Computing Systems}, pages 1--13, 2024.

\bibitem{kunda2021survey}
Douglas Kunda and Mumbi Chishimba.
\newblock A survey of android mobile phone authentication schemes.
\newblock {\em Mobile Networks and Applications}, 26(6):2558--2566, 2021.

\bibitem{lange2024vision}
Tobias L{\"a}nge, Philipp Matheis, Reyhan D{\"u}zg{\"u}n, Melanie Volkamer, and Peter Mayer.
\newblock Vision: Towards fully shoulder-surfing resistant and usable authentication for virtual reality, 2024.

\bibitem{lastpasswhite}
LastPass.
\newblock Lastpass security principles.
\newblock \url{https://www.lastpass.com/resources/tools/lastpass-security-principles-technical-whitepaper}, 2024.

\bibitem{lastpass_passkey}
LastPass.
\newblock Start your passwordless authentication journey.
\newblock \url{https://www.lastpass.com/features/passwordless-authentication}, 2024.
\newblock Visited on 27/11/2024.

\bibitem{lastpass2024}
{LastPass Blog}.
\newblock Lastpass on meta quest: Providing simple, fast, and secure logins across the metaverse.
\newblock \url{https://blog.lastpass.com/posts/lastpass-on-meta-quest-providing-simple-fast-and-secure-logins-across-the-metaverse}, 2024.
\newblock Accessed: 2024-12-09.

\bibitem{lastpasswebtested}
{LastPass, Inc.}
\newblock My lastpass vault.
\newblock \url{https://lastpass.com/vault/}, 2024.
\newblock Accessed: 2025-03-20.

\bibitem{lazar2017research}
Jonathan Lazar, Jinjuan~Heidi Feng, and Harry Hochheiser.
\newblock {\em Research methods in human-computer interaction}.
\newblock Morgan Kaufmann, 2017.

\bibitem{luo2020oculock}
Shiqing Luo, Anh Nguyen, Chen Song, Feng Lin, Wenyao Xu, and Zhisheng Yan.
\newblock Oculock: Exploring human visual system for authentication in virtual reality head-mounted display.
\newblock In {\em 2020 Network and Distributed System Security Symposium (NDSS)}, 2020.

\bibitem{lyastani2018better}
Sanam~Ghorbani Lyastani, Michael Schilling, Sascha Fahl, Michael Backes, and Sven Bugiel.
\newblock Better managed than memorized? studying the impact of managers on password strength and reuse.
\newblock In {\em 27th USENIX Security Symposium (USENIX Security 18)}, pages 203--220, 2018.

\bibitem{ma2011virtual}
Dengzhe Ma, J{\"u}rgen Gausemeier, Xiumin Fan, and Michael Grafe.
\newblock {\em Virtual reality \& augmented reality in industry}.
\newblock Springer, 2011.

\bibitem{mathis2020knowledge}
Florian Mathis, Hassan~Ismail Fawaz, and Mohamed Khamis.
\newblock Knowledge-driven biometric authentication in virtual reality.
\newblock In {\em Extended Abstracts of the 2020 CHI Conference on Human Factors in Computing Systems}, pages 1--10, 2020.

\bibitem{mathis2020rubikauth}
Florian Mathis, John Williamson, Kami Vaniea, and Mohamed Khamis.
\newblock Rubikauth: Fast and secure authentication in virtual reality.
\newblock In {\em Extended Abstracts of the 2020 CHI Conference on Human Factors in Computing Systems}, pages 1--9, 2020.

\bibitem{mathis2021fast}
Florian Mathis, John~H Williamson, Kami Vaniea, and Mohamed Khamis.
\newblock Fast and secure authentication in virtual reality using coordinated 3d manipulation and pointing.
\newblock {\em ACM Transactions on Computer-Human Interaction (ToCHI)}, 28(1):1--44, 2021.

\bibitem{mayer2022users}
Peter Mayer, Collins~W Munyendo, Michelle~L Mazurek, and Adam~J Aviv.
\newblock Why users (don't) use password managers at a large educational institution.
\newblock In {\em 31st USENIX Security Symposium (USENIX Security 22)}, pages 1849--1866, 2022.

\bibitem{mchugh2012interrater}
Mary~L McHugh.
\newblock Interrater reliability: the kappa statistic.
\newblock {\em Biochemia medica}, 22(3):276--282, 2012.

\bibitem{multiaccount}
Meta.
\newblock Add additional meta horizon profiles to your meta quest.
\newblock \url{https://www.meta.com/en-gb/help/quest/articles/accounts/multiple-accounts-and-app-sharing/add-additional-accounts-on-oculus-quest-2-or-quest/}, 2024.
\newblock Visited on 02/09/2024.

\bibitem{lastpassvr}
{Meta / LastPass}.
\newblock Lastpass on meta quest.
\newblock \url{https://www.meta.com/experiences/lastpass/8032154120191335/}, 2022.
\newblock Accessed: 2025-03-20.

\bibitem{microsoft_iris}
Microsoft.
\newblock Windows hello for business.
\newblock \url{https://learn.microsoft.com/en-us/windows/security/identity-protection/hello-for-business/}, 2024.
\newblock Visited on 12/11/2024.

\bibitem{inductiveCoding}
Matthew~B Miles and A~Michael Huberman.
\newblock {\em Qualitative data analysis: An expanded sourcebook}.
\newblock Sage, 1994.

\bibitem{moline1997virtual}
Judi Moline.
\newblock Virtual reality for health care: a survey.
\newblock {\em Virtual reality in neuro-psycho-physiology}, pages 3--34, 1997.

\bibitem{munyendo2023}
Collins~W. Munyendo, Peter Mayer, and Adam~J. Aviv.
\newblock "i just stopped using one and started using the other": Motivations, techniques, and challenges when switching password managers.
\newblock CCS '23, page 3123–3137, New York, NY, USA, 2023. Association for Computing Machinery.

\bibitem{murtza2017heuristic}
Rabia Murtza, Stephen Monroe, and Robert~J Youmans.
\newblock Heuristic evaluation for virtual reality systems.
\newblock In {\em Proceedings of the Human Factors and Ergonomics Society Annual Meeting}, volume~61, pages 2067--2071. SAGE Publications Sage CA: Los Angeles, CA, 2017.

\bibitem{naranjo2020scoping}
Jose~E Naranjo, Diego~G Sanchez, Angel Robalino-Lopez, Paola Robalino-Lopez, Andrea Alarcon-Ortiz, and Marcelo~V Garcia.
\newblock A scoping review on virtual reality-based industrial training.
\newblock {\em Applied Sciences}, 10(22):8224, 2020.

\bibitem{nielsenheuristics}
Jakob Nielsen.
\newblock 10 usability heuristics for user interface design.
\newblock \url{https://www.nngroup.com/articles/ten-usability-heuristics/}, 2024.
\newblock Visited on 02/08/2024.

\bibitem{nielsen1990heuristic}
Jakob Nielsen and Rolf Molich.
\newblock Heuristic evaluation of user interfaces.
\newblock In {\em Proceedings of the SIGCHI conference on Human factors in computing systems}, pages 249--256, 1990.

\bibitem{noah2023proposal}
Naheem Noah, Peter Mayer, and Sanchari Das.
\newblock A proposal to study shoulder-surfing resistant authentication for augmented and virtual reality: Replication study in the us.
\newblock In {\em Companion Publication of the 2023 Conference on Computer Supported Cooperative Work and Social Computing}, pages 317--322, 2023.

\bibitem{oesch2021}
Sean Oesch, Anuj Gautam, and Scott Ruoti.
\newblock The emperor’s new autofill framework:a security analysis of autofill on ios and android.
\newblock In {\em Proceedings of the 37th Annual Computer Security Applications Conference}, ACSAC '21, page 996–1010, New York, NY, USA, 2021. Association for Computing Machinery.

\bibitem{oesch2020}
Sean Oesch and Scott Ruoti.
\newblock That was then, this is now: A security evaluation of password generation, storage, and autofill in {Browser-Based} password managers.
\newblock In {\em 29th USENIX Security Symposium (USENIX Security 20)}, pages 2165--2182. USENIX Association, 2020.

\bibitem{palan2018prolific}
Stefan Palan and Christian Schitter.
\newblock Prolific. ac—a subject pool for online experiments.
\newblock {\em Journal of Behavioral and Experimental Finance}, 17:22--27, 2018.

\bibitem{pallavicini2016virtual}
Federica Pallavicini, Luca Argenton, Nicola Toniazzi, Luciana Aceti, and Fabrizia Mantovani.
\newblock Virtual reality applications for stress management training in the military.
\newblock {\em Aerospace medicine and human performance}, 87(12):1021--1030, 2016.

\bibitem{pearman2017let}
Sarah Pearman, Jeremy Thomas, Pardis~Emami Naeini, Hana Habib, Lujo Bauer, Nicolas Christin, Lorrie~Faith Cranor, Serge Egelman, and Alain Forget.
\newblock Let's go in for a closer look: Observing passwords in their natural habitat.
\newblock In Bhavani Thuraisingham, David Evans, Tal Malkin, and Dongyan Xu, editors, {\em Proceedings of the 2017 {ACM} {SIGSAC} Conference on Computer and Communications Security, {CCS} 2017, Dallas, TX, USA, October 30 - November 03, 2017}, pages 295--310. {ACM}, 2017.

\bibitem{pearman2019people}
Sarah Pearman, Shikun~Aerin Zhang, Lujo Bauer, Nicolas Christin, and Lorrie~Faith Cranor.
\newblock Why people (don't) use password managers effectively.
\newblock In {\em Fifteenth Symposium on Usable Privacy and Security (SOUPS 2019)}, pages 319--338, 2019.

\bibitem{pfeuffer2019behavioural}
Ken Pfeuffer, Matthias~J Geiger, Sarah Prange, Lukas Mecke, Daniel Buschek, and Florian Alt.
\newblock Behavioural biometrics in vr: Identifying people from body motion and relations in virtual reality.
\newblock In {\em Proceedings of the 2019 CHI Conference on Human Factors in Computing Systems}, pages 1--12, 2019.

\bibitem{prolific}
Prolific.
\newblock Prolific.
\newblock \url{https://app.prolific.com/}, 2024.
\newblock Visited on 24/03/2024.

\bibitem{pwa}
Meta Quest.
\newblock Get started developing progressive web apps.
\newblock \url{https://developer.oculus.com/pwa/}, 2024.
\newblock Visited on 10/09/2024.

\bibitem{ray2021older}
Hirak Ray, Flynn Wolf, Ravi Kuber, and Adam~J. Aviv.
\newblock Why older adults (don't) use password managers.
\newblock In Michael~D. Bailey and Rachel Greenstadt, editors, {\em 30th {USENIX} Security Symposium, {USENIX} Security 2021, August 11-13, 2021}, pages 73--90. {USENIX} Association, 2021.

\bibitem{redditrequest}
{Reddit Community}.
\newblock 1password in vr.
\newblock \url{https://www.reddit.com/r/1Password/comments/b7p2i0/1password_in_vr/}, 2019.
\newblock Accessed: 2025-03-20.

\bibitem{riyadh2024usable}
HTMA Riyadh, Divyanshu Bhardwaj, Adrian Dabrowski, and Katharina Krombholz.
\newblock Usable authentication in virtual reality: Exploring the usability of pins and gestures.
\newblock In {\em International Conference on Applied Cryptography and Network Security}, pages 412--431. Springer, 2024.

\bibitem{rose2023overcoming}
Markus R{\"o}se, Emiram Kablo, and Patricia Arias-Cabarcos.
\newblock Overcoming theory: Designing brainwave authentication for the real world.
\newblock In {\em Proceedings of the 2023 European Symposium on Usable Security}, pages 175--191, 2023.

\bibitem{sadik2024large}
John Sadik and Scott Ruoti.
\newblock A large-scale survey of password entry practices on non-desktop devices.
\newblock {\em arXiv preprint arXiv:2409.03044}, 2024.

\bibitem{nielsencw}
Kim Salazar.
\newblock Evaluate interface learnability with cognitive walkthroughs.
\newblock \url{https://www.nngroup.com/articles/cognitive-walkthroughs/}, 2022.
\newblock Visited on 16/08/2024.

\bibitem{seiler2019don}
Sunyoung Seiler-Hwang, Patricia Arias-Cabarcos, Andr{\'e}s Mar{\'\i}n, Florina Almenares, Daniel D{\'\i}az-S{\'a}nchez, and Christian Becker.
\newblock " i don't see why i would ever want to use it" analyzing the usability of popular smartphone password managers.
\newblock In {\em Proceedings of the 2019 ACM SIGSAC Conference on Computer and Communications Security}, pages 1937--1953, 2019.

\bibitem{silver2014password}
David Silver, Suman Jana, Dan Boneh, Eric Chen, and Collin Jackson.
\newblock Password managers: Attacks and defenses.
\newblock In {\em 23rd USENIX Security Symposium (USENIX Security 14)}, pages 449--464, 2014.

\bibitem{simmons2021systematization}
James Simmons, Oumar Diallo, Sean Oesch, and Scott Ruoti.
\newblock Systematization of password manageruse cases and design paradigms.
\newblock In {\em Proceedings of the 37th Annual Computer Security Applications Conference}, pages 528--540, 2021.

\bibitem{statista_vr_stats}
Statista.
\newblock Virtual reality (vr) statistics.
\newblock \url{https://www.statista.com/topics/2532/virtual-reality-vr/}, 2022.
\newblock Accessed: 2024-12-09.

\bibitem{stephenson2022sok}
Sophie Stephenson, Bijeeta Pal, Stephen Fan, Earlence Fernandes, Yuhang Zhao, and Rahul Chatterjee.
\newblock Sok: Authentication in augmented and virtual reality.
\newblock In {\em 2022 IEEE Symposium on Security and Privacy (SP)}, pages 267--284. IEEE, 2022.

\bibitem{stock2014protecting}
Ben Stock and Martin Johns.
\newblock Protecting users against xss-based password manager abuse.
\newblock In {\em Proceedings of the 9th ACM symposium on Information, computer and communications security}, pages 183--194, 2014.

\bibitem{suzuki2023pinchkey}
Mei Suzuki, Ryo Iijima, Kazuki Nomoto, Tetsushi Ohki, and Tatsuya Mori.
\newblock Pinchkey: A natural and user-friendly approach to vr user authentication.
\newblock In {\em Proceedings of the 2023 European Symposium on Usable Security}, pages 192--204, 2023.

\bibitem{cnet}
Attila Tomaschek.
\newblock Best password manager in 2024.
\newblock \url{https://www.cnet.com/tech/services-and-software/best-password-manager/}, 2024.
\newblock Visited on 16/08/2024.

\bibitem{lastpass2022}
Karim Toubba.
\newblock Notice of security incident.
\newblock \url{https://blog.lastpass.com/posts/notice-of-recent-security-incident}, 2022.
\newblock Visited on 14/11/2024.

\bibitem{ur2015added}
Blase Ur, Fumiko Noma, Jonathan Bees, Sean~M Segreti, Richard Shay, Lujo Bauer, Nicolas Christin, and Lorrie~Faith Cranor.
\newblock " i added'!'at the end to make it secure": Observing password creation in the lab.
\newblock In {\em Eleventh symposium on usable privacy and security (SOUPS 2015)}, pages 123--140, 2015.

\bibitem{yang2024can}
Zhuolin Yang, Zain Sarwar, Iris Hwang, Ronik Bhaskar, Ben~Y Zhao, and Haitao Zheng.
\newblock Can virtual reality protect users from keystroke inference attacks?
\newblock In {\em 33rd USENIX Security Symposium (USENIX Security 24)}, pages 2725--2742, 2024.

\bibitem{yu2016exploration}
Zhen Yu, Hai-Ning Liang, Charles Fleming, and Ka~Lok Man.
\newblock An exploration of usable authentication mechanisms for virtual reality systems.
\newblock In {\em 2016 IEEE Asia Pacific Conference on Circuits and Systems (APCCAS)}, pages 458--460. IEEE, 2016.

\bibitem{zimmermann2020password}
Verena Zimmermann and Nina Gerber.
\newblock The password is dead, long live the password--a laboratory study on user perceptions of authentication schemes.
\newblock {\em International Journal of Human-Computer Studies}, 133:26--44, 2020.

\end{thebibliography}

\appendix
\section{Surveys}
\subsection{Consent} 
\label{app:screening_consent}
\textbf{Information:}
Thank you for your interest in our research work. In this research project we want to expose authentication challenges and needs that Virtual Reality (VR) users from different domains face. This is a three-part study and this survey will cover the first part. You will not necessarily go through all phases. You will be invited to the next part if selected.

Survey 1 - In this short survey we want to investigate the primary reasons for which Virtual Reality (VR) systems are used as well as identify the fields you are associated with, so that we can better comprehend your perspective. 

Survey 2 - From this base, we would like to conduct a second, larger survey about usable authentication in VR for users from different fields. If selected, you will be asked to complete a survey about your experiences in authenticating on VR devices.

\textbf{Procedure and Participation:}
Survey 1 will take approximately 3-5 minutes to complete. We offer 1 EUR / 0.87 GBP for completing this first survey. 

Survey 2 will take approximately 15-20 minutes to complete. We offer 5 EUR / 4.29 GBP for completing this first survey.

If you have accessed the surveys via Prolific, you will receive a code for your payment. The participation in any phase of this study is voluntary and you can choose to withdraw from the study at any time. Only complete surveys will receive compensation.

\textbf{Data Collection and Processing:}
If you did not access the surveys via Prolific, we would need to collect your email address, so we can eventually contact you for the next part. Regarding the compensation in this case, before payment, we will check whether you have answered the questions completely and attentively. Participants who have not done so will not receive a payout. We would need to collect the following information which are necessary, following the University’s procedure. Personal data that you submit for the payout will not be associated with your decisions in the study. The information will be collected in a separate survey and stored separately. This information will not be used for any other purpose and will be deleted immediately after.
Other than this information, the record of your survey responses does not contain any personally-identifying information about you and please do not enter any personal information, even from others, in the open text fields, unless a specific survey question explicitly asks for it. Art. 4 para. 1 of the General Data Protection Regulation (GDPR) defines personal data as all information about an identifiable individual. A person is identifiable if they can be identified by a name, ID number, location data, online identifier, physical, physiological, genetic, mental, economic, cultural, or social identity.
All data will remain anonymous. The survey will be executed on an GDPR-compliant instance of LimeSurvey, hosted by the University’s server, therefore all data are stored securely on the server. The data will be processed locally using statistical software, and the collection of background data, such as gender, is solely for analyzing statements diversely. No conclusions about individuals will be drawn from the provided information. Results will be disclosed anonymously in tables or graphics, preventing individual recognition. The interviews will be audio-recorded only and the audio data will be stored locally. The data will then be transcribed anonymously and the audios will be removed while the transcriptions are kept in a local storage, which is only accessible to academic researchers.

\textbf{Risks and Benefits:}
Your participation in this study does not involve any risks to you beyond those of everyday life. Taking part in this research study may not benefit you personally, but we may learn new things that could help others.


\subsection{Screening Survey Questionnaire}
\label{app:screening_questionnaire}
We have replicated the following questions for the screening survey from Stephenson et al. \cite{stephenson2022sok}: Q1-Q3, Q5, Q6, Q9, Q12, Q13 while we have added Q4, Q7, Q8, Q10, Q11, and Q14.
\subsubsection{Virtual Reality Usage}
\begin{enumerate}[label=Q\arabic*, start=1]
\item\label{scr_q1} How frequently do you use a VR device?\\
\textcolor[rgb]{0.5,0.5,0.5}{(Answer choices:  $\circ$ Less than 1 time/month $\circ$ 1 - 5 times/month $\circ$ 6 - 10 times/month $\circ$ 11 - 15 times/month $\circ$ more than 15 times/month)}
\item\label{scr_q2} Which VR device do you use? If you use more than one, select the one you use more frequently. \\ \textcolor[rgb]{0.5,0.5,0.5}{(Answer choices: $\circ$ Meta Quest (any model) $\circ$ HTC VIVE (any model) $\circ$ PICO (any model) $\circ$ Valve Index $\circ$ PSVR (any model) $\circ$ HP Reverb (any model) $\circ$ Varjo (any model) $\circ$ VR Glasses for Smartphones (e.g., Google Cardboard) $\circ$ Other: (free text))}
\item\label{scr_q3} Do you own the device?\\  \textcolor[rgb]{0.5,0.5,0.5}{(Answer choices: $\circ$ Yes $\circ$ No, my friend owns it $\circ$ No, my employer owns it $\circ$ No, somebody else owns it (please specify): (free text))}
\item\label{scr_q4} When do you use VR?\\ \textcolor[rgb]{0.5,0.5,0.5}{(Answer choices: $\Box$ At work $\Box$ In my freetime $\Box$ At the institution I'm studying in)}
\item\label{scr_q5} Some activities people are using VR for are listed below. If you use VR for any of those purposes, please indicate it by checking all options that apply to you.\\  \textcolor[rgb]{0.5,0.5,0.5}{(Answer choices: 
$\circ$ Gaming
$\circ$ Web browsing
$\circ$ Watching videos and other media purposes
$\circ$ Fitness
$\circ$ Virtual tours
$\circ$ Metaverse
$\circ$ E-Learning
$\circ$ Therapy sessions
$\circ$ Medical training
$\circ$ Stress management
$\circ$ Industrial training
$\circ$ Flight simulations
$\circ$ Military training
$\circ$ Prototyping and production planning
$\circ$ Architecture, engineering and construction
$\circ$ Attending courses
$\circ$ Meetings/Video conferences)}
\item\label{scr_q6} For what other purposes, if any, do you use VR?\\  \textcolor[rgb]{0.5,0.5,0.5}{(Free text)}
\item\label{scr_q7} Which are the three apps that you use the most on the VR device? Please provide the name and a short description of the app.\\  \textcolor[rgb]{0.5,0.5,0.5}{(Answer choices: $\Box$ App 1: (free text) $\Box$ App 2: (free text) $\Box$ App 3: (free text)}
\item\label{scr_q8} How is your overall experience while using a VR device?\\  \textcolor[rgb]{0.5,0.5,0.5}{(Answer choices: $\circ$ Very satisfying $\Box$ Satisfying $\circ$ Neither satisfying nor unsatisfying $\circ$ Unsatisfying $\circ$ Very unsatisfying)}
\end{enumerate}

\subsubsection{Authentication in VR}
Authentication refers to the act of verifying your identity through various means such as a password, biometric information, a paired device, or other similar techniques.
\begin{enumerate}[label=Q\arabic*, start=9]
    \item\label{scr_q9} Did you ever have to authenticate on a VR device?\\  \textcolor[rgb]{0.5,0.5,0.5}{(Answer choices: $\circ$ Yes $\circ$ No (please specify the reason in the next question))} 
    \item\label{scr_q10} \textit{If Q9 was answered with `No'.} Why did you not have to authenticate?\\  \textcolor[rgb]{0.5,0.5,0.5}{(Free text)}
\end{enumerate}

\subsubsection{Demographics}
\begin{enumerate}[label=Q\arabic*, start=11]
    \item\label{scr_q11} What is your age range?\\ \textcolor[rgb]{0.5,0.5,0.5}{( $\circ$ 18-24 $\circ$ 25-34  $\circ$ 35-44  $\circ$ 45-54 $\circ$ 55-64 $\circ$ 65 or older  $\circ$ Prefer not to answer)}
    \item\label{scr_q12} What is your gender identity?\\
  \textcolor[rgb]{0.5,0.5,0.5}{($\circ$  Female $\circ$  Male $\circ$  Non-binary $\circ$  Prefer not to answer) $\circ$ Self-described as: (free text))}
  \item\label{scr_q13} What is the highest level of school you have completed? \\ \textcolor[rgb]{0.5,0.5,0.5}{($\circ$ Some High-School, no diploma $\circ$ High-School or equivalent $\circ$ Some college, no degree $\circ$ Associate degree $\circ$ Bachelor’s degree $\circ$ Master’s degree  $\circ$  Doctoral degree $\circ$ Prefer not to answer $\circ$ Other: (free text))}
   \item\label{scr_q14} Which of the following best describes your educational background or job field? \\
\textcolor[rgb]{0.5,0.5,0.5}{($\circ$ I have an education in, or work in, the field of computer science, engineering, or IT. $\circ$ I do not have an education in, or work in, the field of computer science, engineering, or IT. $\circ$ Prefer not to answer)}
\item\label{scr_q15} Which of the following best describes the sector you primarily work in? \\  \textcolor[rgb]{0.5,0.5,0.5}{(Answer choices: 
$\circ$ Agriculture, Food and Natural Resources
$\circ$ Architecture and Construction
$\circ$ Arts
$\circ$ Business Management \& Administration
$\circ$ Education \& Training
$\circ$ Finance
$\circ$ Government \& Public Administration
$\circ$ Medicine
$\circ$ Hospitality \& Tourism
$\circ$ Information Technology
$\circ$ Legal
$\circ$ Policing
$\circ$ Military
$\circ$ Manufacturing
$\circ$ Marketing \& Sales
$\circ$ Retail
$\circ$ Science, Technology, Engineering \& Mathematics
$\circ$ Social Sciences
$\circ$ Transportation, Distribution \& Logistics
$\circ$ Other: (free text))}
\end{enumerate}

\subsection{Main survey Questionnaire}
\label{app:mainsurvey_questionnaire}

\subsubsection{General Experience with Authentication in VR}
We have extracted the following questions for the survey from Stephenson et al. \cite{stephenson2022sok}: Q1-Q3, Q5-8, while we have added Q4 and Q9-Q21 ourselves.
\begin{enumerate}[label=Q\arabic*, start=1]
    \item\label{q1} In the previous survey, you stated that you had to authenticate while using a VR device. When did you have to authenticate on the device(s)? \\ \textcolor[rgb]{0.5,0.5,0.5}{(Answer choices: $\Box$ During device setup $\Box$ Every time when opening an app $\Box$ While setting up an app/after opened the first time $\Box$ Before a purchase was made $\Box$ Before every device usage $\Box$ Other: (free text))} 

    \item\label{q2} Please choose the option(s) that describe your experience with the following authentication mechanisms for each item.\\
    \begin{tabular}{p{1.735cm}*{5}{c}}
          & \footnotesize \textbf{\makecell{Don't \\ know}} & \footnotesize\textbf{\makecell{Haven't \\ used it}} & \footnotesize\textbf{\makecell{For VR}} & \footnotesize\textbf{\makecell{Outside\\ VR}} & \\
        \noalign{\hrule height 0.1pt} 
        PIN & $\Box$ & $\Box$ & $\Box$ & $\Box$\\
        Password & $\Box$ & $\Box$ & $\Box$ & $\Box$\\
        Via pattern & $\Box$ & $\Box$ & $\Box$ & $\Box$  \\
        With token & $\Box$ & $\Box$ & $\Box$ & $\Box$  \\
        \makecell[l]{Pairing with \\another device}& $\Box$ & $\Box$ & $\Box$ & $\Box$  \\
        Eye tracking& $\Box$ & $\Box$ & $\Box$ & $\Box$ \\
        Iris scan & $\Box$ & $\Box$ & $\Box$ & $\Box$  \\
        Face recognition & $\Box$ & $\Box$ & $\Box$ & $\Box$  \\
        Fingerprint & $\Box$ & $\Box$ & $\Box$ & $\Box$  \\
        Body movement & $\Box$ & $\Box$ & $\Box$ & $\Box$  \\
    \end{tabular}
    \item\label{q3} Which other method(s) did you use for authenticating in VR?\\ \textcolor[rgb]{0.5,0.5,0.5}{(Free text)}
    \item\label{q4} How satisfied are you with the convenience of the following authentication mechanisms in VR? Please rate each of the items on a scale from 1 to 5 (very unsatisfied to very satisfied).\\
    \begin{tabular}{l*{6}{c}}
          & \textbf{1} & \textbf{2} & \textbf{3} & \textbf{4} & \textbf{5}\\
        \toprule
        Entering PIN & $\Box$ & $\Box$ & $\Box$ & $\Box$ & $\Box$\\
        Entering password & $\Box$ & $\Box$ & $\Box$ & $\Box$ & $\Box$\\
        Via pattern & $\Box$ & $\Box$ & $\Box$ & $\Box$ & $\Box$ \\
        With a token & $\Box$ & $\Box$ & $\Box$ & $\Box$ & $\Box$ \\
        \makecell[l]{Pairing with another device} & $\Box$ & $\Box$ & $\Box$ & $\Box$ & $\Box$ \\
        Eye tracking & $\Box$ & $\Box$ & $\Box$ & $\Box$ & $\Box$ \\
        Iris scan & $\Box$ & $\Box$ & $\Box$ & $\Box$ & $\Box$ \\
        Face recognition & $\Box$ & $\Box$ & $\Box$ & $\Box$ & $\Box$ \\
        Fingerprint & $\Box$ & $\Box$ & $\Box$ & $\Box$ & $\Box$ \\
        Body movement & $\Box$ & $\Box$ & $\Box$ & $\Box$ & $\Box$\\
    \end{tabular}
\end{enumerate}

\subsubsection{Method-related Experiences}

\textit{Questions 5-8 were repeated for each method were the answer for Q2 was `Used for VR'.}
\begin{enumerate}[label=Q\arabic*, start=5]
    \item\label{q5} How easy/hard is it for you to use [method] on the VR device?\\ \textcolor[rgb]{0.5,0.5,0.5}{(Answer choices: $\circ$ Very easy to use $\circ$ Somewhat easy to use $\circ$ Neither easy nor hard $\circ$ Somewhat hard to use $\circ$ Very hard to use)}
    \item\label{q6} Please explain your answer for the above question.\\ \textcolor[rgb]{0.5,0.5,0.5}{(Free text)}
    \item\label{q7} How secure did you feel while using [method] for authentication on the VR device?\\\textcolor[rgb]{0.5,0.5,0.5}{(Answer choices: $\circ$ Very secure $\circ$ Somewhat secure $\circ$ Neither secure nor insecure $\circ$ Somewhat insecure $\circ$ Not secure)}
    \item\label{q8} Please explain your answer for the above question.\\ \textcolor[rgb]{0.5,0.5,0.5}{(Free text)}
\end{enumerate}

\subsubsection{Account creation}
\begin{enumerate}[label=Q\arabic*, start=9]
    \item\label{q9} Did you ever have to create an account or register for an app/website/system that required setting a password on the VR device?  \\ \textcolor[rgb]{0.5,0.5,0.5}{(Answer choices: $\circ$ Yes $\circ$ No)}
    \item\label{q10} For what purpose did you have to create an account?\\ \textcolor[rgb]{0.5,0.5,0.5}{(Free text)} 
    \item\label{q11} Do you create your passwords differently when using VR than in other devices?  \\ \textcolor[rgb]{0.5,0.5,0.5}{(Answer choices: $\circ$ Yes $\circ$ No)}
    \item\label{q12} Please explain your answer for the above question by elaborating the reasons.\\ \textcolor[rgb]{0.5,0.5,0.5}{(Free text)} 
    \item\label{q13} How was your overall experience regarding the account creation process? \\  \textcolor[rgb]{0.5,0.5,0.5}{(Answer choices: $\circ$ Very satisfying $\circ$ Satisfying $\circ$ Neither satisfying nor unsatisfying $\circ$ Unsatisfying $\circ$ Very unsatisfying)}
    \item\label{q14} Please explain your answer for the above question and specify what challenges you encountered when setting up the account. \\ \textcolor[rgb]{0.5,0.5,0.5}{(Free text)} 
\end{enumerate}

\subsubsection{Password Managers}
A Password Manager (PM) could help you authenticating throughout different devices, systems and apps. It stores your passwords and other login credentials and can generate strong passwords. There are external password managers (e.g. LastPass) as well as built-in browser password managers (e.g. Chrome's Passwordmanager) which are able to fill in your credentials automatically wherever needed.

Password managers are tools that can securely handle passwords for you. They can remember your passwords, generate new ones, and even sync them across devices. There are various types of password managers:
\begin{itemize}
    \item If you allow your web browser to save your passwords, you are using your browser’s password manager.
    \item Third-party application password managers are software that you install directly onto your devices or a service you can access on the web.
    \item Your operating system can serve as a password manager as well. For example, the Keychain functionality on MacOS can remember passwords in and out of your browser.
\end{itemize}

\begin{enumerate}[label=Q\arabic*, start=15]
    \item\label{q15} Based on our description, which password managers are you currently using (not only considering VR)? \\ \textcolor[rgb]{0.5,0.5,0.5}{(Answer choices: $\Box$ I don’t use any password manager $\Box$ Password Manager as a third-party application (e.g., Keeper, Dashlane, Lastpass) $\Box$ Browser Password Manager (e.g., Google Chrome, Mozilla Firefox) $\Box$ System Password Manager (e.g., Keychain, Samsung Pass))}
    \item\label{q16} Please name the reasons why or why you are not using password managers. \\ \textcolor[rgb]{0.5,0.5,0.5}{(Free text)} 
    \item\label{q17} For which device(s) are you using a PM? \\ \textcolor[rgb]{0.5,0.5,0.5}{(Answer choices: $\Box$ Smartphone $\Box$ Tablet $\Box$ Laptop/PC $\Box$ VR $\Box$ Other: (Free text))} 
\end{enumerate}

\subsubsection{User Preferences and Needs}
\begin{enumerate}[label=Q\arabic*, start=18]
    \item\label{q18} Do you think the authentication processes in VR can be improved?\\ \textcolor[rgb]{0.5,0.5,0.5}{(Answer choices: $\circ$ Yes $\circ$ No)}
    \item\label{q19} Please explain your answer for the above question. \\ \textcolor[rgb]{0.5,0.5,0.5}{(Free text)} 
    \item\label{q20} \textit{If answer to Q17 was not `VR'.} Would you consider using a Password Manager for Virtual reality? \\ \textcolor[rgb]{0.5,0.5,0.5}{(Answer choices: $\Box$ Yes, I would like to use a PM directly on the VR device $\Box$ Yes, but outside VR on a smartphone or PC $\Box$ No)} 
    \item\label{q21} Please explain your answer for the above question. \\ \textcolor[rgb]{0.5,0.5,0.5}{(Free text)} 
\end{enumerate}

\subsubsection{Follow Up}
\begin{enumerate}[label=Q\arabic*, start=22]
    \item\label{q22} Please let us know if you would be interested in taking part in an interview. The session would take around 30 minutes. \\ \textcolor[rgb]{0.5,0.5,0.5}{$\circ$ Yes $\circ$ No} 
\end{enumerate}


\begin{table*}[htbp]
\caption{Detailed participant demographics and VR background. {Participants could select multiple usage purposes, so percentages represent the proportion of participants who reported using VR for each purpose, and they may add up to more than 100\%.}}
\centering
\footnotesize
\begin{tabular}{@{}ll@{}c||ll@{}c||ll@{}c@{}c}
\toprule
                  &&\textbf{\textit{n = 126}}&&&\textbf{\textit{n = 126}}&&&\textbf{\textit{n = 126}}\\
                  \midrule
                  
\multirow{4}{*}{\rotatebox[origin=c]{90}{\textbf{Gender}}} 
& Male & 65\% & \multirow{6}{*}{\rotatebox[origin=c]{90}{\textbf{Devices}}}  & Meta Quest (any) & 53\% & \multirow{17}{*}{\rotatebox[origin=c]{90}{\textbf{Usage Purposes}}} & Gaming & 83\%\\
& Female& 32\% && PSVR (any) & 17\% && Videos/other Media & 50\% \\
& Non-binary& 2\% &&  HTC Vive (any) &10\%&&Virtual tours & 40\%\\
&No Answer  &1\% & & HP Reverb (any) &9\%&&Fitness &29\% \\
&&&&Other&6\%&& Metaverse & 29\%\\
&&&&Valve Index&5\%&&E-Learning & 27\%\\
\cmidrule{1-6}
\multirow{5}{*}{\rotatebox[origin=c]{90}{\textbf{Age}}}  
                    & 18-24 &31\% &\multirow{5}{*}{\rotatebox[origin=c]{90}{\textbf{Frequency}}}& < once/month & 16\% && Web browsing & 25\%\\
                 &  25-34& 49\% &  & 1-5 times/month & 42\%&&Stress management & 19\%\\                 
                  &  35-44& 14\% && 6-10 times/month & 25\%&& Flight simulation & 10\%\\
                  &  45-54& 4\% && 11-15 times/month & 9\%&&Attending courses & 10\%\\
                  &  55-64& 2\%&& > 15 times/month & 9\%&& Architecture/engineering/construction& 9\%\\
\cmidrule{1-6}
\multirow{6}{*}{\rotatebox[origin=c]{90}{\textbf{Education}}}  
&High school&17\% &\multirow{5}{*}{\rotatebox[origin=c]{90}{\textbf{Ownership}}}&Self & 56\%&& Therapy & 9\%\\
&Bachelor’s degree &43\% &  &By friend & 29\% &&Meetings/conferences & 8\% \\
&Master’s degree&19\% &&By employer & 9\% && Prototyping/production planning & 7\%\\
&Doctorate&2\%&&\makecell[l]{By school/institution}& 4\%&& Industrial training&6\%\\
&Some college&17\%&& Other &3\%&&Medical training & 6\% \\
&Associate degree&2\%&&&&&Military training & 1\%\\
\midrule
\multirow{3}{*}{\rotatebox[origin=c]{90}{\textbf{IT}}}   
&Yes& 53\%\\
&No &45\%\\
&Prefer not to say& 2\%\\
\bottomrule
\end{tabular}
  \label{tab:demographic}
\end{table*}

\begin{table*}[htbp]
\caption{{Number of participants and purposes per usage meta-category.}}
\centering
\footnotesize
\color{black}
\begin{tabular}{@{}llc@{}}
\toprule
\textbf{Meta-Category} & \textbf{Purposes} & \textbf{Participants (n)} \\
\midrule
Education              & Attending courses, E-Learning & 31 \\
\hline
Health and Wellness    & Fitness, Therapy sessions, Stress management & 29 \\
\hline
Productivity           & \makecell[l]{Military training, Industrial training, Medical training,\\ Prototyping and production planning, Architecture/\\engineering and construction, Flight simulation,\\ Meetings/Video conferences} & 36 \\
\hline
Entertainment          & \makecell[l]{Gaming, Watching videos and other media purposes,\\ Virtual tours, Metaverse, Web browsing} & 30 \\
\bottomrule
\end{tabular}
\label{tab:metacategories}
\end{table*}
\color{black}
\begin{table*}[htbp]
\caption{Most used VR apps (reported by users) and their required authentication mechanisms (based on researcher testing). Numbers in brackets indicate the number of mentions. A "-" indicates no authentication required. An asterisk (*) means login is mandatory, while no asterisk indicates login is optional.}
\centering
\footnotesize
\begin{tabular}{@{}ll||ll@{}}
\toprule
\textbf{Meta Quest device} & \textbf{Authentication} & \textbf{PSVR device} & \textbf{Authentication}  \\
\midrule
Beat Saber (17) & -  & YouTube (7) & \makecell[l]{Device pairing} \\
YouTube VR (11) & Token & Beat Saber (5) & - \\
Netflix (4) & *Token/Password & \makecell[l]{Gran Turismo (2)} & - \\
VR Chat (4) & \makecell[l]{*Password/Account pairing} &  &  \\
Google Earth VR (4) & - & & \\
\midrule
\textbf{HTC Vive device} & \textbf{Authentication} & \textbf{HP Reverb device} & \textbf{Authentication} \\
\midrule
YouTube (4) & Password & YouTube (5) & Password \\
Google Earth VR (4) & - & Browser (2) & \makecell[l]{Password/OTP/SSO/Federated login}\\
VR Chat (2) & \makecell[l]{*Password/Account pairing} & &  \\
\bottomrule
\end{tabular}
\label{tab:devices}
\end{table*}

\section{Cognitive Walkthrough Tasks}\label{app:cw_tasks}
The tasks marked with an asterisk (*) have been adapted from the original tasks by Simmons et al. \cite{simmons2021systematization}.
\begin{enumerate}
    \item \textbf{*First time setup:} The evaluator was required to set up the password manager. This included downloading the manager (if applicable), installing it, log in with an existing PM account, and finishing any remaining setup tasks. Evaluators avoided using external documentation, focusing instead on how the manager’s UI supported this use case. 
    \item \textbf{Registering a credential in the manager:} The evaluator needed to register a credential (an account name and password) in the manager. They were free to complete this task in any way they wanted so long as they did not leave the manager’s interface. 
    \item \textbf{Log in with an unstored credential:} The evaluator would log into a test website we created with credentials not already stored in the manager. This would trigger the manager to suggest saving the credentials. The task was complete once the credentials were saved. 
    \item \textbf{Updating a credential in the manager:} The evaluator needed to update a credential previously stored in the manager. They were free to complete this task in any way they wanted so long as they did not leave the manager’s interface.
    \item \textbf{Login with updated credentials:} The evaluator would log into a test website with credentials different than those stored in the manager, this requires the extra task of updating the credentials on the test website before. This would trigger the manager to suggest saving the updated credentials. The task was complete once the updated credentials were saved. 
    \item \textbf{Removing a credential stored in the manager:} The evaluator would use the manager to remove a target credential.
    \item \textbf{Log in with credentials stored in the manager:} The
    evaluator would log into a test website that has associated
    credentials stored in the manager.  The task was complete once the
    evaluator finished the autofill process and was logged into
    the website.
    \item \textbf{*Log in from a VR without a PM:} The evaluator would
    log into a test website using their VR device. This device did not have a manager installed, and the evaluator was not
    allowed to install one. Instead, they would need to view the
    credential in the desktop manager or mobile device, then manually enter it into the VR device.
    \item \textbf{Create an account:} The evaluator would need to create a new online account, making sure to use a generated password. What approach they took in generating the password was left up to the evaluator, though they did examine the default settings and modified them if they wished to.
    \item \textbf{Lock manager:} The evaluator would need to lock the manager.
    \item \textbf{Unlock manager:} The evaluator would need to unlock the manager.
    \item \textbf{*Ensure safe settings:} The evaluator would need to ensure that user interaction was required before credentials
would be autofilled or viewed. This included identifying the appropriate setting and changing it if necessary.
    \item \textbf{Store non-credential data in manager:} The evaluator
would be tasked with storing a phone number and
address in the manager. They would then be asked to enter
this information into a website. They were free to enter
the information using autofill if supported.
\end{enumerate}

\section{Heuristics}\label{app:heuristics}
The 10 usability heuristics by Nielsen \cite{nielsenheuristics}. Heuristics \ref{item:immersion}, \ref{item:glitch}, and \ref{item:switch} were derived from \cite{murtza2017heuristic}.

\begin{enumerate}
    \item \textbf{Visibility of system status:} The design should always keep users informed about what is going on, through appropriate feedback, within a reasonable amount of time. 
    \item \textbf{Match between the system and the real world:} The design should speak the users' language. Use words, phrases, and concepts familiar to the user, rather than internal jargon. Follow real-world conventions, making information appear in a natural and logical order.
    \item \textbf{User control and freedom:} Users often perform actions by mistake. They need a clearly marked "emergency exit" to leave the unwanted action without having to go through an extended process.
    \item \textbf{Consistency and standards:} Users should not have to wonder whether different words, situations, or actions mean the same thing. Follow platform and industry conventions.
    \item \textbf{Error prevention:}
    Good error messages are important, but the best designs carefully prevent problems from occurring in the first place. Either eliminate error-prone conditions, or check for them and present users with a confirmation option before they commit to the action.
    \item \textbf{Recognition rather than recall:}
    Minimize the user's memory load by making elements, actions, and options visible. The user should not have to remember information from one part of the interface to another. Information required to use the design (e.g. field labels or menu items) should be visible or easily retrievable when needed.
    \item \textbf{Flexibility and efficiency of use:}
    Shortcuts — hidden from novice users — may speed up the interaction for the expert user such that the design can cater to both inexperienced and experienced users. Allow users to tailor frequent actions.
    \item \textbf{Aesthetic and minimalist design:}
    Interfaces should not contain information that is irrelevant or rarely needed. Every extra unit of information in an interface competes with the relevant units of information and diminishes their relative visibility.
    \item \textbf{Help Users Recognize, Diagnose, and Recover from Errors:} Error messages should be expressed in plain language (no error codes), precisely indicate the problem, and constructively suggest a solution.
    \item \textbf{Help and Documentation:}
    It’s best if the system doesn’t need any additional explanation. However, it may be necessary to provide documentation to help users understand how to complete their tasks.
    \item \textbf{VR-specific: Immersion:} \label{item:immersion}
    The system should immerse the user in virtual reality, specific to visual realism. Survey respondents reported issues with
    poor screen resolution, low frame rates, lens flare, lack of
    spatial sounds, etc. that all contributed to a less immersive VR
    experience.
    \item \textbf{VR-specific: Glitchiness:} \label{item:glitch}
    The system should promote a streamlined experience by
    keeping systematic glitches low. Survey respondents reported
    issues with glitches, including crashes, errors, high buffering
    times, and other software malfunctions.
    \item \textbf{VR-specific: Minimize switching between actual and virtual world:} \label{item:switch}
    The system should be able to rely on itself for all usage; that
    is, keep all necessary user tasks and information within VR,
    instead of creating tasks that the user may only be able to
    execute when VR headset is taken off or information that can
    only be accessed by taking headset off.
\end{enumerate}
\section{Definitions of Evaluation Criteria}\label{app:criteria}

\textbf{Deployability Criteria:}
\begin{enumerate}
    \item \textbf{OS-Supported:} A PM/method is built into the operating systems of the device, therefore is already installed on the device and no extra step for installation is required.
    \item \textbf{Platform-Agnostic:} A PM/method could also be used on a
    computer or smartphone with no additional hardware.
    For example, a PM/method that requires the use of controllers
    is not platform-agnostic. A PM/method is \textit{quasi-Platform-Agnostic} if another version of the PM/method is available on other platforms.
    \item  {\textbf{Mature-for-VR:}} A PM/method has been implemented and deployed for VR  {for actual authentication purposes beyond research}. A  {method} is \textit{quasi-Mature} if a PM/method is implemented and deployed on other devices/for other platforms, but not exactly this version.
    \item \textbf{Low-Power-Consumption:} A PM/method does not perform
    any type of signal processing. A PM/method may also
    fulfill the criterion if it performs signal processing but
    is proven to have a negligible effect on the battery life
    of the device. 
\end{enumerate}

\textbf{Usability Criteria:}
\begin{enumerate}
    \item \textbf{Efficient-to-Use:} The time the user must spend authenticating with a PM/method is lower than without  {or acceptably short}. The criterion is fulfilled  {for a PM} if autofill-feature is available. A PM is \textit{quasi-Efficient-to-Use} if it offers copy-pasting-credential feature but not autofill.
    \item \textbf{Physically-Effortless:} The authentication process, or interacting with the PM/method does not involve explicit actions requiring physical effort. A PM/method
is \textit{quasi-Physically-Effortless} if the user’s effort is limited
to a single movement, comparable to a button press.
    \item \textbf{Memorywise-Effortless:} Users of the PM/method do not have to remember any secrets at all. A PM/method is \textit{quasi-
    Memorywise-Effortless} if users have to remember one
    secret for everything (such as the  {primary} password).
    \item \textbf{Easy-to-Learn:} A PM/method is familiar and is not complicated to explain.
    \item \textbf{Nothing-to-Carry:} Users do not need to carry an additional physical object (electronic device, mechanical key,
    piece of paper) to use the PM/method. A PM/method is \textit{quasi-Nothing-to-Carry} if the object is one that they would carry everywhere all the time anyway, such
as their mobile phone, but not if it is their computer
(including tablets).
    \item \textbf{Infrequent-Errors:} The task that users must perform to
    log in usually succeeds when performed by a legitimate
    and honest user. In other words, the PM/method is not so
    hard to use or unreliable that genuine users are routinely
    rejected. 
    \item \textbf{Acceptable-in-Public:} A majority of users would feel
    comfortable using the PM/method in any public place. The
    PM/method therefore must not require large, visible actions
    or speaking. If the PM/method only requires gestures the user
    would already do to interact with the device, that also fulfills this criterion. We are not considering the use of the  {primary} password for this scenario but assume it was entered before and the user already has access to the PM in public.
\end{enumerate}

\textbf{Accessibility Criteria:}
\begin{enumerate}
    \item \textbf{Accessible-Visual:} The PM/method does not require a user
to be able to see or the use of eye biometrics.
    \item \textbf{Accessible-Hearing:} The PM/method does not require a user to hear.
    \item \textbf{Accessible-Speech:} The PM/method does not require a user
to speak.
    \item \textbf{Accessible-Mobility:} The PM/method does not require physical movements, besides speech. The use of the PM/method is also either completely passive or requires only movement that the user is likely to already be doing. This is similar to \textit{Physically-Effortless}, but it permits speech interaction instead of requiring any physical button presses.
    \item \textbf{Accessible-Cognitive:} The PM/method does not require the user to memorize a secret or perform other cognitive tasks such as counting. A PM is considered \textit{quasi-Accessible-Cognitive} if  {the user needs to remember a secret once, then stays logged in or if the secret is optional.}
\end{enumerate}

\textbf{Security Criteria:}
\begin{enumerate}

    \item \textbf{Resilient-to-Physical-Observation:} The PM/method does not
    require the use of gestures which would be visible to
    observers. A PM is \textit{quasi-Resilient-to-Physical-Observation} if a secret for accessing the PM or saved passwords is optional and needs to be activated first.
    \item \textbf{Protects-User-Privacy:} The PM/method does not utilize any
sensitive user data. Passwords are not considered sensitive user data.
    \item \textbf{Multi-Factor:} The PM/method involves multiple factors
/authentication layers by design/by default. A PM is considered \textit{quasi-Multi-Factor} if an additional layer is only added when the device or location is unknown.
    \item  {\textbf{Resilient-to-Targeted-Impersonation:} It is not possible for an acquaintance (or skilled investigator) to impersonate a specific user by exploiting
knowledge of personal details (birth date, names of relatives etc.).}
    \item  {\textbf{Resilient-to-Internal-Observation:} An attacker cannot
impersonate a user by intercepting the user’s input from inside the user’s device (e.g., by keylogging
malware) or eavesdropping on the cleartext communication between prover and verifier.}
    \item  {\textbf{Resilient-to-Leaks-from-Other-Verifiers:} Nothing
that a verifier could possibly leak can help an attacker impersonate the user to another verifier.
This penalizes PMs/methods where insider fraud at one provider, or a successful attack on one back-end,
endangers the user’s accounts at other sites.}
    \item  {\textbf{Resilient-to-Phishing:} An attacker who simulates
a valid verifier cannot collect credentials that can later be used to impersonate the user to the actual verifier. This penalizes PMs/methods allowing phishers to get victims to authenticate to lookalike sites and later use the harvested credentials against the genuine sites.}
    \item  {\textbf{Resilient-to-Theft:} If the PM/method uses a physical
object for authentication, the object cannot be used
for authentication by another person who gains possession of it. We still grant \textit{quasi-Resilient-to-
Theft} if the protection is achieved with the modest
strength of a PIN, even if attempts are not ratecontrolled,
because the attack does not easily scale
to many victims.}
    \item  {\textbf{No-Trusted-Third-Party:} The PM/method does not rely
on a trusted third party (other than the prover and the verifier) who could, upon being attacked
or otherwise becoming untrustworthy, compromise the prover’s security or privacy.}
    \item  {\textbf{Requiring-Explicit-Consent:} The authentication
process cannot be started without the explicit consent of the user.}
    \item  {\textbf{Unlinkable:} Even if verifiers collaborate, they cannot determine, based solely on the authenticator, whether the same user is authenticating to both. To rate this benefit, we disregard linkability introduced by other mechanisms (same user ID, same IP address, etc).}
    \item  {\textbf{Resilient-to-Throttled-Guessing:} The resilience of a PM/method against an attacker whose rate of guessing is throttled by the verifier. While user-chosen secrets can be resilient, they need to be of sufficient complexity. Hence, four-digit PINs are not considered resilient. For iris scan-method, we consider the same as in \cite{stephenson2022sok}: According to \cite{microsoft_iris, appleID}, companies impose very strict accuracy requirements for biometrics, making it \textit{Resilient-to-Guessing.}}
    \item  {\textbf{Resilient-to-Unthrottled-Guessing:} The resilience of a PM/method against an attacker whose rate of guessing is \textit{not} throttled by the verifier, e.g. offline brute-force attacks. For passwords, we only consider randomly generated passwords truly resilient, and user-chosen passwords as \textit{quasi-resilient} if these are properly checked for strength (i.e. complexity, vulnerability to dictionary attacks and rainbow tables, etc.).}
\end{enumerate}
\onecolumn
\section{Complete Codebook}\label{app:complete_codebook}
\begin{table*}[htbp]
    \centering
    \footnotesize
    \caption{Themes and Frequencies: Reasons why participants never authenticated. Filtered out smartphone VR users before (n=147).}
    \begin{tabular}{p{2cm}lp{8.5cm}cc}
    \hline
        \textbf{Theme} & \textbf{Code} & \textbf{{Quote Example}}&\textbf{N} & \textbf{\%} \\ \hline
        Ownership & ~ & &61 & 41 \\ 
        ~ & by a friend  & ``It isn't my device, my friend sets it up and I use it.''& 23 &  \\ 
        ~ & not owned  & ``i do not own it''&21 &  \\ 
        ~ & by company  & ``Provided by company demo specialists.'' &7 &  \\ 
        ~ & by institution/school  &``It belongs to my university.'' &5 &  \\ 
        ~ & by employer &``I used VR in my workplace so everytime my work enviroment is setup for me to be used.'' &3 &  \\
        ~ & by family member & ``It is my son's device and he lets me use it sometimes just to try out games and so on''&2 & \\  
        \hline       
        Necessity & ~ & &63 & 43 \\ 
        ~ & not necessary & ``It never asked me to''&21 &  \\ 
        ~ & app doesn't need authentication & ``Not required on the apps and games I use''& 21 &  \\  
        ~ & device doesn't need authentication &  ``Psvr has never required me to''&11 &  \\         
        ~ & was automatically done & ``The device is already signed in''& 10 &  \\ 
        \hline
        \makecell[l]{Done by somebody\\ else}& ~& & 21 & 14\\ 
        ~ & somebody else authenticated &  ``the person who owns it did''&21 &  \\ 
        \hline
        Other & ~ & &6 & 4\\ 
        ~ & don't know &  ``I'm not sure - I'm not that familiar with it''&4 &  \\ 
        ~ & unclear  &  ``i didnt''&2 &  \\ 
        \hline
    \end{tabular}
    
\end{table*}
\begin{table*}[htbp]
\centering
\footnotesize
\caption{Themes and Frequencies: Reasons for Perceived Easiness and Security using \textbf{Face Recognition Authentication} (n=34). Multicoding was allowed. Kappa: 0.82 (Easiness), 0.84 (Security).}
    \centering
    \begin{tabular}{lllcc}
        \hline
        \textbf{Theme} & \textbf{Code} & \textbf{{Quote Example}}& \textbf{N} & \textbf{\%} \\ \hline
        Easy & ~ && 27 & 79 \\
        ~ & simple &``Easy and convenient''& 13 & \\
        ~ & automatic track & ``the device automatically scan your face and log you in''&11 & \\
        ~ & unique & ``Simple operation once it is set up and ones face is unique''&3 & \\
        ~ & fast & ``The device instantly recognizable my face''&2 & \\
        ~ & useful &``I love that option, it is very useful and easy.''& 2 & \\
        ~ & no problems encountered & ``it was fair , i had no difficulty navigating around that mode ''&2 & \\
        ~ & well explained &``It's well explained and intuitive to me''& 1 & \\
        ~ & no need to remember secrets &``you do not have to remember any password or pin''& 1 & \\
        ~ & reliable & ``[...] is otherwise reliable and safe to use.''&1 & \\ \hline
        Not easy & ~ & &9 & 26 \\
        ~ & depends on physical appearance &``I've had some issues before with this due to my distinct physical appearance''& 2 & \\
        ~ & bad light & ``Sometimes in bad lighting the face recognition is a bit challenging [...].''&2 & \\
        ~ & struggle in beginning & ``[...] at first it was difficult for me to record my face with the software [...]''&1 & \\
        ~ & hard to set it up & ``Because sometimes i get lost in the setup.''&1 & \\
        ~ & doesn't always detect & ``[...] it sometimes does not recognise my face and I have to retry it a few times.''&1 & \\
        ~ & time consuming & ``This can sometimes take long to recognize the face.''&1 & \\
        ~ & depends on technology & ``Can be tricky if technology is not strong enough''&1 & \\ \hline
        Secure & ~ && 1 & 3 \\
        ~ & secure & ``it is very secured''&1 & \\ \hline
    \end{tabular}
    \caption*{Reasons for Perceived Easiness}
        \centering
        \begin{tabular}{lllcc}
            \hline
            \textbf{Theme} & \textbf{Code} & \textbf{{Quote Example}}& \textbf{N} & \textbf{\%} \\ \hline
            Secure & ~ && 27 & 79 \\
            ~ & no access from outside & ``No one can access your device without your face.''&13 & \\
            ~ & unique & ``Faces are different''&11 & \\
            ~ & secure & ``It's secure''&6 & \\
            ~ & no harms & ``I see no reason to feel insecure while using face recognition''&2 & \\ \hline
            Not secure & ~ && 7 & 21 \\
            ~ & a picture or AI clone can be used &``[...] I feel that they can hack me with some image or photo of me and use my devices''& 2 & \\
            ~ & face similarity &``The security is very secure, unless someone has a similar face to me.''& 2 & \\
            ~ & other methods are more secure &``I feel secure although not as much as with the iris/eye methods.''& 2 & \\
            ~ & company mistrust & ``I am not sure how secure the company is when it comes to data/security breaches.''&1 & \\ \hline
            Usability & ~ && 3 & 9 \\
            ~ & bad light & \makecell[l]{``Sometimes in bad lighting or the camera is a little foggy, \\the face recognition is a bit challenging, [...]''}&1 & \\
            ~ & time consuming & ``It takes a while at times, to recognize the face.''&1 & \\
            ~ & fast &``face recognition is fast ''& 1 & \\ \hline
        \end{tabular}
        \caption*{Reasons for Perceived Security}
\end{table*}


\begin{table*}[htbp]
    \centering
    \footnotesize
    \caption{Themes and Frequencies: Reasons for Perceived Easiness and Security using \textbf{Fingerprint Authentication} (n=21), multicoding was allowed. Kappa: 0.77 (Easiness), 0.86 (Security).}
    \centering
        \begin{tabular}{llp{9cm}cc}
        \hline
            \textbf{Theme} & \textbf{Code} & \textbf{{Quote Example}}& \textbf{N} & \textbf{\%} \\ \hline
            Easy  & ~ && 16 & 80 \\ 
            ~ & simple &``Just place your finger on the device and get authenticated ''& 14 &  \\ 
            ~ & fast & ``[...] it saves me time whenever I need to unlock the device. ''&5 &  \\ 
            ~ & efficient &``It was fast and efficient''& 1 &  \\ 
            ~ & smooth & ``I can quickly enroll their fingerprints, and the subsequent authentication process is smooth and seamless [...]''&1 &  \\ 
            ~ & seamless & ``I can quickly enroll their fingerprints, and the subsequent authentication process is smooth and seamless [...]''&1 &  \\ \hline
            Secure & ~ && 4 & 20 \\ 
            ~ & secure & ``simple and secure''&3 &  \\ 
            ~ & no access without knowledge & ``Anyone wont be able to access it without your knowledge''&1 &  \\ \hline
            Not easy & ~ & &1 & 5 \\ 
            ~ & time consuming &``For some reason fingerprints can take more than one try to work.''& 1 &  \\ \hline  
            Other & ~ & &1 & 5 \\ 
            ~ & none & ``i dont have any experience with fingerprint on vr''&1 &  \\ \hline        
        \end{tabular}
        \caption*{Reasons for Perceived Easiness}
    \centering
        \begin{tabular}{lllcc}
        \hline
            \textbf{Theme} & \textbf{Code} & \textbf{{Quote Example}}&\textbf{N} & \textbf{\%} \\ \hline
            Secure  & ~ & &17 & 85 \\ 
            ~ & unique &``No two fingerprints are the same''&6 &  \\ 
            ~ & no access from outside & ``I felt I only have access to the device''&5 &  \\ 
            ~ & can't be copied &``Because the fingerprint is unique to me and cannot be copied by another user to gain access''& 3 &  \\ 
            ~ & data is secured & ``Your files or software wont get affected because of someone trying to access something''&2 &  \\ 
            ~ & most secure method & ``The safest option''&2 &  \\  \hline
            Not secure & ~ && 2 & 10 \\ 
            ~ & can be hacked &``It's not 100\% bulletproof. I imagine it can be cracked as well.''& 1 &  \\ 
            ~ & can be stolen &``Fingerprints can be 'stolen' ''& 1 &  \\  \hline  
            Other & ~ && 1 & 5 \\ 
            ~ & none &``no experience finger print on VR''& 1 &  \\  \hline
        \end{tabular}
        \caption*{Reasons for Perceived Security}
\end{table*}


\begin{table*}[!ht]
    \centering
    \footnotesize
    \caption{Themes and Frequencies: Reasons for Perceived Easiness and Security using \textbf{Token-based Authentication} (n=16), multicoding was allowed. Kappa: 0.93 (Easiness), 0.85 (Security).}
    \centering
        \begin{tabular}{lllcc}
        \hline
            \textbf{Theme} & \textbf{Code} & \textbf{{Quote Example}}& \textbf{N} & \textbf{\makecell{\%}} \\ \hline
            Easy  & ~ & &9 & 56 \\ 
            ~ & easy to input & ``It is much easier to input the digits from the token''&3 &  \\ 
            ~ & easy &``It was easy as it was a basic token verification which is straight forward ''& 3 &  \\ 
            ~ & \makecell[l]{no need to remember secret} & ``It’s a once off token even if I loose it after using it im’m secured''&1 &  \\ 
            ~ & using app is easy &``I use an app that provides a OTP''& 1 &  \\ 
            ~ & user friendly & ``It is user friendly ''&1 &  \\ \hline
            Somewhat Easy  & ~ && 3 & 19 \\ 
            ~ & unusual method & ``Not as widely used as the other methods''&2 &  \\ 
            ~ & neutral & ``I feel pretty indifferent about this one''&1 &  \\   \hline
            Not easy & ~ && 2 & 13 \\ 
            ~ & time expiration &\makecell[l]{``Sometimes the token might not work due to time has expired \\and after requesting more, it might take a little while.''}& 1 &  \\
            ~ & can get lost & ``Tokens can be faulty or somehow used or lost.''&1 &  \\
            ~ & can be faulty & ``Tokens can be faulty or somehow used or lost.''&1 &  \\  \hline  
            Other & ~ & &2 & 13 \\ 
            ~ & don't know &``I am uncertain''& 1 &  \\
            ~ & none & ``I haven't used a token, or thought it was a password when I saw it...''&1 &  \\ \hline        
        \end{tabular} 
        \caption*{Reasons for Perceived Easiness}
        \centering
        \begin{tabular}{lllcc}
        \hline
            \textbf{Theme} & \textbf{Code} & \textbf{{Quote Example}}&\textbf{N} & \textbf{\%} \\ \hline
            Secure  & ~ & &8 & 50 \\ 
            ~ & secure & ``Using a token is very safe and no one will get your token.''&4 &  \\
            ~ & unique & ``Very secure because it is unique to each token ''&3 &  \\          
            ~ & auto generated secret &``The device automatically generate digits ''& 3 &  \\ 
            ~ & as safe as passwords & ``I presume a token would be as save as a password.''&1 &  \\ \hline
            Not secure & ~ & &7 & 50 \\ 
            ~ & access to outside with token & ``You never know how many devices that token is for or how many people have access to it.''&3 &  \\ 
            ~ & sometimes not working & \makecell[l]{``The tokens do not work sometimes especially when the\\ connection might interrupt or you have left it for too long''}&2 &  \\
            ~ & can be used beforehand & ``Maybe the token can be used beforehand''&1 & \\  
            ~ & time expiration & \makecell[l]{``The tokens do not work sometimes especially when the\\ connection might interrupt or you have left it for too long''}&1 & \\ 
            ~ & less known method & ``Less known to me'&2 & \\ \hline  
            Other & ~ & &2 & 13 \\ 
            ~ & don't know &``I wasn't sure if it's secure''& 1 &  \\ 
            ~ & none &``very hard due to some key for token missing''& 1 &  \\  \hline
        \end{tabular}
        \caption*{Reasons for Perceived Security}
\end{table*}

\begin{table*}[!ht]
    \centering
    \footnotesize
    \caption{Themes and Frequencies: Reasons for Perceived Easiness and Security using \textbf{Pairing Authentication} (n=72), multicoding was allowed. Kappa: 0.71 (Easiness), 0.72 (Security).}
    \centering
        \begin{tabular}{llp{9cm}cc}
        \hline
            \textbf{Theme} & \textbf{Code} & \textbf{{Quote Example}}& \textbf{N} & \textbf{\makecell{\%}} \\ \hline
            Easy  & ~ && 45 & 63 \\ 
            ~ & simple &`Its easy and fast when connecting to other devices''& 24 &  \\ 
            ~ & no problems encountered &`I’ve had no problems and it’s easy to authenticate my vr using my device''& 8 &  \\ 
            ~ & clear process &`It is [...] easy to locate to follow the steps''& 7 &  \\ 
            ~ & fast & `Its easy and fast when connecting to other devices''&7 &  \\ 
            ~ & known method &`It is just like any other device I could use to pair and I was familiar.''& 3 &  \\ 
            ~ & depends on device & `Easy depending on the other device''&2 &  \\ 
            ~ & device always available &`Always have my phone on me so it's incredibly simple''& 2 &  \\ 
            ~ & helpful UI &`The UI simplifies the process''& 1 &  \\ 
            ~ & practical &`very practical''& 1 &  \\ 
            ~ & no technical knowledge required & `a minimal level of technological knowledge is enough to deal with such matters ''&1 &  \\ \hline
            Not easy & ~ && 31 & 43 \\ 
            ~ & time consuming &`Because it’s long and sometime it doesn’t work well''& 7 &  \\
            ~ & technical issue: device not found &`Sometimes the other device is not really easy to find or there is a time limit''& 6 &  \\
            ~ & requires (strong) internet connect & `As long as the signal is strong enough it usually goes fast and easy enough.''&5 &  \\
            ~ & complex process & `Sometimes it is hard to understand the process of pairing.''&3 &  \\
            ~ & bothersome & `It's a little unconvienent, because I have to take off the VR headset [...]''&3 &  \\
            ~ & doesn't work well &`[...] under various circumstances, the expected behavior might not be the one exhibited.''& 2 &  \\
            ~ & devices need to be nearby &`If I don’t have the other device nearby the. i cannot connect''& 2 &  \\
            ~ & not the easiest method &`Not the easiest method''& 2 &  \\
            ~ & additional work & `Additionaly work''&1 &  \\
            ~ & has to take off VR &`It's a little unconvienent, because I have to take off the VR headset [...]''& 1 &  \\
            ~ & guidance necessary &`[...] they can accomplish it with a bit of patience and guidance from the device's instructions.''& 1 &  \\
            ~ & need for another device & `You need to have a different device with You''&1 &  \\
            ~ & devices need to be up to date &\makecell[l]{`Sometimes it is not easy to pair with another devices \\especially if that device doesn’t have the latest software and features.''}& 1 &  \\
            ~ & can't switch paired devices easily &\makecell[l]{`For some apps you may only pair your VR with one device, \\for example if the paired device is my android phone, \\the App struggles when I try to change the paired device to my laptop.''}& 1 &  \\  \hline  
            Other & ~ && 6 & 8 \\ 
            ~ & unclear & `It depends''&6 &  \\ \hline        
        \end{tabular} 
        \caption*{Reasons for Perceived Easiness}
\end{table*}

\clearpage 

\begin{table*}[!ht]
    \centering
    \footnotesize
    \centering
        \begin{tabular}{llp{9cm}cc}
        \hline
            \textbf{Theme} & \textbf{Code} & {\textbf{Quote Example}}& \textbf{N} & \textbf{\%} \\ \hline
            Secure  & ~ & & 47 & 66 \\ 
            ~ & no harm & "I’ve no reason to not feel secure whilst using another device to authenticate my VR device" & 9 & \\
            ~ & feels secure &"Secure enough, when it is paired." & 8 & \\
            ~ & only person controlling devices & "No one can do it except me" & 8 & \\ 
            ~ & other device is protected & "My phone is protected by the PIN and fingerprint, so it's safe" & 7 & \\ 
            ~ & permission is required & "[...] the vr device can be only accessed when accepting the request on my other device" & 6 & \\ 
            ~ & local network & "Secure because it was via cable not wireless." & 3 & \\ 
            ~ & known method & "[...] I was familiar with the process and most device use the same method so it felt safe." & 3 & \\ 
            ~ & multiple devices involved & "Pretty secure to have 2 distinct devices" & 2 & \\ 
            ~ & bluetooth as secure protocol & "I find this authentication method secure because I understand how the bluetooth protocol works, which makes it a strong auth. method." & 2 & \\
            ~ & devices need to be nearby & \makecell[l]{"It is very secure because both devices need to be around \\the same location for it to be done, and confirmation is needed from both\\ devices to pair successfully."} & 2 & \\ \hline
            Not secure & ~ & & 19 & 27 \\ 
            ~ & cautious about pairing with unknown devices & "I felt it might connect to other random devices that are not mine" & 6 &  \\ 
            ~ & personal devices involved & "I don’t feel entirely secure because I don’t know the kind of virus the other device has" & 4 &  \\ 
            ~ & information can be leaked & "not sure if my information wont be hacked" & 4 &  \\
            ~ & feels somewhat not secure & "You struggle a bit at first" & 2 & \\  
            ~ & own devices can be vulnerable & \makecell[l]{"[...] While the process itself doesn't raise major security concerns, \\there might be vulnerabilities depending on\\how the devices handle encryption and authentication protocols. [...]"} & 2 & \\ 
            ~ & bluetooth unsafe & "If its through Bluetooth it could be connected to any Bluetooth device"& 1 & \\ 
            ~ & fake phone numbers can be used &\makecell[l]{"I know some people can fake a phone number but its still quite hard for them \\to get past that"} & 1 & \\ 
            ~ & devices are shared & \makecell[l]{"Pairing I would think is not that secure since everyone \\in my family can pair with their smartphone if they want."} & 1 & \\ 
            ~ & easy access to one device & "I feel the other person might have access to my devise" & 1 & \\ \hline  
            Usability & ~ & & 4 & 6 \\ 
            ~ & fast & "it makes it fast and easy for me to use and enjoyable" & 4 &  \\ 
            ~ & easy & "Its easy to use" & 1 &  \\\hline
            Other & ~ & & 5 & 7 \\ 
            ~ & unclear & "I had doubts that it would reflect faster and that I had to search for it under settings." & 5 &  \\ \hline
        \end{tabular}
        \caption*{Reasons for Perceived Security}
\end{table*}

\begin{table*}[!ht]
    \centering
    \footnotesize
    \caption{Themes and Frequencies: Reasons for Perceived Easiness and Security using \textbf{Unlock Pattern Authentication} (n=28), multicoding was allowed. Kappa: 0.78 (Easiness), 0.72 (Security).}
    \centering
        \begin{tabular}{lllccc}
        \hline
            \textbf{Theme} & \textbf{Code} &  {\textbf{Quote Example}} & \textbf{N} & \textbf{\%} \\ \hline
            Easy  & ~ & & 21 & 75 \\ 
            ~ & simple & "Very easy to use a pattern." & 11 &  \\ 
            ~ & known method & "I use this method on other devices" & 4 &  \\ 
            ~ & fast & "[...] it's easy, fast and simple." & 4 &  \\ 
            ~ & easy to remember & "easy to remember" & 3 &  \\ 
            ~ & easy if pattern is simple & "Just choose an easy pattern and use it [to] get authenticated" & 1 &  \\
            ~ & user friendly & "this authentication approach is designed to be user-friendly [...]" & 1 &  \\ 
            ~ & accessible & "this authentication approach [...] offers a smooth and accessible way to unlock and access VR devices." & 1 &  \\  \hline
            Not easy & ~ & & 9 & 32 \\ 
            ~ & need to be familiar & "You need to be familiar with the input devices in order to draw or select the pattern [...]" & 3 &  \\
            ~ & easy to forget pattern & "[...] I tend to forget my pattern" & 3 &  \\
            ~ & dizziness & "Requires some head movement, which can cause dizzines" & 1 &  \\ 
            ~ & too much work & "It’s too much work" & 1 &  \\ 
            ~ & not flexible & "For me a pattern is not that flexible [...]" & 1 &  \\
            ~ & problem with indication & "Sometimes I had a problem with an indication of the pattern [...]" & 1 &  \\ 
            ~ & complex input in VR & "It was a bit confusing [...] with the complexity of having to virtually draw a pattern." & 1 &  \\ \hline 
            Security & ~ & & 2 & 7 \\ 
            ~ & easy pattern not safe & "creating a easy pattern is not safe [...]" & 2 &  \\  \hline
            Other & ~ & & 1 & 4 \\ 
            ~ & unclear & "its same as using pin" & 1 &  \\ \hline        
        \end{tabular}
        \caption*{Reasons for Perceived Easiness}
        \centering
        \begin{tabular}{lllccc}
        \hline
            \textbf{Theme} & \textbf{Code} &  {\textbf{Quote Example}} & \textbf{N} & \textbf{\%} \\ \hline
            Not secure & ~ & & 15 & 54 \\ 
            ~ & can be copied or known & "the only risk is someone else getting to know it." & 6 &  \\ 
            ~ & observable & "Other people can see my hands movement [...]" & 3 &  \\
            ~ & easy patterns not secure & "because it is easy to create simply to remember patterns which might not be secure" & 3 & \\  
            ~ & not as secure as passwords & "Easier to figure out than a password" & 2 & \\ 
            ~ & not safe longterm & "I don't feel that it's really safe long term." & 1 & \\ 
            ~ & always same pattern not secure & \makecell[l]{"from my experience people tend to use the same patterns [...], \\and in some cases a pattern might not be that secure"} & 1 & \\ \hline 
            Secure  & ~ & & 13 & 46 \\ 
            ~ & secure & "my pattern is secure" & 5 &  \\
            ~ & only one knowing & "No one can crack the pattern only me who set the pattern." & 3 &  \\          
            ~ & not easy to guess & "It’s not that easy to guess my pattern" & 3 &  \\ 
            ~ & unique & "It is a unique pattern, why should I feel insecure?" & 1 &  \\
            ~ & sees no harm & "I see no harm or insecurity while using Patterns in VR" & 1 &  \\ 
            ~ & complex pattern is secure & \makecell[l]{"A bit secure the complexity of using the pattern somehow gave me an impression \\of good security."} & 1 &  \\
            ~ & in combination with other methods & \makecell[l]{"[...] for the highest level of security, combining pattern-based \\authentication with other stronger authentication factors is recommended."} & 1 &  \\
            \hline
            Usability & ~ & & 1 & 4 \\ 
            ~ & simple & "simple and fast" & 1 &  \\ 
            ~ & fast & "simple and faste" & 1 &  \\  \hline
            Other & ~ & & 1 & 4 \\ 
            ~ & unclear & "I do not know whether this option of authentication is secure or not." & 1 &  \\ \hline
        \end{tabular}
        \caption*{Reasons for Perceived Security}
\end{table*}

\begin{table*}[!ht]
    \centering
    \footnotesize
    \caption{Themes and Frequencies: Reasons for Perceived Easiness and Security using \textbf{PIN Authentication} (n=68), multicoding was allowed. Kappa: 0.81 (Easiness) and 0.76 (Security) }
    \centering
        \begin{tabular}{lllccc}
        \hline
            \textbf{Theme} & \textbf{Code} &  {\textbf{Quote Example}} & \textbf{N} & \textbf{\%} \\ \hline
            Easy  & ~ & & 63 & 93 \\ 
            ~ & easy to use & "Is very easy [...]" & 28 &  \\ 
            ~ & easy to input & "Entering pin is fast , convenient [...]" & 24 &  \\ 
            ~ & easy to remember & "I easily remember my pin.. [...]" & 16 &  \\ 
            ~ & fast & "It's a pretty simple and quick process [...]" & 14 &  \\
            ~ & intuitive & "The way to enter the PIN is very intuitive [...]" & 6 &  \\ 
            ~ & short secret & "PINs being short by nature don't pose much of a problem" & 4 &  \\ 
            ~ & easy restoring & \makecell[l]{"It is the best authentication method because should one forget their pin, \\it allows one to retrieve the pin via an app (e.g. oculus)"} & 2 &  \\ 
            ~ & accessible & \makecell[l]{"A pin is always accessible from my paired devices so It just automatically fills in."} & 2 &  \\ 
            ~ & shared PIN & "I use pin given to Me if I'm using another person's headsets" & 1 &  \\ 
            \hline
            Security  & ~ & & 10 & 15 \\ 
            ~ & secure & "I think It's friendly user and protects data" & 5 &  \\ 
            ~ & unsecure & "It’s fast and easy but people can guess it" & 4 &  \\ 
            ~ & different PINs every time & "I use different passwords for each and every thing that I use." & 1 &  \\ 
            \hline
            Not easy & ~ & & 5 & 7 \\ 
            ~ & need to remember & "as long as you remember your pin using it is almost straight forward" & 2 &  \\ 
            ~ & difficult tracking sometimes & "There may be problems in tracking hand movements." & 2 &  \\ 
            \hline 
            Other & ~ & & 2 & 3 \\ 
            ~ & unclear & "method of key movement that was set to enter the pin" & 2 &  \\ 
            \hline        
        \end{tabular}
        \caption*{Reasons for Perceived Easiness}
        \centering
        \begin{tabular}{lllccc}
        \hline
        \textbf{Theme} & \textbf{Code} &  {\textbf{Quote Example}} & \textbf{N} & \textbf{\%} \\ \hline
        Secure  & ~ & & 33 & 49 \\ 
        ~ & secure & "There is alphabets, digits etc so I believe that my password it's very secured" & 10 &  \\
        ~ & only one knowing & "I belive l am the only one who knows my pin and can put it in." & 10 &  \\          
        ~ & not easy to guess & "Somewhat because it's not easy to steal" & 4 &  \\ 
        ~ & selecting different or complex PINs & "It can be only as secure as the PIN is." & 3 &  \\ 
        ~ & change regularly & "I find it secure because I use a pin code which is not easy to guess and it changes regularly" & 3 &  \\ 
        ~ & no harm & "I didn't experience any problems" & 2 &  \\ 
        ~ & can't be hacked & "My pin can not be hacked" & 2 &  \\ 
        ~ & unique & "I have a unique PIN which only is known to me" & 1 &  \\ 
        ~ & attack doesn't stay unnoticed & "No one can break the 4-number code so fast that I won't notice" & 1 &  \\ 
        ~ & environment looks reliable & "the environment for entering the PIN looks reliable"& 1 &  \\ 
        \hline
        Not secure & ~ & & 31 & 46 \\ 
        ~ & easy to guess/break & "It is secured, but pin is easy to brake" & 14 &  \\ 
        ~ & not different from other devices & "because i can't see a difference with other systems, it's safe" & 5 &  \\ 
        ~ & not the safest & "It's not the super safest option out there [...]" & 5 &  \\  
        ~ & other options desirable & "[...] pin usage wouldn't be my favourite option for security." & 4 &  \\ 
        ~ & only digits & "A small amount of digits isnt the safest route to go" & 4 &  \\ 
        ~ & simple security better than none & "even simple security is better than none." & 2 &  \\ 
        ~ & short and vulnerable & "since it is only a short number of digits ,it makes it easy for someone to guess." & 2 &  \\ 
        ~ & not unique enough & "Could be more unique, so more secure " & 1 &  \\ 
        ~ & observable & \makecell[l]{"I do not feel secure because somebody might be watching me\\ entering the PIN code from the external monitor."} & 1 &  \\ 
        \hline     
        Usability & ~ & & 12 & 18 \\ 
        ~ & easy to use & "It's not the super safest option out there but very easy to use." & 8 &  \\ 
        ~ & easy to remember & "use numbers l like and remember" & 4 &  \\  
        \hline
        Other & ~ & & 2 & 3 \\ 
        ~ & unclear & "My opinion" & 2 &  \\ 
        \hline
    \end{tabular}
    \caption*{Reasons for Perceived Security}
    \label{table:pin_secure}
\end{table*}


\begin{table*}[!ht]
    \centering
    \footnotesize
    \caption{Themes and Frequencies: Reasons for Perceived Easiness and Security using \textbf{Password Authentication} (n=86), multicoding was allowed. Kappa: 0.72 (Easiness) and 0.73 (Security)}
    \centering
    \begin{tabular}{lllcc}
        \hline
        \textbf{Theme} & \textbf{Code} &  {\textbf{Quote Example}} & \textbf{N} & \textbf{\%} \\ \hline
        Easy  & ~ & & 54 & 63 \\ 
        ~ & easy to use & "I do not find it difficult to insert my password [...]" & 19 &  \\ 
        ~ & easy to input & "Like a pin only more secure, it's easy to enter." & 12 &  \\ 
        ~ & known method & "I do not find it difficult to insert my password because I am familiar with the password" & 9 &  \\ 
        ~ & easy (conditionally) & "It’s easier provided I remember the password" & 7 &  \\
        ~ & fast & "[...] it does not even consume much time or require are lot to be done." & 7 &  \\ 
        ~ & intuitive & "It’s pretty straightforward" & 5 &  \\ 
        ~ & easy to remember & "same as pin simple password can easily be remembered" & 4 &  \\ 
        ~ & clear instructions & "I mean I just had to follow the guide, nothing too complicated." & 2 &  \\ 
        ~ & autofill & "It is easy because the password autofills itself" & 2 &  \\ 
        ~ & restore option & \makecell[l]{"The password is always easy to remember, and one can always reset it easily if one forgets it."} & 1 &  \\ 
        ~ & stored credentials & "always save my details" & 1 &  \\ 
        ~ & other device was used & "It was average because I used my phone" & 2 &  \\ 
        \hline
        Not easy  & ~ & & 29 & 34 \\ 
        ~ & difficult input with controllers & \makecell[l]{"I use password managers that generate long random passwords.\\ Having to enter these manually is tedious and prone to error"} & 19 &  \\ 
        ~ & hard to enter complex passwords & "Aiming for specific characters on a floating keyboard proved to be quite troublesome [...]" & 11 &  \\ 
        ~ & hard to remember & \makecell[l]{"the systems require a complicated password that contains different characters\\ which might be hard to remember if you do not write it down."} & 4 &  \\ 
        ~ & time consuming & "[...] you have to slowly type in your password otherwise you will make mistakes." & 4 &  \\ 
        ~ & rather choose other methods & "[...] in this case, regarding authentication, I'd be inclined to choose other methods" & 1 &  \\ 
        ~ & technical issue & "[...] But my password as a developer has not worked since they changed to Meta [...]" & 1 &  \\ 
        ~ & hard to use a PM & \makecell[l]{"The password, in my case, isn't very pleasant to use because I use a 3rd party\\ password manager, that needs to be opened and consulted every time I have to log in."} & 1 &  \\ 
        \hline
        Other & ~ & & 5 & 6 \\ 
        ~ & unclear & "I only complicated put all the email used vr controller" & 5 &  \\ 
        \hline
    \end{tabular}
    \caption*{Reasons for Perceived Easiness}
    \label{table:pw_easiness}
\end{table*}

\clearpage 

\begin{table*}[!ht]
    \centering
    \footnotesize
    \centering
    \begin{tabular}{llp{9cm}cc}
        \hline
        \textbf{Theme} & \textbf{Code} &  {\textbf{Quote Example}} & \textbf{N} & \textbf{\%} \\ \hline
        Secure  & ~ & & 60 & 70 \\ 
        ~ & use strong passwords & "I usually create strong passwords." & 21 &  \\
        ~ & secure & "I didn't feel any threat so i felt secure" & 11 &  \\         
        ~ & nobody can access & "Password is only known by myself." & 10 &  \\ 
        ~ & not easy to guess & \makecell[l]{"your password can involve basically any character's from your keyboard\\ and as long as you want it to be not easy to guess"} & 6 &  \\
        ~ & known method & "[...] I don't think vr has any more problems than anywhere else." & 6 &  \\ 
        ~ & no harm & "I never had problem with that method " & 6 &  \\ 
        ~ & change regularly & "I change my password often and use different passwords for different interfaces" & 3 &  \\ 
        ~ & only one knowing the password & "Only I know the password, which makes me feel secure." & 3 &  \\ 
        ~ & different passwords & "I use different passwords for everything and this seems secure to me" & 2 &  \\ 
        ~ & stored securely & "A password is a code that is known by you and when kept safe, a very secure method to use." & 2 &  \\ 
        ~ & unique password & "I feel very secure because I know my password is both unique and strong [...]" & 2 &  \\ 
        ~ & hidden input & "I feel secure as the password is hidden" & 2 &  \\ 
        ~ & will notice bad activity & "You received email when you account is used in the devices" & 1 &  \\ 
        ~ & trust & \makecell[l]{"my trust in the overall VR platform and the specific application requesting\\ the Entering password plays a significant role in my sense of security."} & 1 &  \\ 
        ~ & encrypted & \makecell[l]{"I think I am relatively safe, because I think my banking and SS number,\\ are in a encrypted secured computer, belonging to Meta."} & 1 &  \\ 
        \hline
        Not secure & ~ & & 20 & 23 \\ 
        ~ & not most secure & \makecell[l]{"I think that if the password is complex and long, it is even a good form\\ of security, but there are much better ones."} & 5 &  \\ 
        ~ & can be hacked & "Sometime passwords can be hacked [...]" & 4 &  \\ 
        ~ & observable & \makecell[l]{"Some people tend to want to look over your shoulder as you key\\ in the password so I still feel a bit insecure entering my password"} & 4 &  \\  
        ~ & stored somewhere else & "I have to have my password up on my phone so that's not too secure" & 2 &  \\ 
        ~ & aware that easy passwords are unsafe & "the more simple the password the more insecure it is" & 2 &  \\ 
        ~ & no trust & "I felt somewhat secure because I was fairly new to the tech and couldn't completely trust it" & 2 &  \\ 
        ~ & data in risk when hacked & "Should anyone have access to my pin then my data is in risk" & 1 &  \\ 
        \hline       
        Other & ~ & & 6 & 7 \\ 
        ~ & unclear & "I was not comfortable with the security features as they were not blatantly stated." & 6 &  \\ 
        \hline
    \end{tabular}
    \caption*{Reasons for Perceived Security}
    \label{table:pw_security}
\end{table*}

\begin{table*}[!ht]
    \centering
    \footnotesize
    \caption{Themes and Frequencies: Reasons for using password managers, multicoding allowed (n = 102). Kappa: 0.71}
    \begin{tabular}{lllccc}
    \hline
        \textbf{Theme} & \textbf{Code} &  {\textbf{Quote Example}} & \textbf{N} & \textbf{\%} \\ \hline
        Convenience  & ~ & & 90 & 88 \\ 
        ~ & no need to remember credentials & \makecell[l]{"There are many passwords that l have to remember so using password managers\\ becomes easy when l have forgotten some of them"} & 45 &  \\
        ~ & easier login process & "This is an example quote" & 14 &  \\          
        ~ & overview of accounts & "This is an example quote" & 13 &  \\ 
        ~ & time saving & "saves time when logging in" & 12 &  \\
         ~ & easy to use & "Also, it aids in fast and easy login process." & 9 &  \\ 
         ~ & autofill & "I don't like retyping passwords when entering apps [...]" & 9 &  \\ 
         ~ & password storing & \makecell[l]{"Because I find it convenient to have my passwords stored\\ in a built-in software as part of the browsers I'm using frequently."} & 7 &  \\
         ~ & synching & "I use the same browser on all devices.  I always have all my passwords with me" & 7 &  \\
         ~ & convenient & "Mostly convenience" & 6 &  \\
         ~ & for backups & "For back-ups." & 3 &  \\
         ~ & easy access to frequently used sites & "I like being able to acces commonly used sites quickly" & 2 &  \\
         ~ & makes life easier & "I use it to make my life easier." & 2 &  \\
         ~ & sharing credentials & \makecell[l]{"They allow my family to log out and in to my accounts while\\ I am absent in case they need to do anything for me or them"} & 1 &  \\
         ~ & reliable & "I find them reliable" & 1 &  \\
          \hline
        Security & ~ & & 28 & 27 \\ 
        ~ & stored securely  & \makecell[l]{"I used a password manager because of the convenience of \\having all my passwords stored in a safe platform."} & 11 &  \\ 
        ~ & password generation & "They are useful for creating strong randomised passwords [...]" & 9 &  \\
        ~ & protects in general & "It's simple way to keep online accounts safe" & 5 & \\  
        ~ & secure & "Easy and secure" & 3 & \\ 
        \hline       
        Other reasons & ~ & & 3 & 3 \\ 
        ~ & activated by default & "They kind of already have been there and I didn't turn them off" & 2 &  \\ 
        ~ & mandatory by company & \makecell[l]{"My company's policy requires to have the internal\\ Password Manager account activated and use it."} & 1 &  \\ 
        \hline
        Other & ~ & & 2 & 2\\ 
        ~ & unclear & "Its the ones I know" & 2 &  \\ 
        \hline
    \end{tabular}
\end{table*}

\begin{table*}[!ht]
    \centering
    \footnotesize
    \caption{Themes and Frequencies: Reasons for not using password managers (n = 24), multicoding allowed. Kappa: 0.71 (same question as above)}
    \begin{tabular}{lllccc}
    \hline
        \textbf{Theme} & \textbf{Code} & {\textbf{Quote Example}} & \textbf{N} & \textbf{\%} \\ \hline
        Security & ~ & & 17 & 71 \\ 
        ~ & trust issues  & \makecell[l]{"I feel that if someone can get access to my password\\ manager they will have access to other sensitive information"} & 15 &  \\ 
        ~ & uses own practices & "I have few creative passwords, so I believe that there is no need for password managers." & 7 &  \\        
        ~ & sees no risk & "I dont think that it could be hacked or anything like that, so I don't see a reason to be too secure" & 2 & \\ 
        ~ & considering using a PM & \makecell[l]{"I wasn't convinced by them until now.\\ I am getting closer and closer to changing my view, hearing\\ around how often data leakage has been happening lately"} & 1 & \\              
        \hline       
        Other reasons & ~ & & 7 & 29 \\ 
        ~ & lack of awareness & "It’s my first time hearing about it" & 3 &  \\ 
        ~ & sees no necessity & "I am the only one using the device." & 2 &  \\  
        ~ & cost & "I don't trust in the password managers and it's too pricey to me." & 1 &  \\         
        ~ & no time for set up & "I dont have time to set up the PM and migrate all my passwords" & 1 &  \\    
        \hline
        Other & ~ & & 1 & 1\\ 
        ~ & unclear & "I am not using any password manager." & 1 &  \\ 
        \hline
    \end{tabular}
    
\end{table*}


\begin{table*}[!ht]
    \centering
    \footnotesize
    \caption{Themes and Frequencies: Reasons for why or why not authentication process in VR needs to be improved, multicoding allowed. N = 126; Yes = 77; No = 49. The Within-percentages for all themes, except for \textit{No improvements} and \textit{Other}, are calculated based on the number of participants who responded affirmatively to the question about whether improvements could be made. Kappa: 0.83}
    \begin{tabular}{llp{7cm}cccc}
    \hline
        \textbf{Theme} & \textbf{Code} &  {\textbf{Quote Example}} & \textbf{N} & \textbf{\makecell{\% Within}} &\textbf{\makecell{\% Total}} \\ \hline
        Usability & ~ & & 28 & 36 &22 \\ 
        ~ & make it easier  & "To make it simpler to log onto." & 7 &&  \\ 
        ~ & make it more user-friendly & \makecell[l]{"It can be something more intuitive and in a friendly environment\\ to transmit security for passwords but easy to use"} & 6 &&  \\
        ~ & faster authentication & "If the process is somehow faster [...]" & 6 && \\  
        ~ & more accuracy & "The verification system is not accurate and technology needs to grow" & 3 && \\ 
        ~ & accessibility & "Maybe in the future with more accessible options" & 2&& \\
        ~ & better input methods & \makecell[l]{"The biggest problem in vr is you have to type, I don't\\ know if we can improve that."} & 2 && \\
        ~ & frustration & \makecell[l]{"There is still some [frustration] when trying\\ to authenticate using VR Apps"} & 2&& \\
        ~ & larger virtual keyboard & \makecell[l]{"[...] The virtual keyboard should be large and easy to see"} & 1&& \\
        \hline
        \makecell[l]{Biometric as \\alternative} & ~ & & 23 &30& 18 \\ 
        ~ & iris scan  & \makecell[l]{"Iris scanning [...] could be the way to go"} & 8 &&  \\ 
        ~ & fingerprint & \makecell[l]{"Users might feel more secure if the VR device\\ supports [...] biometric authentication (e.g., fingerprint [...])."} & 8 &&  \\
        ~ & face recognition & "Include face ID in more apps and processes to make it more secure." & 7 && \\  
        ~ & voice recognition & "Introducing more immersive and biometric authentication methods tailored to VR, such as voice recognition or hand gestures, could enhance security and user experience." & 5 &&  \\ 
        ~ & include biometric features & \makecell[l]{"Provides some good options, though would like the more secure\\ biometric options more readily available"} & 6 & &\\ 
        ~ & improve body movement & \makecell[l]{"I believe that body or eye movement could\\ be improved to work more correctly and precisely."} & 3&& \\
        ~ & DNA authentication & "We could move to more hard to copy techniques like DNA" & 2 && \\
        ~ & eye tracking & \makecell[l]{"I believe that body or eye movement could\\ be improved to work more correctly and precisely."} & 1&& \\
        \hline      
        Security & ~ & & 15 &19& 12 \\ 
        ~ & security can be improved  & \makecell[l]{"We could develop more VR-specific \\methods [...] to make it more [...] secure."} & 10 &&  \\ 
        ~ & MFA or MMA & \makecell[l]{"VR devices should support more secure authentication methods, \\such as [...] multi-factor authentication."} & 4 & & \\
        ~ & use more secure alternatives & \makecell[l]{"There can be alternative input methods, that [...]\\implement regular security audits and updates."} & 3 && \\ 
        \hline  
        Innovation & ~ & & 13 & 17&10 \\ 
        ~ & innovation necessary & "They could introduce something new" & 10 & & \\ 
        ~ & VR specific solutions & \makecell[l]{"All apps and services should have VR friendly options\\ since VR is becoming more and more common."} & 7 & & \\  
        ~ & VR still new technology & "Its still new thing there so it can only be better " & 5 & & \\    
        \hline
        More alternatives & ~ & & 7 &9& 6\\ 
        ~ & more PINs/patterns & "More options for pattern or PIN after initial account set up." & 2 & & \\ 
        ~ & tokens & \makecell[l]{"Users might feel more secure if the VR device supports [...]\\ virtual security tokens."} & 2 & & \\ 
        ~ & pairing & \makecell[l]{"I think it can be massively improved if you could do it\\ all on your phone then connect your vr headset to your device"} & 2 & & \\ 
         ~ & \makecell[l]{move away from social media\\ account pairing} & "Maybe move away from social media accounts" & 1&&  \\ 
        \hline
        \end{tabular}
\end{table*}

\begin{table*}[!ht]
    \centering
    \footnotesize
    \caption{Themes and Frequencies (cont.): Reasons for why or why not authentication process in VR needs to be improved, multicoding allowed. N = 126; Yes = 77; No = 49. The Within-percentages for all themes, except for \textit{No improvements} and \textit{Other}, are calculated based on the number of participants who responded affirmatively to the question about whether improvements could be made. Kappa: 0.83}
    \begin{tabular}{llp{6.8cm}ccc}
    \hline
        \textbf{Theme} & \textbf{Code} &  {\textbf{Quote Example}} & \textbf{N} & \textbf{\makecell{\% Within}} &\textbf{\makecell{\% Total}} \\ \hline
        Password managers & ~ & & 3 & 4&2\\ 
         ~ & use of password managers & "Some [...] devices don’t allow PM’s" & 3&&  \\ 
        \hline
        General & ~ & & 7 & 9&6\\ 
         ~ & processes can be improved & "As more and more users use VR, the process should get easier" & 7&&  \\ 
        \hline
        Transparency & ~ & & 1 & 1&1\\ 
         ~ & make methods more transparent & "I think it is good but could be more transparent" & 1 &&  \\ 
        \hline
        No improvements & ~ & & 42 & 86&33 \\ 
         ~ & current state good enough & "The various authentication processes in VR are efficient." & 38 &&  \\ 
         ~ & no over-complication needed & "I don't think it need to be complicated. " & 3 &&  \\ 
         ~ & no over-protection needed & "For leisurely purposes I do not think over protection is necessary." & 1 &&  \\ 
        \hline
        Other & ~ & & 11 & $-$&9 \\ 
         ~ & don't know & "I am not sure how" & 8 &  &\\ 
        ~ & unclear & "No questions" & 5 & & \\ 
        \hline
    \end{tabular}
    \label{table:pms_usage_cont}
\end{table*}

\begin{table*}[!ht]
    \centering
    \footnotesize
    \caption{Themes and Frequencies: Reasons for considering using a PM for VR. Question was asked to participants who are not using PMs for VR (n=92)  {if they would consider using a PM directly on the VR devices, outside VR, or not at all.} Kappa: 0.71}
   \begin{tabular}{llp{7.5cm}ccc}
    \hline
        \textbf{Theme} & \textbf{Code} & {\textbf{Quote Example}} & \textbf{N} & \textbf{\%} \\ \hline
        Directly on VR device & ~ & & 38 & 41 \\ 
        ~ & easiness & "It would [...] make the process of pairing and authentication quick and easy." &12 & \\
        ~ & automate login process  & \makecell[l]{"All my passwords are captured and use the VR to\\ automatically make use of them"} & 8 &  \\ 
        ~ & fast & "To save time" & 7 &  \\
        ~ & synching & \makecell[l]{"With an apple headset, all my stored keys would be instantly\\ added from keychain to the VR as well"} & 6 &  \\
        ~ & no need to remember & "To help me remember my passwords" & 5 &  \\
        ~ & convenient & "It's more convenient" & 4& \\
        ~ & to prevent a lot of typing in VR & \makecell[l]{"To avoid having to type passwords I'd consider using a \\native PM on the VR device."} & 3& \\
        ~ & positive experience & "As I use PM elsewhere, I would be happy to use it in VR too" & 3& \\
        ~ & no need to go back to main device & "It will be easier to login and authenticate, without going to the main device." & 2 & \\  
        ~ & storing passwords & "its easier to recover your passwords" & 2& \\
        ~ & support the VR user & "It would be of great help" & 1 & \\ 
        ~ & if PM available & "It's very tedious to not have it built in so it would help a lot" & 1 & \\ 
        \hline
        Outside & ~ & & 21 & 23 \\ 
        ~ & easier outside & "I think it would be easier just to confirm on a smartphone" & 5 & \\ 
        ~ & feels more secure with other devices & "It will be much secure this way." & 4 &  \\         
        ~ & trust issues with PMs on VR & "Not sure of the trustworthiness of it being on VR" & 2 &  \\         
        ~ & improves security on VR & "I would use it because then my passwords are more secure and in one safe spot." & 2& \\
        ~ & outside VR for better input methods & "Even tho vr is great, as of now it does not easy to type [...]" & 1 & \\ 
        ~ & smartphone preferred  & "If i could sign in on my phone that is on my person before i use the vr headset. " & 1& \\ 
        ~ & prefers not to mix devices & "not mixing" & 1& \\
        ~ & redundant on VR & \makecell[l]{"I do not want to bother using PM on the VR because it might\\ be redundant action on this type of device."} & 1& \\
        ~ & not so many passwords on VR & "Not that many passwords to remember on the oculus." & 1& \\
        ~ & faster outside & "It just simply easier and faster to use a password manager outside of it." & 1&\\
        \hline      
        Not considering & ~ & &30 & 33 \\ 
        ~ & no need  & "I don’t see the need" & 11 &  \\ 
        ~ & doesn't trust PMs in general & "I do not trust" & 10 &  \\
        ~ & own practices preferred & "I just prefer to memorize my passwords" & 4 & \\ 
        ~ & fear of complexity & "I think it would be somewhat more difficult to set up in VR" & 2& \\
        ~ &  {primary} passwords complex to enter & \makecell[l]{"Password for the password manager should be long, so entering\\ it on the VR headset would be infuriating"} & 1& \\
        ~ & continuous authentication is safer & "It would be better for VR to check everytime who is logging" & 1 & \\
        ~ & satisfied with current solutions & "I am happy with what I have at the moment" & 1& \\
        ~ & no familiarity with PMs & "I am not familiar with password manager." & 1& \\
        \hline  
        Both considerable & ~ & & 3 & 3 \\ 
        ~ & no need to remember & \makecell[l]{"It would provide me with secure passwords which I\\ wouldn't have to remember myself"} & 1 &  \\
        \hline
        Other & ~ & & 8 &  9\\ 
        ~ & unclear & "I just don't feel it's something I'd want to do at this moment." & 9 &  \\ 
        \hline
    \end{tabular}
\end{table*}

\begin{table*}[!ht]
    \centering
    \footnotesize
     \caption{Themes and Frequencies: Reasons for why or why not choosing passwords differently for VR. Question was posed to participants who indicated they had previously created an account that required setting a password on a VR device (n = 51). Kappa: 0.77}
    \begin{tabular}{lllccc}
    \hline
        \textbf{Theme} & \textbf{Code} & {\textbf{Quote}} & \textbf{N} & \textbf{\%} \\ \hline
        Not differently & ~ & & 32 &  63\\ 
        ~ & pw easy to remember & "I keep my password same so not to forget them" & 18 &  \\ 
        ~ & same practice as usual & "It was a similar interface to other devices" & 10 &  \\
        ~ & has set of passwords & "I usually use multiple passwords between websites" & 3 &  \\
        ~ & already has strong passwords & \makecell[l]{"Because if the combination is strong enough it will secure\\ your info across different operating systems and devices"} & 2 &  \\
        ~ & no need to use a different one & "I don't see a reason to why it would be different" & 9 &  \\
        ~ & security not important & "I don't really require secure methods to access these apps" & 1 &  \\
        \hline
        Differently & ~ & & 16 & 31 \\ 
        ~ & security practice & "To make sure that my VR activities are private." & 6 &  \\ 
        ~ & easier passwords & "I tend to choose shorter, easier-to-type passwords that I can remember well. " & 5 &  \\  
        ~ & different policies & "I used a different password based on the requirements" & 3 &  \\
        ~ & shared device or account & \makecell[l]{"I have different passwords for different devices because\\ certain people close to me may use some of them with me"} & 2& \\
        \hline
        Other & ~ & & 5 &  10\\ 
        ~ & unclear reason & "I did not create a different password" & 5 &  \\ 
        \hline
    \end{tabular}
   
\end{table*}

\begin{table*}[!ht]
    \centering
    \footnotesize
    \caption{Themes and Frequencies: Reasons for perceived satisfaction of creating an account in VR. Question was posed to participants who indicated they had previously created an account that required setting a password on a VR device (n = 51).  Kappa: 0.84}
    \begin{tabular}{llp{8cm}ccc}
    \hline
        \textbf{Theme} & \textbf{Code} & {\textbf{Quote Example}} & \textbf{N} & \textbf{\%} \\ \hline
        Satisfied & ~ & & 33 & 65 \\ 
        ~ & easy to use & "easy to do it" & 13 &  \\ 
        ~ & no difficulties encountered & "I didn't experience any problems." & 12 &  \\  
        ~ & clear instructions & "The setup had clear and concise instructions which did not take long to action" & 5 &  \\
        ~ & secures data & "Simple, makes apps more secure" & 4 & \\
        ~ & known process & "It is the usual process so it is not that difficult" & 3 & \\
        ~ & uses same password for everything & "I use one password for everything and I did not encounter any challenges" & 1& \\
        ~ & satisfied, when no personal data involved & \makecell[l]{"No need for me to share personal banking information which I\\ am always worried of doing"} & 1& \\
        ~ & quick with SSO & \makecell[l]{"Simple and quick registration process even when using\\ external portals - google, facebook."} & 1 &  \\
        ~ & quick with pairing & "If it is connected to the device, it uses the information provided on the device." & 1 &  \\
        \hline
        Not satisfied & ~ & & 19 &  37\\ 
        ~ & controller annoyance & "Writing in vr is a nuisance" & 8 &  \\ 
        ~ & process took too long & "It's mostly annoying to write the emails and the passwords because it takes a long time" & 7 &  \\
        ~ & difficult process on VR & "It was a fun little challenge that required a lot of guidance" & 3 &  \\
        ~ & no options to choose from & "Most accounts they do not have a vast options to choose from [...]" & 1 &  \\
        ~ & mandatory account & "I think needing a facebook/meta account to use a VR headset is wrong" & 1 &  \\
        ~ & privacy concerns & "[...] a person form the community tried to help, but asked to many personal identifiers." & 1 &  \\
        ~ & buggy process & "My main issue was bugs, either on the interface interaction or the process overall" & 1 &  \\
        \hline
        Neither nor & ~ & & 2 & 4 \\ 
        ~ & somewhat satisfied & "It was okay" & 1 &  \\ 
        ~ & no reason mentioned & "It was as expected" & 2 &  \\  
        \hline
        Other & ~ & & 1 &  2\\ 
        ~ & unclear & "We didn't have a serial number since it was company issued" & 1 &  \\ 
        \hline
    \end{tabular}
\end{table*}

\onecolumn
\section{Password Manager Screenshots}\label{app:pm_screenshots}


\begin{figure}[htbp]
  \centering
  \begin{minipage}{0.9\textwidth}
    \centering
    \begin{minipage}{0.3\textwidth}
      \centering
      \includegraphics[width=\linewidth]{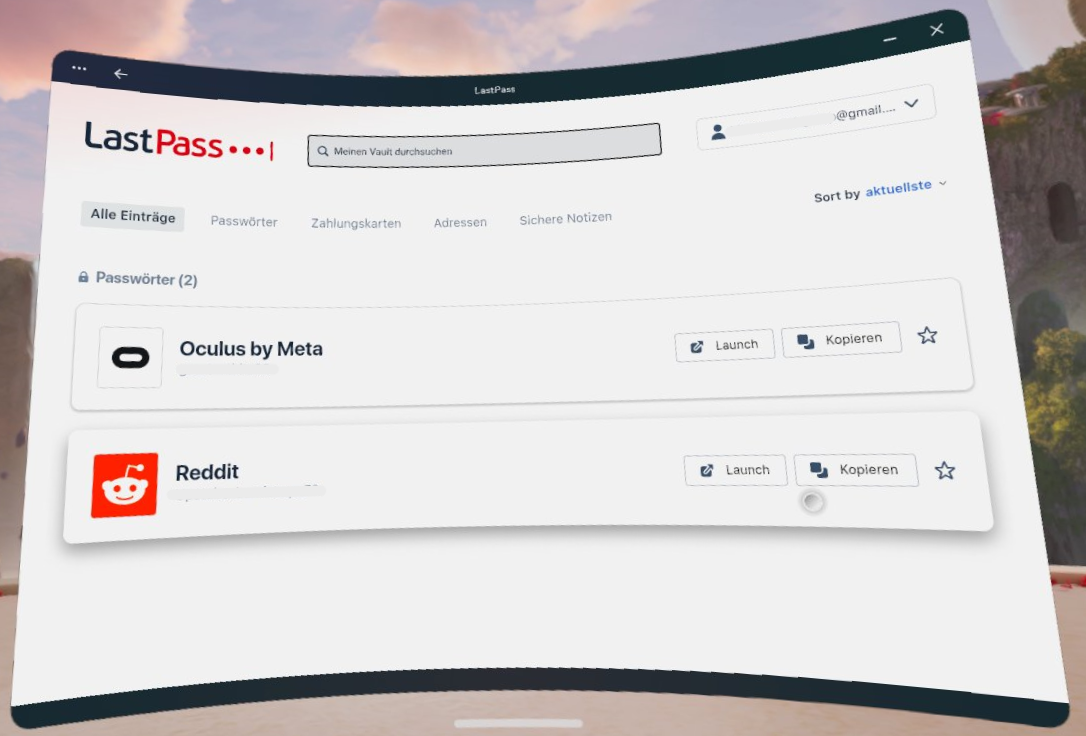} \\ 
      \caption*{\small (a) LastPass VR app \cite{lastpassvr}}
    \end{minipage}
    \begin{minipage}{0.3\textwidth}
      \centering
      \includegraphics[width=\linewidth]{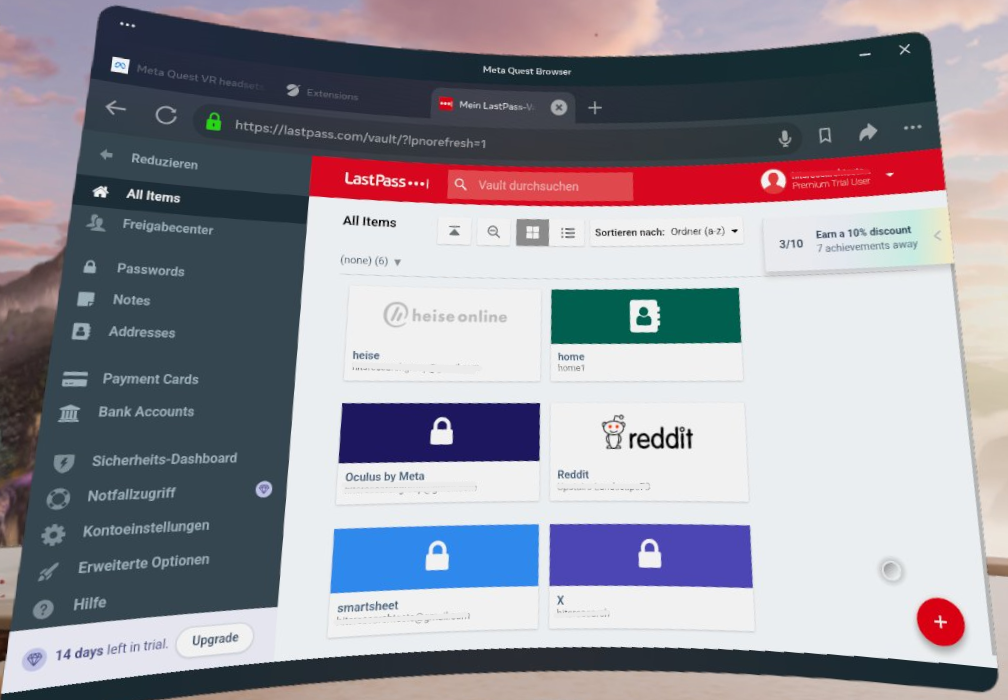} \\ 
      \caption*{\small (b) LastPass web \cite{lastpasswebtested}}
    \end{minipage}
    \begin{minipage}{0.3\textwidth}
      \centering
      \includegraphics[width=\linewidth]{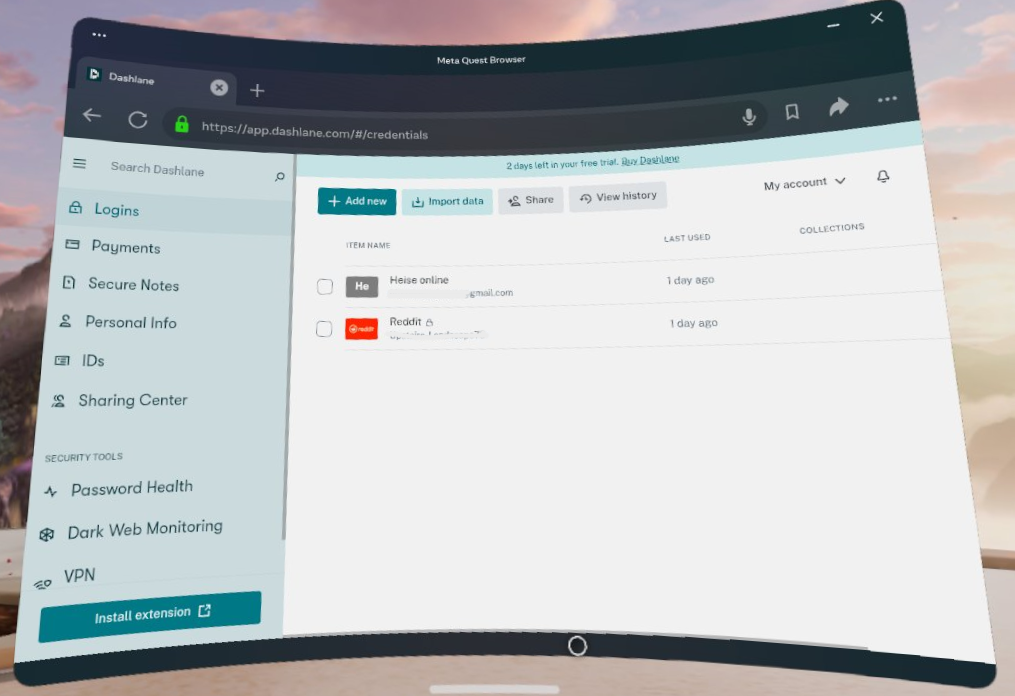} \\ 
      \caption*{\small (c) Dashlane web \cite{dashlanetested}}
    \end{minipage}
  \end{minipage}
  
  \begin{minipage}{0.9\textwidth}
    \centering
    \begin{minipage}{0.3\textwidth}
      \centering
      \includegraphics[width=\linewidth]{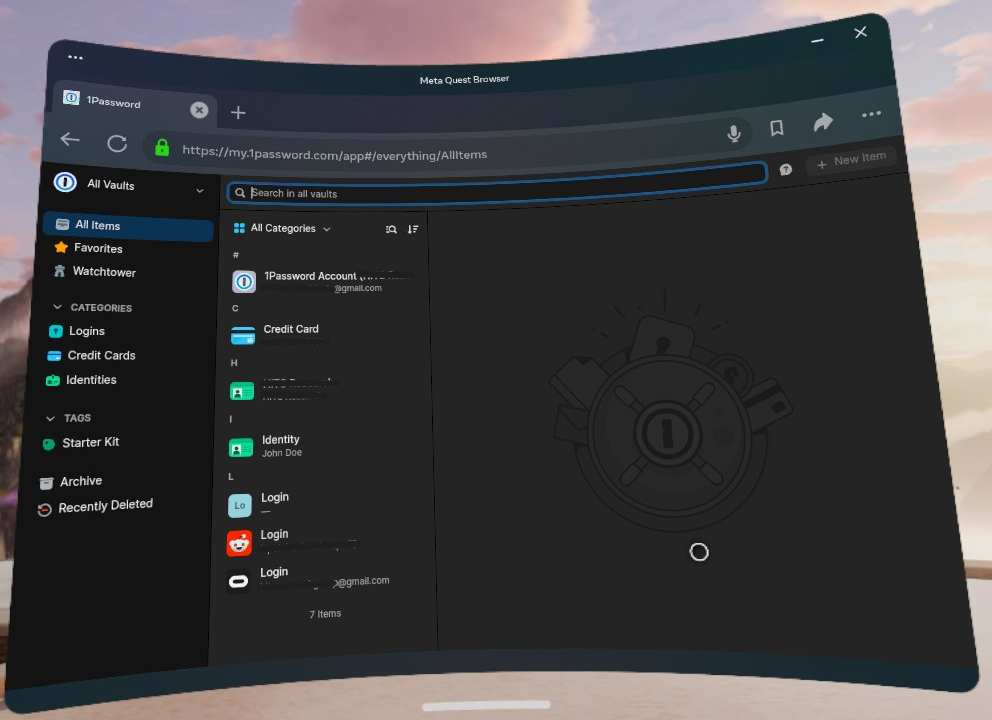} \\ 
      \caption*{\small (d) 1Password web \cite{1passwordtested}}
    \end{minipage}
    \begin{minipage}{0.3\textwidth}
      \centering
      \includegraphics[width=\linewidth]{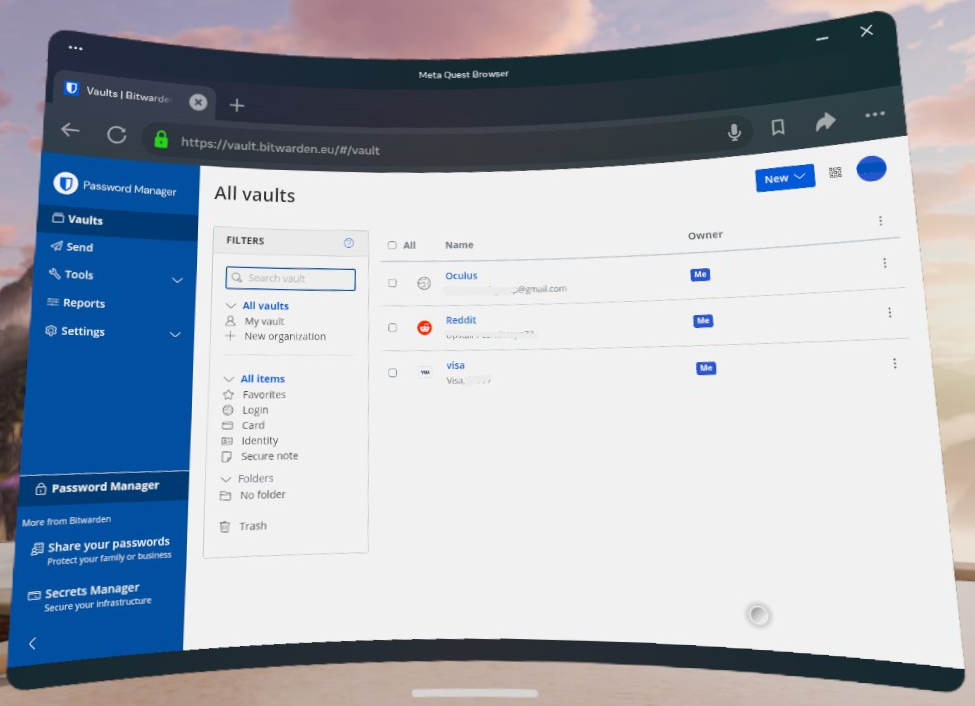} \\ 
      
      \caption*{\small (e) Bitwarden web \cite{bitwardentested}}
    \end{minipage}
    \begin{minipage}{0.3\textwidth}
      \centering
      \includegraphics[width=\linewidth]{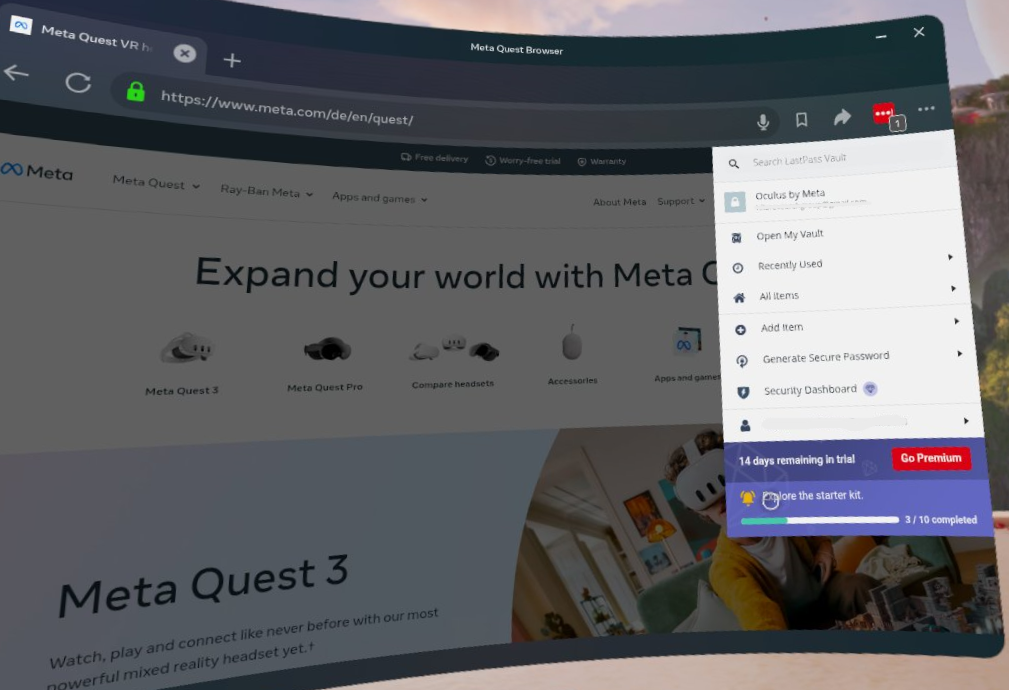} \\ 
      \caption*{\small (f) LastPass extension}
    \end{minipage}
  \end{minipage}
  
  \begin{minipage}{0.9\textwidth}
    \centering
    \begin{minipage}{0.3\textwidth}
      \centering
      \includegraphics[width=\linewidth]{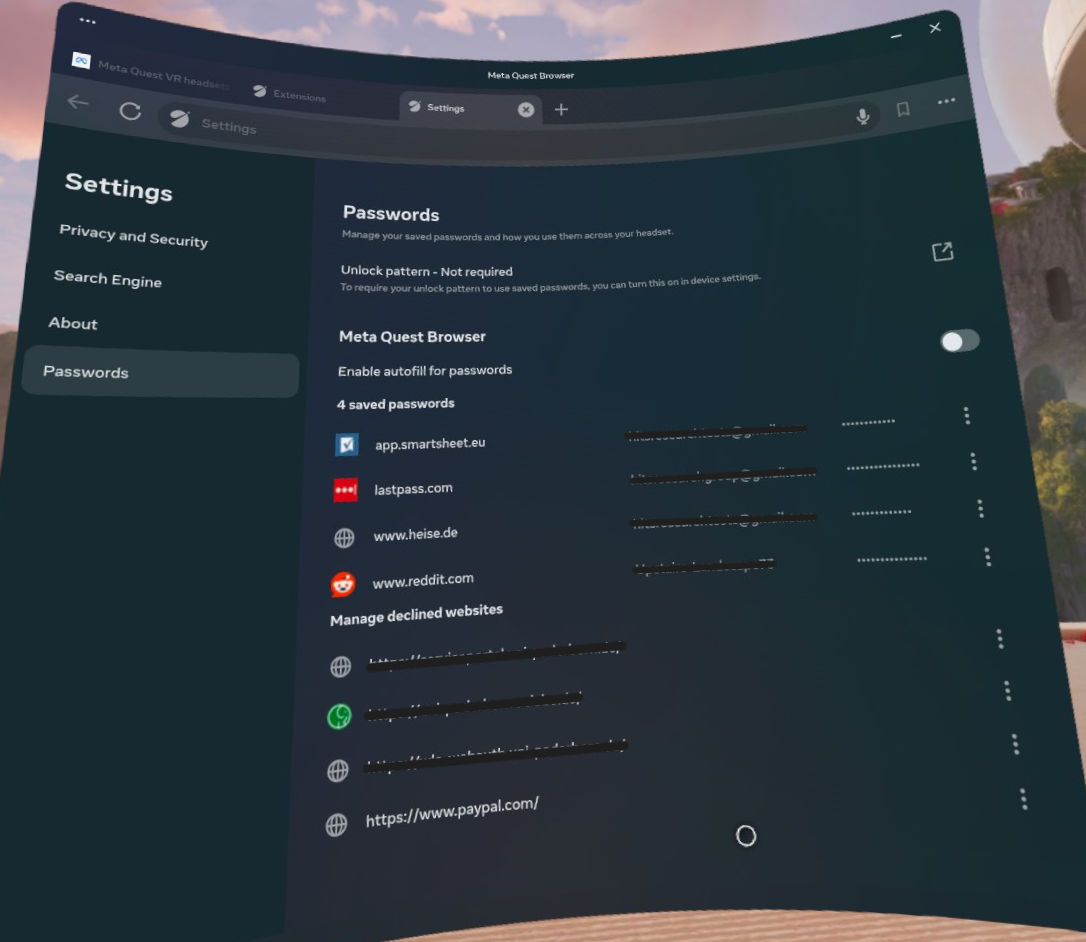} \\ 
      \caption*{\small (g) Meta Quest browser PM}
    \end{minipage}
  \end{minipage}
  
  \caption{The seven password managers in VR that were reviewed by experts.}
    \label{fig:pms_grid}
\end{figure}

\begin{figure}[htb]
    \centering
    \includegraphics[width=1\linewidth]{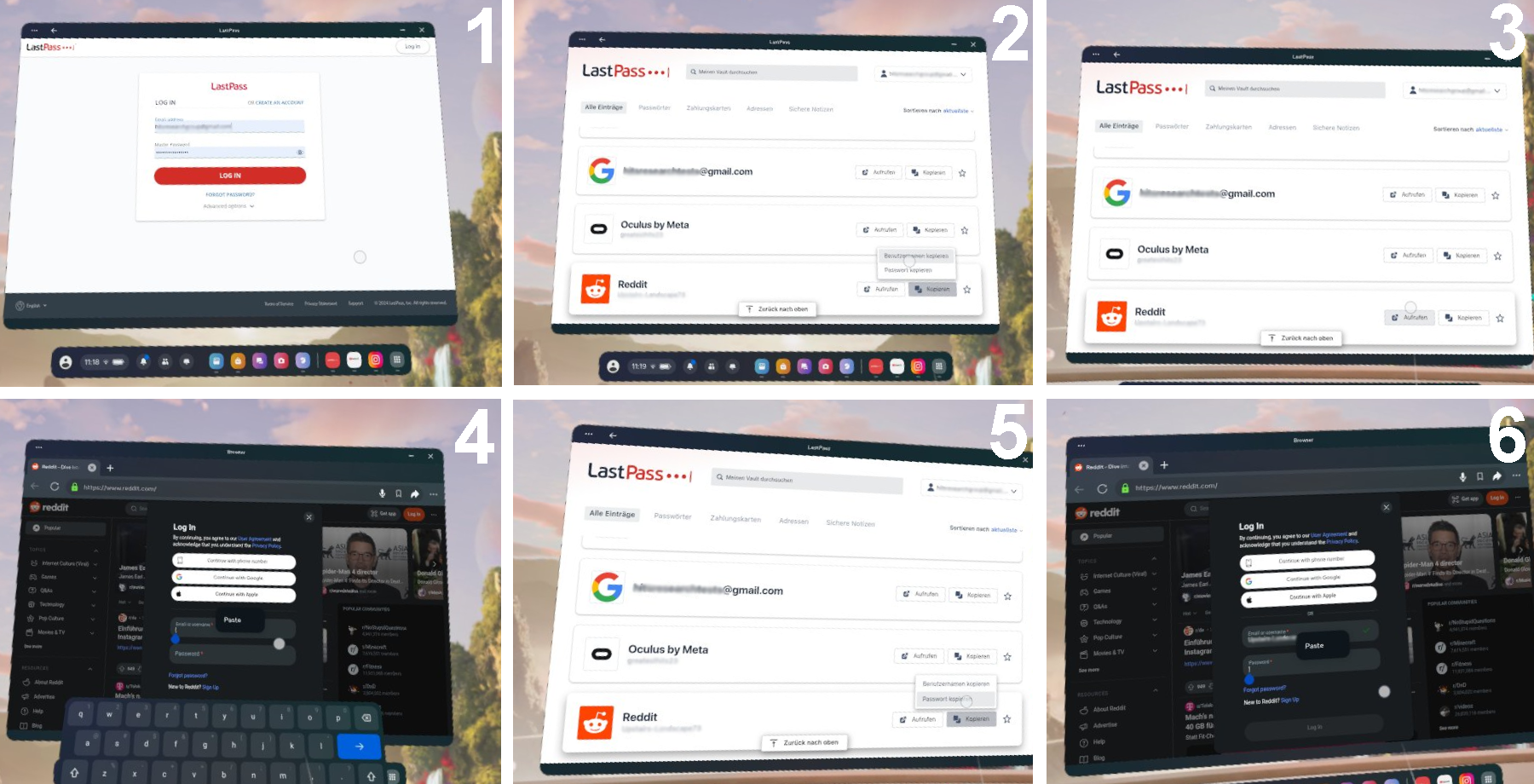}
    \caption{Steps required to authenticate into Reddit using the LastPass VR app. Autofill is not available.
    1) Logging into the password manager with credentials. 2) Vault overview and copying the username. 3) Launching the website. 4) Opening the website in the browser and pasting in the username. 5) Going back to LastPass to copy password. 6) Switching back to the browser for pasting in the password and log in.}
    \label{fig:flow_lastpass}
\end{figure}

\begin{figure}[htb]
    \centering
    \includegraphics[width=0.75\linewidth]{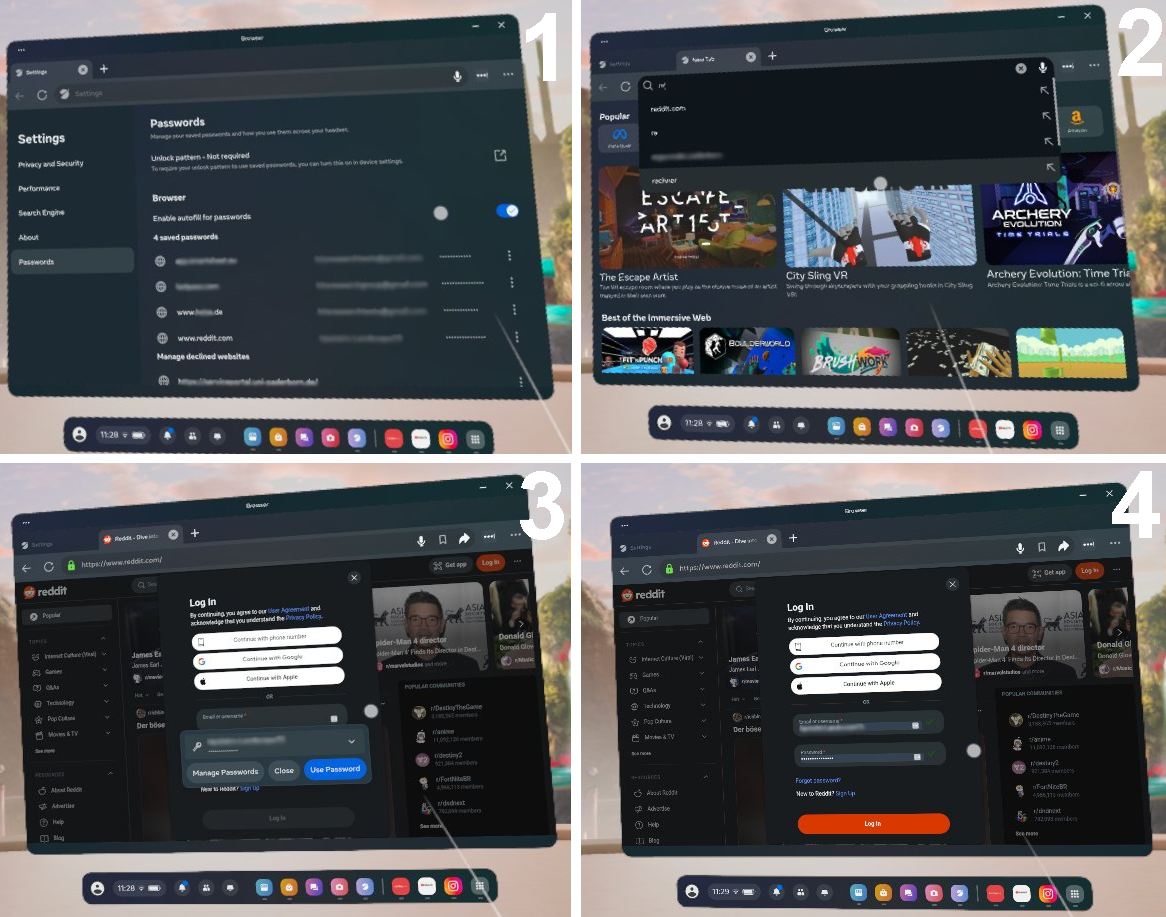}
    \caption{Steps required to authenticate into Reddit using the Meta Quest Browser password manager. Autofill is available.
    1) Activating the password manager in the browser settings, no login required. 2) Manually entering the URL in the browser to launch the website. 3) Clicking into the username field to get the autofill request by the PM. 4) Once the stored credentials are selected, the fields are filled, and the user is ready to log in.}
    \label{fig:flow_browser}
\end{figure}

\begin{figure}[htb]
    \centering
    \includegraphics[width=1\linewidth]{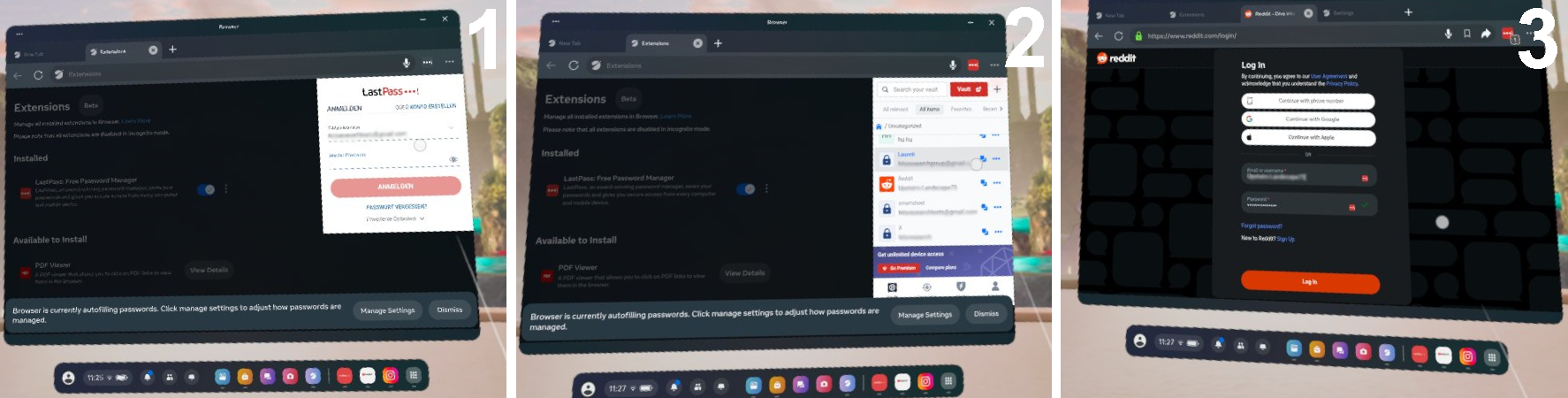}
    \caption{Steps required to authenticate into Reddit using the LastPass browser extension. Autofill is available.
    1) Logging into the extension with credentials, after clicking on the extension icon in the browser menu bar. 2) Vault overview and launching the website of the stored credentials. 3) The website is launched with the credentials already autofilled in the fields. }
    \label{fig:flow_extension}
\end{figure}

\section{Password Manager Versions}
\label{app:pmversions}

    \begin{tabular}{ccccccc}
        \toprule
            \bfseries\makecell{LastPass\\VR app \cite{lastpassvr}} & \bfseries\makecell{Meta Quest\\browser PM} & \bfseries\makecell{LastPass\\extension} & \bfseries\makecell{LastPass\\web \cite{lastpasswebtested}} & \bfseries\makecell{Bitwarden\\web \cite{bitwardentested}} & \bfseries\makecell{Dashlane\\web \cite{dashlanetested}} & \bfseries\makecell{1Password\\web \cite{1passwordtested}}\\
        \midrule
            1.0 & \makecell{34.4.0.51.164.632504017\\(Meta Quest Browser)} & 4.127.0 &22/11/2024 & 22/11/2024 & 22/11/2024&22/11/2024\\
        \bottomrule
    \end{tabular}

\end{document}